\documentclass[fleqn,usenatbib]{mnras}

\usepackage{newtxtext,newtxmath}

\usepackage[T1]{fontenc}

\DeclareRobustCommand{\VAN}[3]{#2}
\let\VANthebibliography\thebibliography
\def\thebibliography{\DeclareRobustCommand{\VAN}[3]{##3}\VANthebibliography}

\usepackage{graphicx}	
\usepackage{amsmath}	
\usepackage[utf8]{inputenc}
\usepackage{parskip}
\usepackage[pdftex,dvipsnames]{xcolor}
\usepackage{cite}
\usepackage{appendix}
\usepackage{comment}
\usepackage{changepage}
\usepackage[pagewise]{lineno}
\usepackage{natbib}
\usepackage{xspace} 
\usepackage{soul} 
\usepackage{hyperref}
\usepackage{eso-pic}

\AddToShipoutPictureBG*{%
  \AtPageUpperLeft{%
    \hspace{0.72\paperwidth}%
    \raisebox{-4.5\baselineskip}{%
      \makebox[0pt][l]{\textnormal{DES-2022-0715}}
}}}%

\AddToShipoutPictureBG*{%
  \AtPageUpperLeft{%
    \hspace{0.72\paperwidth}%
    \raisebox{-5.5\baselineskip}{%
      \makebox[0pt][l]{\textnormal{FERMILAB-PUB-22-877-PPD}}
}}}%

\newcommand{\balrog}{\textsc{Balrog}~}
\newcommand{\maglim}{\textsc{MagLim}~}
\newcommand{\redmagic}{\textsc{redMaGiC}~}
\newcommand{\healpix}{\textsc{HealPix}~}

\newcommand{\cosmos}{\texttt{COSMOS2015}~}
\newcommand{\paucosmos}{\texttt{PAUS+COSMOS}~}
\newcommand{\highz}{high-$z$~}
\newcommand{\sdir}{\textsc{3sDir}\xspace}
\newcommand{\Alowz}[1]{A^{#1}_{{\rm low}-z}}
\newcommand{\Delz}[1]{\Delta z^{#1}}
\newcommand{\Sigz}[1]{\sigma_{z}^{#1}}
\newcommand{\meanz}{\langle z \rangle}

\title[High-$z$ in DES]{\centering \LARGE 
The Dark Energy Survey Year 3 high redshift sample: \\Selection, characterization and analysis of galaxy clustering}

\author[DES Collaboration]{
\parbox{\textwidth}{
\Large
C.~S{\'a}nchez,$^{1,2}$\thanks{Corresponding author: carles.sanchez.alonso@gmail.com}
A.~Alarcon,$^{3}$\thanks{Corresponding author: alexalarcongonzalez@gmail.com}
G.~M.~Bernstein,$^{2}$
J.~Sanchez,$^{4}$
S.~Pandey,$^{2}$
M.~Raveri,$^{5}$
J.~Prat,$^{6,7}$
N.~Weaverdyck,$^{8,9}$
I.~Sevilla-Noarbe,$^{10}$
C.~Chang,$^{6,7}$
% ...
\begin{center} \vspace{1mm}(DES Collaboration) \end{center} 
\vspace{2mm}
\small{\emph{Author affiliations are shown at the end of this paper.}}
}
}

\pubyear{2022}

\begin{document}
\label{firstpage}
\pagerange{\pageref{firstpage}--\pageref{lastpage}}
\maketitle

\begin{abstract}

\vspace{-3mm}
\begin{adjustwidth}{15pt}{15pt}
The fiducial cosmological analyses of imaging galaxy surveys like the Dark Energy Survey (DES) typically probe the Universe at redshifts $z<1$. This is mainly because of the limited depth of these surveys, and also because such analyses rely heavily on galaxy lensing, which is more efficient at low redshifts. In this work we present the selection and characterization of high-redshift galaxy samples using DES Year 3 data, and the analysis of their galaxy clustering measurements. In particular, we use galaxies that are fainter than those used in the previous DES Year 3 analyses and a Bayesian redshift scheme to define three tomographic bins with mean redshifts around $z\sim0.9$, $1.2$ and $1.5$, which significantly extend the redshift coverage of the fiducial DES Year 3 analysis. These samples contain a total of about 9 million galaxies, and their galaxy density is more than 2 times higher than those in the DES Year 3 fiducial case. We characterize the redshift uncertainties of the samples, including the usage of various spectroscopic and high-quality redshift samples, and we develop a machine-learning method to correct for correlations between galaxy density and survey observing conditions. The analysis of galaxy clustering measurements, with a total signal-to-noise $S/N \sim 70$ after scale cuts, yields robust cosmological constraints on a combination of the fraction of matter in the Universe $\Omega_m$ and the Hubble parameter $h$, $\Omega_m h = 0.195^{+0.023}_{-0.018}$, and 2-3\% measurements of the amplitude of the galaxy clustering signals, probing galaxy bias and the amplitude of matter fluctuations, $b \sigma_8$. A companion paper (\emph{in preparation}) will present the cross-correlations of these \highz samples with CMB lensing from \emph{Planck} and SPT, and the cosmological analysis of those measurements in combination with the galaxy clustering presented in this work. 
\\ \\
\end{adjustwidth}

\end{abstract}

\vspace{2mm}



\section{Introduction}
\label{sec:intro}

The combination of large-scale structure (LSS) and weak gravitational lensing (WL) constitutes one of the main avenues to study cosmology and to stress test the standard cosmological model. In recent years, several imaging surveys such as the Hyper Suprime-Cam (HSC\footnote{\href{https://hsc.mtk.nao.ac.jp/ssp/}{\texttt{hsc.mtk.nao.ac.jp/ssp/}}}), the Kilo-Degree Survey (KiDS\footnote{\href{http://kids.strw.leidenuniv.nl}{\texttt{kids.strw.leidenuniv.nl}}}), and the Dark Energy Survey (DES\footnote{\href{https://www.darkenergysurvey.org}{\texttt{darkenergysurvey.org}}}), analyzing data from more than 100 million galaxies, have used galaxy weak lensing to produce cosmological constraints that rival in precision those from CMB experiments like Planck (see \citealt{2019PASJ...71...43H,2021A&A...646A.140H,y3-3x2ptkp} and references therein). These analyses have reported tensions between the amplitude of structures at late time and the predictions from the CMB (the so-called “$S_8$ tension”). However, the majority of these analyses probe the Universe at low redshifts, $z<1$. There exist at least three reasons for this. First, due to the faint nature of high redshift galaxies, it is difficult for imaging surveys to characterize such populations, both in terms of redshift distributions and also in terms of mapping the effect of spatially varying observing conditions on the selection function. Second, it is challenging to measure shapes of high-redshift sources for galaxy lensing at sufficient signal-to-noise. And third, even if those galaxy sources could be defined, their lensing signals are still most sensitive to mass structure at $z<1$. On the other hand, if one can get around the first of these issues and characterize high-redshift lens galaxy samples, then the use of CMB lensing will provide a solution for the second and third problems. 



The definition and characterization of galaxy samples at higher redshifts would enable a more optimal combination with CMB lensing, whose sensitivity peaks around $z=2$ and drops significantly at redshifts $z<1$. In this way, a combination of galaxy clustering and CMB lensing at high redshift would be key to cosmology in several ways. On the one hand, the regime at redshifts $z \geq 1.5$ remains largely unexplored by galaxy surveys in the context of the $S_8$ tension, and various alternative dark energy models predict deviations from the standard model at high redshifts \citep{Bull2021}, which could be tested in this way. On the other hand, being able to make this measurement is important to constrain large-scale observables like primordial non-Gaussianity, which would open the window to the physics of the early inflationary period sourcing the large-scale structures we see in the Universe today \citep{Schmittfull2018}. Furthermore, CMB lensing is subject to different systematic errors than galaxy lensing---the former measurement is not affected by intrinsic alignments or galaxy blending, and the redshift of the CMB is well known as opposed to the case of galaxy sources. 

There exist numerous previous analyses that have explored the combination of galaxy clustering and CMB lensing to probe cosmology at redshifts $z<1$ \citep{DESY1_6x2pt_cosmo,2020JCAP...05..052M,2021MNRAS.501.1481H,2021MNRAS.502..876A,2022arXiv220312440C}. Some analyses have also used the combination to probe cosmology at higher redshifts. In particular, the analysis of the unWISE sample \citep{unWISEcatalog,unWISE1,unWISE2} provided such measurements in three broad redshift bins, the last one with a median redshift around $z=1.5$. Also, the HSC survey has explored much higher redshift regimes using dropout galaxies over smaller areas \citep{HSC_dropouts_I_luminosity, HSC_dropouts_II_clustering}, probing the Universe at the $4<z<7$ regime  \citep{2021arXiv210315862M}. 


For the particular case of the Dark Energy Survey, the analysis of Year 3 (Y3) data has so far used two different lens galaxy samples, \textsc{MagLim} and \textsc{redMaGiC} \citep{y3-2x2ptaltlensresults,y3-2x2ptbiasmodelling,y3-3x2ptkp}. The \textsc{MagLim} sample is a magnitude-limited galaxy selection, split into six redshift bins using the Directional Neighborhood Fitting (DNF) algorithm \citep{DeVicente2016}, and the first four bins of the sample, covering an approximate redshift range $0<z<1$, were used as the fiducial lens sample in the DES Y3 analysis. The \textsc{redMaGiC} \citep{Rozo2015} is a sample of bright Luminous Red Galaxies (LRGs), covering a similar redshift range in five redshift bins, and was used in Y3 as an alternative lens sample. 


In this work we push the limits of the DES Y3 data to explore the regime at redshift $z>1$. To this end, we select and characterize ``high-redshift'' (high-$z$) samples of galaxies in the DES wide-field survey.  This includes the estimation of the redshift distributions of the samples and their uncertainties, corrections for variations in completeness across the survey footprint due to varying observing conditions, and characterization of the lens magnification coefficients of the samples.  The definition and characterization of these \highz samples differs from the process used for the fiducial DES Y3 lens samples \citep{y3-2x2ptaltlensresults,y3-2x2ptbiasmodelling} in several ways:
\begin{enumerate}
    \item We start from a fainter galaxy selection, already excluding all lens galaxies used in the DES Y3 fiducial analysis.
    \item Both the selection and redshift characterization of the samples are based on a Bayesian scheme using \emph{Self-Organizing Maps} (SOMs), and we use a new SOM algorithm, better suited for lower S/N galaxies (different than that used in \citealt{y3-sompz}).
    \item We use a different redshift marginalization scheme, explicitly accounting for uncertainties in low-redshift tails of the redshift distributions. 
    \item We use a non-linear, machine-learning-based approach to account for correlations in the galaxy number density with survey observing properties.
\end{enumerate}

Steps (i) and (ii) are the ones responsible for the selection of high redshift galaxies, while steps (iii) and (iv) are necessary because of that faint, high redshift selection. The definition and characterization of the \highz sample in this work is followed by the analysis of the clustering measurements of the galaxies in the sample. The clustering measurements are used to place constraints on the cosmological model, in particular as the shape of the clustering signal is sensitive to the scale of matter - radiation equality in the mass power spectrum, which in turn depends on a combination of the matter density $\Omega_m$ and the Hubble constant $h$, close to the direction $\Omega_m h$ (see e.g.~\citealt{2021PhRvD.103b3538P}).

The \highz samples defined in this work, given their redshift range and sky density, will make excellent lens galaxy samples for CMB lensing. In this way, this paper will be followed by a companion paper (\emph{in preparation}) that will present the cross-correlations between these \highz samples and CMB lensing from Planck \citep{2020A&A...641A...8P} and the South Pole Telescope\citep{2011PASP..123..568C}, and use the combination of clustering and CMB lensing to place constraints on the cosmological model using information from high redshift. 

This paper is organized as follows. Section \ref{sec:data} describes the different data products used for the analysis. Section \ref{sec:method} describes the redshift inference scheme and the method to select tomographic bins. Section \ref{sec:weights} describes the way we correct for correlations between galaxy density and survey observing properties. Section \ref{sec:redshift_uncertainty} presents the characterization of redshift uncertainties, and the parametrization we use to marginalize over them in the clustering analysis. Section \ref{sec:magnification} describes the characterization of lens magnification for the \highz samples. Finally, section \ref{sec:clustering} presents the measurements and analysis of galaxy clustering, and we conclude in Section \ref{sec:summary}. 

\section{Data}
\label{sec:data}

In this section we describe and motivate the different data samples to be used in this work. We begin with the DES Y3 wide-field data, which will contain our \highz samples, and then describe other data sets needed for the characterization of those samples: the DES deep-field data, and the external data used for redshift characterization. 

\subsection{DES wide-field data}
\label{sec:wide}

The \highz samples are subsets of the DES Year 3 Gold catalogue of photometric objects \citep{y3-gold}, which has a total of nearly 400 million objects in  about 5000 sq.~deg.~of area, covering the entire DES footprint. After removing stars
and applying quality cuts (following \citealt{y3-gold}), the catalog consists of $\sim$ 227 million galaxies. For these objects, we use single-object-fitting (SOF) photometry in the $griz$ bands, which have magnitude limit (defined as the average SOF magnitude at $S/N = 10$) of 23.8, 23.6, 23.0 and 22.4, respectively. We apply an initial $i$-band magnitude ``pre-selection'' of $22 < i < 23.5$. The lower limit of this cut removes bright galaxies that are unlikely to be at redshifts $z>1$, and the faint limit excludes the region of magnitude space where the DES Y3 Gold catalog becomes highly incomplete.
Please note that, even with the $i<23.5$ cut, this selection includes galaxies measured with $S/N < 10$ in the $i$ band, pushing the limits of the DES Y3 sample, and therefore the completeness of the sample has significant spatial variations. The characterization of that spatial completeness is a key aspect of this work, and is described in Section \ref{sec:weights}. 

For the pre-selected sample, we apply the standard DES Y3 mask, which includes masking of astrophysical foregrounds (e.g.~bright stars and large nearby galaxies) and of regions with recognized data processing issues, as described in \citet{y3-gold}. Given that we are pushing the limits of DES Y3 photometry, we apply some additional conservative cuts on the mask to avoid regions where our completeness corrections would be less reliable: we remove the 3\% of the footprint area with the highest stellar density, the 3\% with the highest (worst) $g$-band seeing, and then we remove the worst 10\% area in photometric depth, exposure times and sky brighteness in each of the $griz$ bands, some of which are correlated. After applying this mask, the $22<i<23.5$ pre-selected galaxy sample has a total of 77 million galaxies in 2621 sq.~deg.~of area. For comparison, the fiducial DES Y3 analysis uses 4143 sq.~deg. of total area. 

The analysis presented here will be followed by a companion paper (\emph{in preparation}) that will combine the clustering measurements shown here with CMB lensing measurements from the \emph{Planck} satellite and the South Pole Telescope (SPT). Due to SPT data being available only in the south region of the DES Y3 footprint, we will split the sample in this work into two independent regions, "North" (DEC $> -39^{\rm o}$) and "South" (DEC $< -40^{\rm o}$), and test for the consistency of the two. For that test, we choose to leave a separation of 1 degree between the two regions, which corresponds to the maximum angular separation used later on in the galaxy clustering measurements. A similar separation of the DES footprint was made in the analyses studying CMB lensing for the fiducial DES Y3 sample \citep{DESY1_6x2pt_cosmo, DESY1_dxkCMB, DESY1_kxkCMB, DESY1_6x2pt_methods}.

\subsection{DES deep-field data and artificial wide-field data}
\label{sec:data_deep}

The scheme for redshift selection and characterization, described in detail in Section \ref{sec:method}, makes extensive use of DES deep-field data, described extensively in \citet{y3-deepfields}. In short, we use four \textit{deep} fields, named E2, X3, C3, and COSMOS (COS), covering areas of 3.32, 3.29, 1.94, and 1.38 square degrees, respectively (see Fig. 2 in \citealt{y3-sompz} for a visual description). After masking regions with artefacts such as cosmic rays, artificial satellites, meteors, asteroids, and regions of saturated pixels, 5.2 square degrees of overlap with the UltraVISTA and VIDEO near-infrared (NIR) surveys \citep{McCracken2012,Jarvis2013} remain. This yields 2.8M detections with measured $ugrizJHK_s$ photometry with limiting magnitudes 24.64, 25.57, 25.28, 24.66, 24.06, 24.02, 23.69, and 23.58, substantially fainter than the faintest galaxies in the sample of source galaxies. In this work we frequently refer to this sample and its photometry as \textit{deep (field)} data.

So far we have described the wide-field DES data to be used over the full footprint and a set of deep-field photometry over a smaller area. In order to establish the relationship between these two data sets we use the \balrog \citep{Suchyta2016} software, which injects simulated galaxies based on the DES deep fields into real images from DES wide-field observations. For this analysis, \balrog was used to inject model galaxies, with profiles fit to deep-field galaxies, into the wide-field footprint \citep{y3-balrog}. After injecting galaxies into images, the output is analyzed by the DES Y3 photometric pipeline \citep{2018PASP..130g4501M}. Each deep-field galaxy is injected multiple times at different positions in the footprint.
The resulting matched catalogue of 3,194,291 injection-realization pairs, which contains both deep and wide photometric information, is a key part of our redshift calibration scheme since it quantitatively connects the two photometric spaces. This catalogue will be referred to as the \textit{Deep/\balrog} Sample, and contains a total
of 432,657 unique deep-field galaxies having at least 1 \balrog realization that passes the wide-field selection criteria.

\begin{figure}
 \centering
 \includegraphics[width=0.5\textwidth]{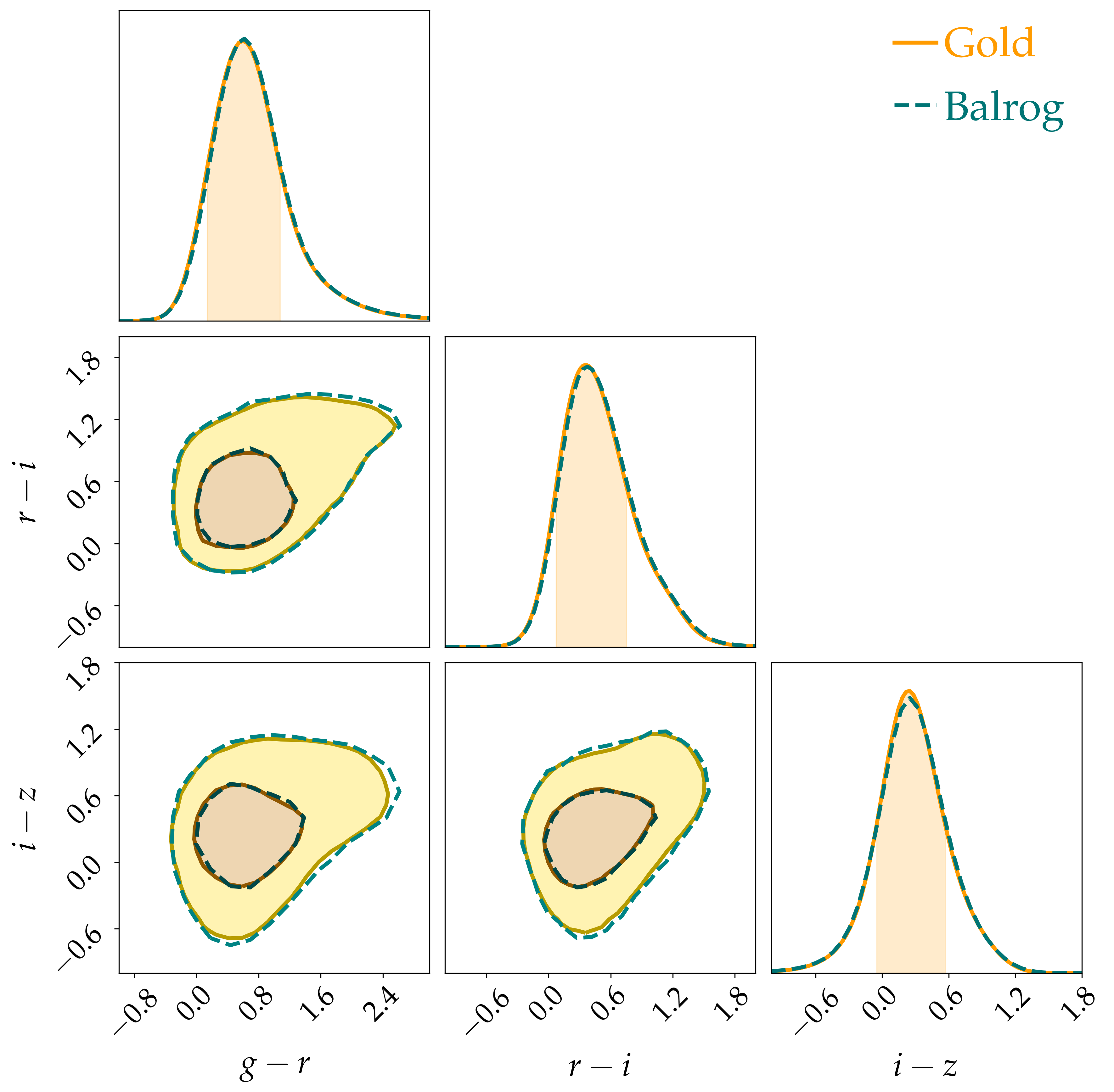}
 \caption{The distribution of photometric colors in the DES wide-field $griz$ bands, after a pre-selection cut of $22<i<23.5$, using the data (Gold) and the artificial data (Balrog). }
 \label{fig:gold_balrog}
\end{figure}

Because we will use the \textsc{Balrog} sample to establish the relationship between wide and deep photometry in DES Y3, it is important that \balrog wide-field detections follow similar photometric distributions to the actual DES Y3 wide-field data in the Gold sample. Figure \ref{fig:gold_balrog} shows the distribution of colors in the DES Y3 photometry for the data (Gold) and for the artificial realizations of deep galaxies (Balrog) for the pre-selected sample described in \S\ref{sec:wide} ($22<i<23.5$). As desired, the color distributions of data and artificial realizations of deep galaxies are in excellent agreement. 

\subsection{Redshift data} \label{sec:redshiftdata}

Our analysis relies on the use of galaxy samples with known redshift and deep-field photometry. To this end, we use catalogues of both high-resolution spectroscopic and multi-band photometric redshifts, and we develop an experimental design that allows us to test uncertainty in our redshift calibration due to biases in these samples. The spectroscopic catalogue we use contains both public and private spectra from the following surveys: zCOSMOS \citep{Lilly09zcosmos}, C3R2 \citep{Masters2017,C3R2_DR2, C3R2_DR3}, VVDS \citep{vvds}, and VIPERS \citep{vimos}. We use two multi-band photo-$z$ catalogues from the COSMOS field \citep{Scoville2007_COSMOS}:  the \cosmos 30-band photometric redshift catalogue \citep{Laigle2016}, which includes 30 broad, intermediate, and narrow bands covering the UV, optical, and IR regions of the electromagnetic spectrum, and the \paucosmos 66-band photometric redshift catalogue \citep{Alarcon2020} from the combination of PAU Survey data \citep{paucam,eriksen2019} in 40 narrow-band filters and 26 \cosmos bands excluding the mid-infrared. We build a redshift calibration sample in the deep fields from the overlapping redshift information we find in these surveys. We prioritize information coming from spectroscopic surveys (S), then \paucosmos (P) and finally \cosmos (C), and we call this redshift sample SPC\footnote{An identical notation was used in \citet{y3-sompz}.}.

\section{Redshift methodology}
\label{sec:method}

This section describes our redshift inference scheme, which allows us to select and characterize samples of \highz galaxies using the data described in the previous section. The next sections will describe the characterization of the uncertainties in the angular and redshift distributions of these \highz samples. 

We work under the framework presented in \citet{Sanchez2018}, in which galaxy “types” are defined by observed properties rather than rest-frame properties, and we call them \textit{phenotypes}. We will use the low-noise, several-band photometry available in the deep fields to define our phenotypes, and we will discretize such photometry using a \textit{Self-Organizing Map} (SOM, \citealt{Kohonen1982,Masters2015}). In this way, every cell in the Deep SOM will be a phenotype, and we will index them with $c$. This approach, proposed initially in \citet{Sanchez2018}, has now been successfully used in several analyses both using simulations \citep{Buchs2019,HBM_clustering} and real data \citep{y3-sompz,y3-2x2ptaltlenssompz}.  

We also discretize the wide-field photometry into a SOM, with wide cells indexed by $\hat{c}$. With this discretized mapping of deep and wide photometric spaces, we can estimate the redshift distribution of a given wide cell $\hat{c}$, passing a wide selection $\hat{s}$, by marginalizing over deep-field information $c$: 
%
\begin{equation}
p(z|\hat{c},\hat{s}) = \sum_{c} p(z|c, \hat{c}, \hat{s}) p(c|\hat{c},\hat{s}).
\label{eq:wide_pz}
\end{equation}

The first term on the right contains information about the redshift of deep phenotypes, while the second term connects the deep and wide photometric spaces. Having the expression for the redshift distribution of a wide cell, we can construct a sample of galaxies by joining wide cells $\hat{c}$ into tomographic bins $\hat{b}$, and their redshift distribution will simply become the sum of its constituents weighted by the occupation of wide cells: 

\begin{align}
    p(z|\hat{b}, \hat{s}) &= \sum_{\hat{c} \in \hat{b}} p(z|\hat{c}, \hat{s}) p(\hat{c}|\hat{s}, \hat{b})\label{eq:wide_nz2} \\
    &\propto \sum_{\hat{c} \in \hat{b}} \sum_{c} p(z|c, \hat{c}, \hat{s}) p(c|\hat{c},\hat{s})p(\hat{c}|\hat{s})\label{eq:wide_nz3}\\
    &\approx \sum_{\hat{c} \in \hat{b}} \sum_{c} p(z|c, \hat{b}, \hat{s}) p(c|\hat{c},\hat{s})p(\hat{c}|\hat{s}). \label{eq:wide_nz}
\end{align}
Going from (\ref{eq:wide_nz2}) to (\ref{eq:wide_nz3}) we use the fact that $p(\hat c | \hat b, \hat s) = p(\hat c | \hat s) / \left(\sum_{\hat{c} \in \hat{b}} p(\hat c | \hat s)\right) \,\mathrm{for}\,\hat{c} \in \hat{b}$, and and in the last line we approximate $p(z|c, \hat{c}, \hat{s}) \approx p(z|c, \hat{b}, \hat{s}.)$  The need for conditioning on bin membership rather than wide-cell measurement (going from Eq.~\ref{eq:wide_nz3} to Eq.~\ref{eq:wide_nz}), and the accuracy of this approximation, will be investigated in Section~\ref{sec:redshift_uncertainty} and Appendix~\ref{sec:appendix_pz_bce}. 
The final expression computes the redshift distribution of tomographic bins made of wide-field SOM cells. We use different samples to estimate the different terms in it, as we describe next:
\begin{enumerate}
    \item $p(\hat{c}|\hat{s})$ is computed from our wide sample, which consists of all galaxies in the DES Year 3 Gold catalog passing the pre-selection performed in Section \ref{sec:data} ($22<i<23.5$).
    \item $p(c|\hat{c},\hat{s})$ is computed from our Deep and \balrog Samples, which consist of all detected and selected \balrog realisations of the galaxies in the Deep Sample. We call this term the \textit{transfer function}.
    \item $p(z|c, \hat{b}, \hat{s})$ is computed from the Redshift Sample subset of the Deep Sample, for which we have reliable redshifts, 8-band deep photometry, and wide-field \balrog realisations\footnote{This term could, in principle, be computed from the overlapping photometry of the deep and wide fields. However the region where these samples overlap is small and it is not representative of the observing conditions found across the whole survey footprint, which are much more well sampled by making use of \balrog.}.
\end{enumerate}

The redshift scheme followed in this work is similar to that used in \citet{y3-sompz} for the selection and characterization of weak lensing source galaxy samples, but there exist some important differences:
\begin{itemize}
\item We perform a pre-selection cut on our sample of $22<i<23.5$, to remove bright galaxies at low redshift and low $S/N$ faint galaxies, cutting the bright end of the $18.5<i<23.5$ used in \citet{y3-sompz}.
    \item In this work we use DES $griz$ wide photometry, while the analysis in \citet{y3-sompz} uses $riz$ information only. 
    \item We also use a different SOM algorithm, improved to better handle the classification of lower-$S/N$ galaxies. This will be described in detail in \S\ref{sec:deep_som}.
    \item The tomographic bins for this work are selected using both the mean redshifts of the Wide SOM cells and also their estimated low-redshift fraction, to avoid having large low-$z$ tails in the tomographic bins. The selection in \citet{y3-sompz} relies only on mean redshift information. 
\end{itemize}

\subsection{The Deep SOM}
\label{sec:deep_som}

\begin{figure*}
 \centering
 \includegraphics[width=0.98\textwidth]{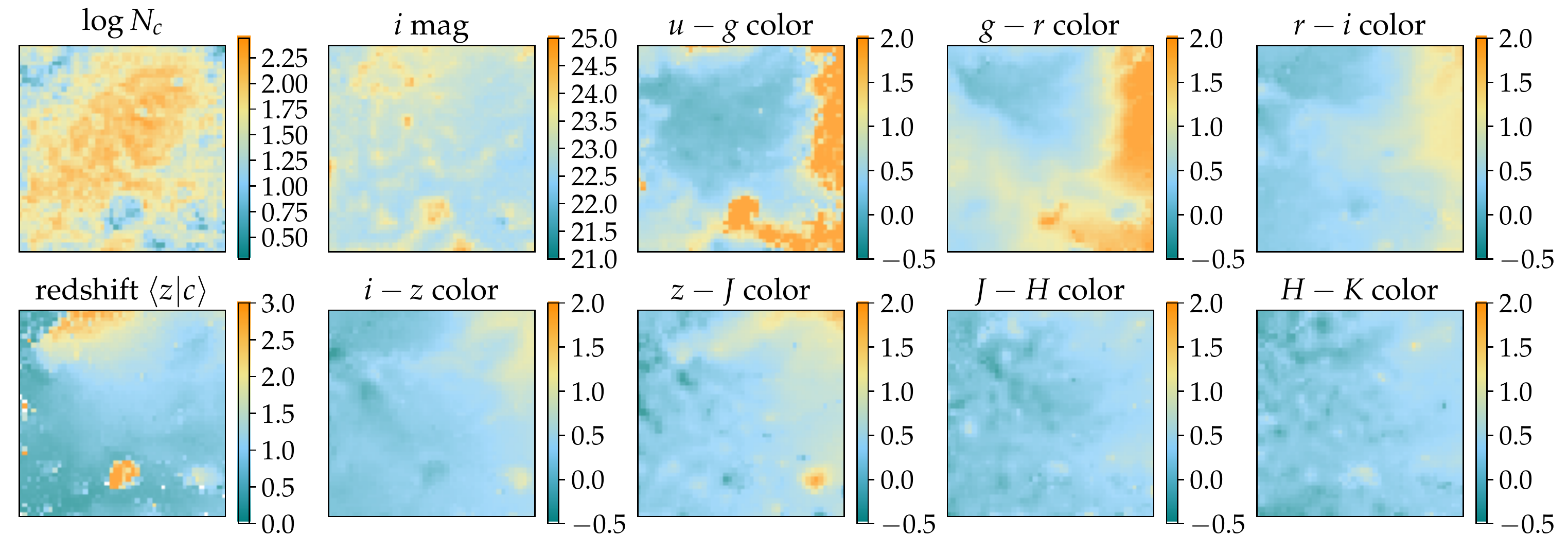}
 \caption{Visualization of various properties of the Deep SOM described in \S\ref{sec:deep_som}. In particular, we show the photometric properties of the map, namely the mapping of $i$-band magnitude and seven of the photometric colors, using the DES deep galaxy sample described in \S\ref{sec:data_deep}. We also show the SOM galaxy occupation, $N_c$ and the redshift mapping of the SOM using SPC redshift galaxies matched to DES deep photometry. A zoomed-in version of the Deep SOM redshift mapping is given in Fig.~\ref{fig:meanc_terrain}.}
 \label{fig:deep_som}
\end{figure*}

In this work we a use \textit{Self-Organizing Map} (SOM) to characterize and discretize the deep photometric space, described in \S\ref{sec:data_deep}. The SOM algorithm uses unsupervised learning to project the 8-dimensional deep photometric data ($ugrizJHK_s$) onto a lower-dimensional grid, in our case a 2-dimensional grid, while attempting to preserve the topology of the 8-dimensional space. This means that similar objects in the 8-D space will be grouped together in the SOM, enabling a visual understanding of features, especially in a 2-D SOM. Each of the cells in the Deep-SOM 2-D grid will be considered a galaxy \textit{phenotype} in our scheme.

There is considerable flexibility in the implementation of the SOM algorithm. We alter the SOM algorithm from that used in previous DES analyses \citep[such as][]{y3-sompz, y3-2x2ptaltlenssompz} with the purpose of improving the classification of galaxies of the low- and modest-S/N photometry used in this work.
This is done by altering the distance metric used by the SOM algorithm to incorporate flux uncertainties. We also allow magnitude (or flux) information, not just colors, to be used in redshift estimation, and we do not impose periodic boundary conditions on the map. This SOM algorithm was introduced and is described in detail in the Appendix of \citet{Sanchez2020}.  

There is also flexibility in the size of the SOM. A larger number of SOM cells can improve the representative power of the map, and hence can be used to describe more complex spaces and resolve finer redshift distinctions. Using too many cells can, however, cause overfitting, with the map modeling noisy features of the data. The Deep SOM in this work uses a $48\times48$ SOM. For comparison purposes, the Deep SOM describing the DES Year 3 space in \citet{y3-sompz} was $64\times64$ in size. We use a smaller SOM size since the wide-field pre-selection cut of $22<i<23.5$ we apply to our sample reduces the volume of our wide-field photometric space, and our Deep SOM
only uses deep galaxies whose \textsc{Balrog} injections have passed this criteria at least once  (see \S\ref{sec:data_deep}).  

Figure \ref{fig:deep_som} shows several properties of the Deep SOM used in this work\footnote{A previous version of the Deep SOM shown here was originally showing significant areas with no available redshift information. After investigation of the issue, this was found to be due to some stellar contamination in the Deep sample. Using that SOM, and the stellar-galaxy separation of \citet{Laigle2016}, the areas of significant stellar contamination were removed, and a new SOM was trained on the clean sample (see Appendix \ref{sec:appendix_stars} for more details). That new SOM trained on the clean sample is the fiducial SOM shown in Fig.~\ref{fig:deep_som}, and used throughout this work.}. It is worth noting that the particular structure of the map depends on randomized initial conditions and training, but the overall topological structure will be similar across different runs. The figure shows different photometric properties of the SOM, mapping colors and $i$-band magnitude. The $u-g$ color mapping shows how most of the map has a near-constant value of $u-g$, but there are well-defined areas showing strong positive (red) values of $u-g$, corresponding to breaks in the spectrum of galaxies such as the Lyman and Balmer breaks (these behavior is also seen in other SOM analyses such as \citealt{Masters2015}).  The $z-J$ color shows a different structure across the map, showing variation across the regions where $u-g$ was constant and close to zero. We also show the mapping of $i$-band magnitude across the map. In this case, it is worth noting that even though our target sample has a selection of $22<i<23.5$, galaxies fainter than $i=23.5$ have a non-zero probability of being selected in our sample because of noise fluctuations. Since we are including in the Deep SOM all deep galaxies whose artificial injections make the selection at least once, that means that we include some galaxies as faint as $i\simeq25$. 

Figure \ref{fig:deep_som} also shows the Deep SOM galaxy occupation, $n(c)$, the density of galaxies as a function of position in the deep photometric space probed by the SOM. Perhaps most importantly, the lower left panel shows the redshift mapping of the Deep SOM. For this panel we use the subset of deep galaxies that have a match in the SPC redshift sample (described in \S\ref{sec:redshiftdata}), and compute the mean redshift of the galaxies occupying each SOM cell. This plot depicts a smooth mapping of redshift in the SOM, reasonably smoother than the mapping of some colors or magnitudes, even though redshift information is never used in the SOM training. 

Since we are mainly concerned about high redshift in this work, it is interesting to explore the regions of the map that correspond to that regime. There exist two main areas of \highz galaxies in the SOM. There is a first \highz region in the upper part of the SOM, with a smooth gradient to middling redshifts in the central part of the map. Figure \ref{fig:deep_som} shows the upper \highz region to have a small $u-g$ color (no break between the $u$ and $g$ bands), with positive and smoothly varying $z-J$ color, and faint magnitudes in the $i$ band. There is a second ``island'' in the lower center of the SOM where very-\highz galaxies live, surrounded by low redshift galaxies. This region has large (red) $u-g$ color and also large (faint) $i$-band magnitude, i.e. is the part of photometric space where we encounter Lyman-break galaxies at high redshift. It also hosts faint Balmer-break galaxies at low redshift, and these two galaxy populations are known to present important degeneracies in the color-redshift relation. That degeneracy is also responsible for a large redshift scatter in that part of the SOM. Finally, regarding the redshift mapping of the Deep SOM, it is important to point out that the vast majority of cells in the map contain galaxies from the SPC redshift sample, with only a four cells (out of 2304) containing no redshift information. In \S\ref{sec:redshift_uncertainty}, when we characterize the redshift uncertainties in the defined tomographic bins, we will use the \textsc{Balrog} sample to estimate how the tomographic bin photometric spaces map into the Deep SOM, and quantify the (small) impact of deep galaxies in cells with no redshift information. 

\subsection{The Wide SOM}
\label{sec:wide_som}

We now turn to characterizing the DES wide space, using the same SOM algorithm as for the deep space. We now use $griz$ DES wide photometry as described in \S\ref{sec:wide} to construct a Wide SOM having $22\times22$ cells. By comparison, the Wide SOM describing the DES Year 3 space in \citet{y3-sompz} was $32\times32$ in size and was constructed using $riz$ photometry (because the $g$-band was not used for galaxy selection in the weak lensing analysis). We use a smaller SOM size due to the pre-selection cut of $22<i<23.5$ applied to our wide-field sample. Figure \ref{fig:wide_som} shows the photometric properties of the Wide SOM, including the mapping of $i$-band magnitude and the three observed colors. 

Given the characterization of galaxy phenotypes in the Deep SOM and its redshift mapping using the SPC redshift sample, we can use the \textsc{Balrog} sample to characterize the redshift mapping of the Wide SOM using Eq.~(\ref{eq:wide_pz}). This equation yields a probability density function for the redshift of each Wide SOM cell, using the redshift mapping of the Deep SOM with the SPC redshift sample and the transfer function between Wide and Deep spaces characterized with the Balrog sample. This is shown in the lower left panel of Fig.~\ref{fig:wide_som}, where we can see a good separation between low- and \highz regions in the Wide SOM, and now we can use this redshift mapping of the Wide SOM to perform the selection of our redshift bins.

\subsection{Selecting tomographic bins}
\label{sec:tomo_selection}

\begin{figure}
 \centering
 \includegraphics[width=0.48\textwidth]{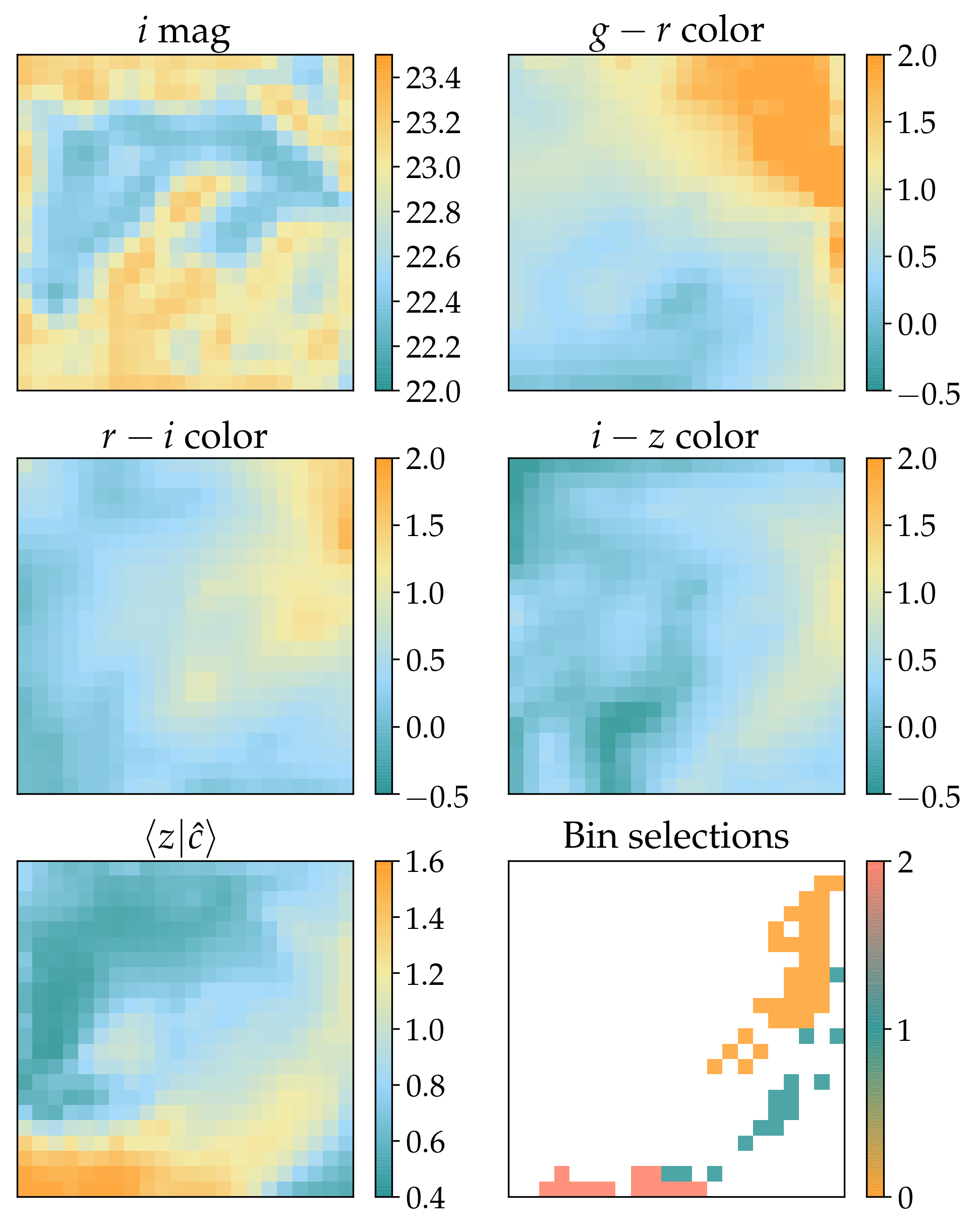}
 \caption{Visualization of various properties of the Wide SOM described in \S\ref{sec:wide_som}. In particular, we show the photometric properties of the map, namely the mapping of $i$-band magnitude and three of the wide photometric colors, using the DES wide galaxy sample described in \S\ref{sec:data}. The bottom left panel shows the redshift mapping of the Wide SOM, using SPC redshift galaxies matched to DES deep photometry and the \textsc{Balrog} transfer function between deep and wide photometry, as described in Eq.~(\ref{eq:wide_pz}) and \S\ref{sec:wide_som}.  The bottom right panel shows the cells of the Wide SOM that constitute the three tomographic bins used in this work, following the procedure described in \S\ref{sec:tomo_selection} and Fig.~\ref{fig:cell_selection}.} 
 \label{fig:wide_som}
\end{figure}
\begin{figure}
 \centering
 \includegraphics[width=0.45\textwidth]{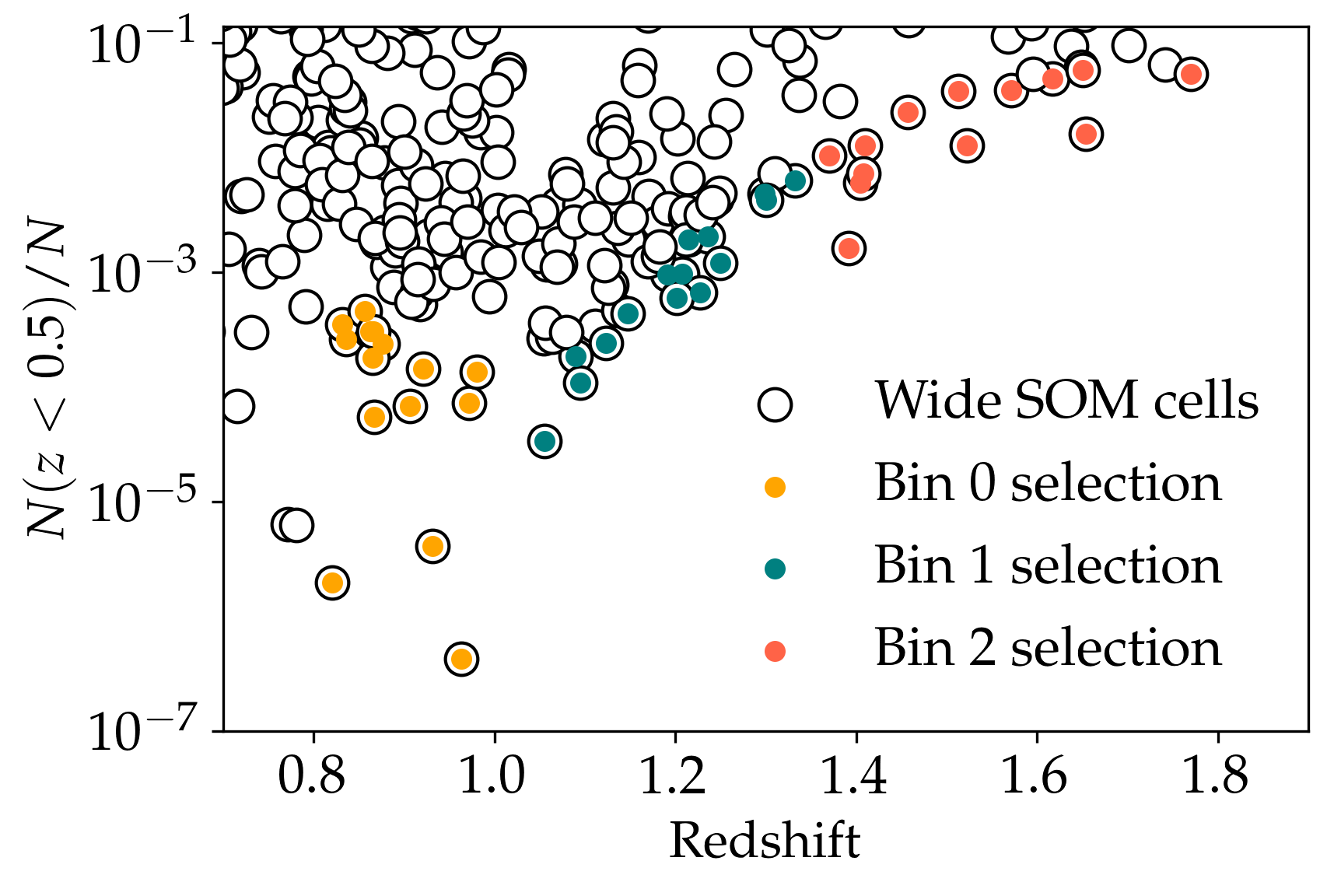}
 \caption{Visualization of the tomographic bin selection as groups of Wide SOM cells, as described in \S\ref{sec:tomo_selection}. The plot shows the estimated low redshift ($z<0.5$) vs mean redshift for each Wide-SOM cell with mean redshift above 0.7. Cells selected for \highz bin 0 are marked in orange, cells for bin 1 are marked in blue, and cells selected for bin 2 are marked in red. }
 \label{fig:cell_selection}
\end{figure}

\begin{figure*}
 \centering
 \includegraphics[width=0.8\textwidth]{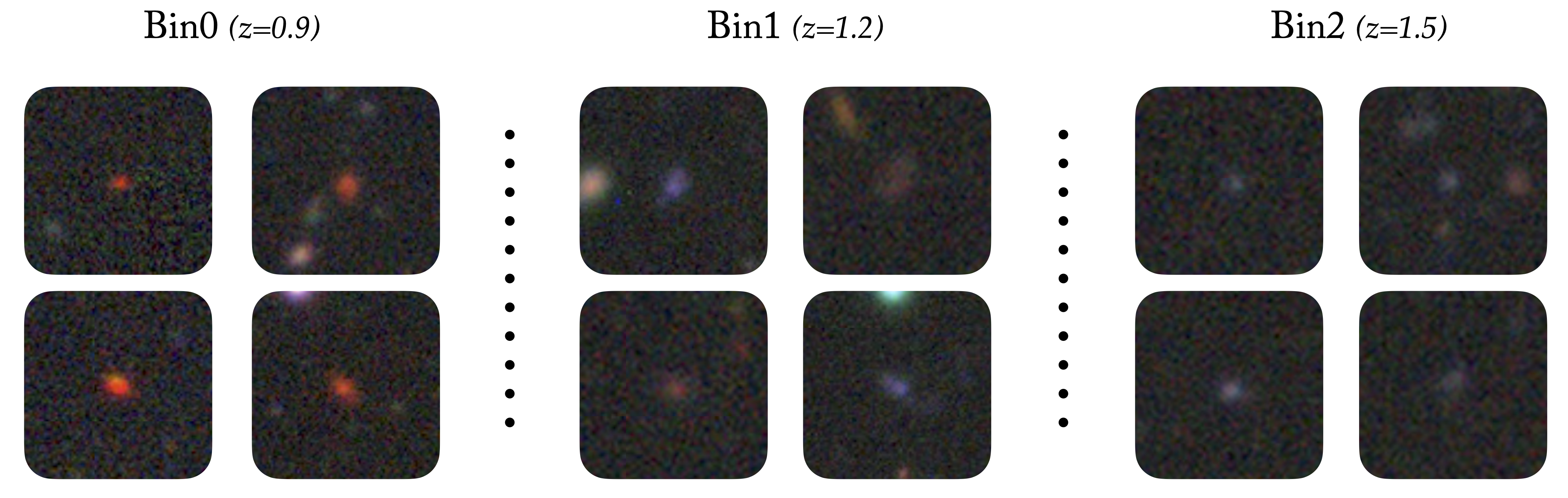}
 \caption{Visualization of color images of random galaxies from each of the three redshift bins defined in \S\ref{sec:tomo_selection}. As apparent from Fig.~\ref{fig:wide_som}, the first bin is made predominantly of red galaxies and then the selection moves to bluer and fainter galaxies for the second and third bin. }
 \label{fig:images}
\end{figure*}

\begin{figure}
 \centering
 \includegraphics[width=0.45\textwidth]{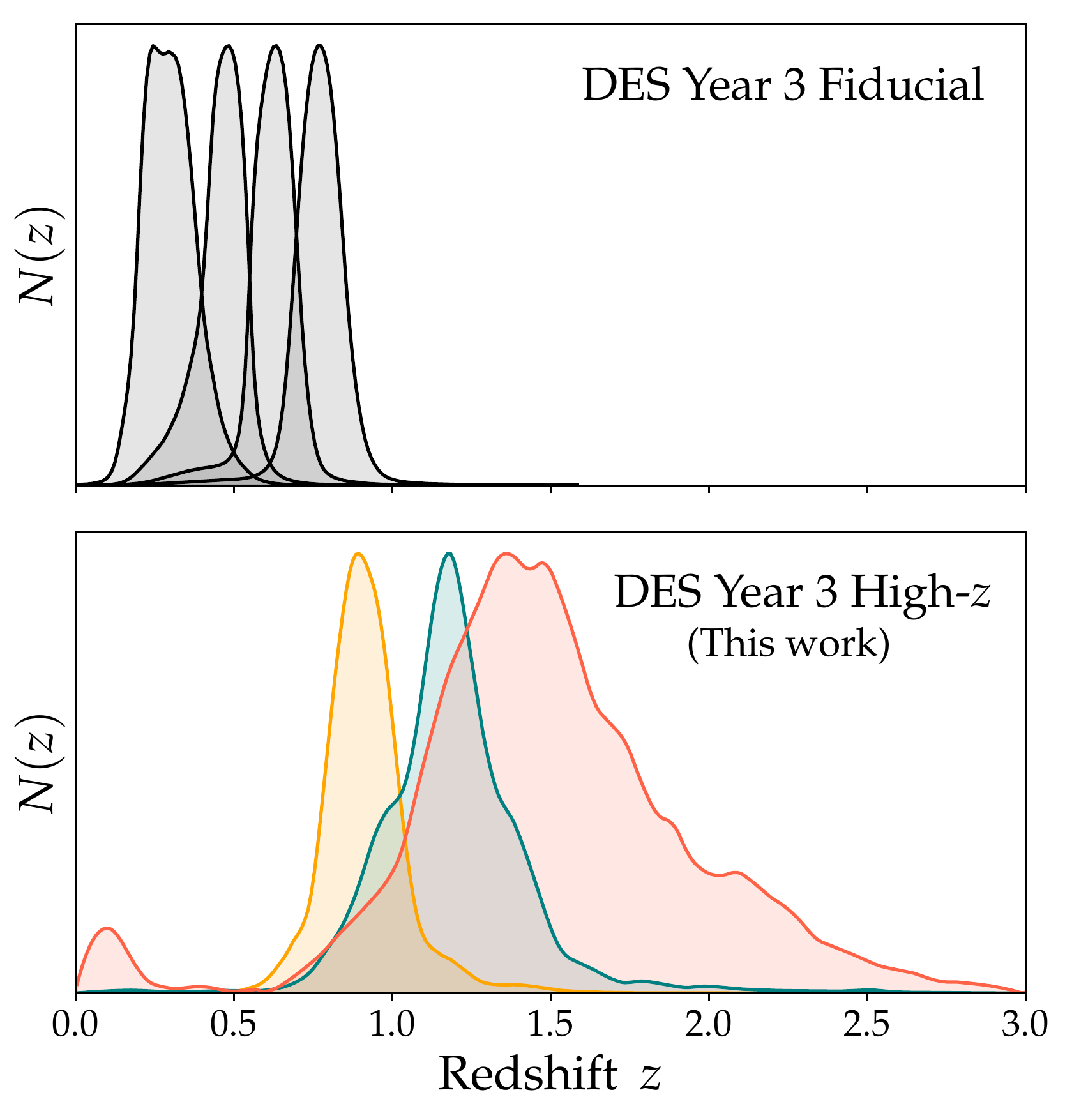}
 \caption{Comparison of the redshift distributions used in the fiducial DES Year 3 lens galaxy sample (\maglim, upper panel) with the redshift distributions of the three tomographic bins defined in this work (\S\ref{sec:tomo_selection}, bottom panel). The three \highz redshift bins defined in this work considerably extend the lens redshift range probed by the DES Year 3 data sample. The number of galaxies, galaxy density and mean redshift of these samples can be found in Table \ref{tab:samples}.  }
 \label{fig:nz_comparison}
\end{figure}

Since each Wide galaxy can be placed in a cell of the Wide SOM, and we have an estimate of the redshift distribution $n(z|\hat c)$ within each Wide-SOM cell, 
we can construct tomographic bins as groups of Wide SOM cells. With the goal of constructing tomographic bins at high redshift with the least possible low-redshift contamination, we compute the mean redshift of each Wide SOM cell and the fraction of the redshift distribution at low redshift $z<0.5$. We choose to define 3 tomographic bins at mean redshifts around 0.9, 1.2 and 1.5 that minimize the low redshift contamination, as described in Fig. \ref{fig:cell_selection}. Using this procedure, the resulting cells in the Wide SOM that make up each redshift bin are depicted in the lower right panel of Fig.~\ref{fig:wide_som}. From that representation, we see how the first redshift bin comes from the upper right part of the Wide SOM and hence contains galaxies with strong (red) $u-g$ and $g-r$ colors, and as the selection moves to the second and third redshift bins the corresponding galaxies will have smaller (blue) $u-g$ colors and fainter $i$-band magnitudes (the average $i$-band magnitude for bins 0, 1, 2 is 22.6, 22.9 and 23.1, respectively). To visualize these trends directly, Fig.~\ref{fig:images} shows a small random sample of galaxy images images from each of the redshift bins, which confirm the characteristics of each bin inferred from the Wide SOM in Fig.~\ref{fig:wide_som}. 

\begin{table}
\centering
\caption{Summary description of the lens galaxy samples defined using DES Year 3 data, as a comparison to the samples defined in this work. The fiducial lens sample in the DES Year 3 analysis consists of the first four \textsc{MagLim} bins. The other two \maglim bins and the \redmagic sample bins are marked in red as they were not part of the fiducial analysis. The table shows $N_\text{gal}$ as the number of galaxies in each redshift bin, $n_\text{gal}$ as the galaxy number density in units of gal/arcmin$^{2}$, and $\left< z \right>$ as the mean redshift of each bin. }
\label{tab:samples}
\vspace{1mm}
DES Year 3 Fiducial \textsc{MagLim} sample \\
\vspace{0.5mm}
\renewcommand{\arraystretch}{1.02}
\begin{tabular}{cccc}
\hline
\hline          
\text{Redshift bin} &   \text{$N_\text{gal}$} &  \text{$n_\text{gal}$} & \text{$\left< z \right>$} \\
\hline  
0 & \, 2 236 473 \, & \, 0.150 & \, 0.30 \\
1 & \, 1 599 500 \, & \, 0.107 & \, 0.46\\
2 & \, 1 627 413 \, & \, 0.109 & \, 0.62\\
3 & \, 2 175 184 \, & \, 0.146 & \, 0.77\\
{\color{BrickRed}4} & \, 1 583 686 \, & \, 0.106 & \, 0.89\\
{\color{BrickRed}5} & \, 1 494 250 \, & \, 0.100 & \, 0.97\\
\hline  
\end{tabular}
\\
\vspace{2mm}
DES Year 3 \textsc{redMaGiC} sample \\
\vspace{0.5mm}
\begin{tabular}{cccc} 
        \hline
        \hline  
$\text{Redshift bin}$  &  $N_\text{gal}$ &  \text{$n_\text{gal}$}& \text{$\left< z \right>$}\\ 
\hline
{\color{BrickRed}0} & \, 330 243 \, & \, 0.022 & \, 0.27 \\
{\color{BrickRed}1} & \, 571 551 \, & \, 0.038& \, 0.43 \\
{\color{BrickRed}2} & \, 872 611 \, & \, 0.058 & \, 0.58\\
{\color{BrickRed}3} & \, 442 302 \, & \, 0.029 & \, 0.73\\
{\color{BrickRed}4} & \, 377 329 \, & \, 0.025 & \, 0.85\\
\hline
\end{tabular}
\\
\vspace{2mm}
DES Year 3 High-$z$ sample (\textbf{This work}) \\
\vspace{0.5mm}
\begin{tabular}{cccc}
        \hline
        \hline
\text{Redshift bin} &   \text{$N_\text{gal}$} &  \text{$n_\text{gal}$ } & \text{$\left< z \right>$} \\
\hline
0 \, & \, 3 929 803 \, & \, 0.416 & \, 0.90 \\
1 \, & \, 2 551 780 \, & \, 0.270 & \, 1.21 \\
2 \, & \, 2 397 667 \, & \, 0.254 & \, 1.49 \\
\hline
\end{tabular}
\renewcommand{\arraystretch}{1.0}
\end{table}

It is notable that the wide-SOM cells $\hat c$ selected for the \highz samples largely \emph{exclude} galaxies from cells $c$ in the second ``island'' of \highz galaxies in the Deep SOM, which contains the Lyman-break galaxies (LBGs).  This is likely because the absence of $u$-band data in the wide sample makes it difficult to localize wide-field galaxies into this Deep SOM island.  Hence the DES Y3 \highz sample defined in this paper is notably orthogonal to many previous \highz catalogs which emphasized LBGs at $z>2.$

Given these tomographic bin selections as lists of Wide SOM cells, we can now use Eq.~(\ref{eq:wide_nz}) to estimate the redshift distribution of each of these bins. Figure \ref{fig:nz_comparison} shows the three resulting redshift distributions, and compares them with the four tomographic bins of the fiducial DES Year lens galaxy sample, the so-called \textsc{MagLim} sample \citep{y3-2x2ptaltlensresults}. As apparent from that figure, the three tomographic bins defined in this work significantly extend the redshift range probed by the DES Year 3 Fiducial lens galaxy sample. Besides extending the redshift range, the three tomographic bins from this work also provide larger number of galaxies and galaxy number densities than the \maglim fiducial DES lens sample, and also the \textsc{redMaGiC} galaxy sample \citep{y3-2x2ptbiasmodelling} (see Table \ref{tab:samples}). The characterization of the uncertainties associated with these three redshift distributions, and the way we will parametrize such uncertainties, will be described in detail in Section~\ref{sec:redshift_uncertainty}. 


\section{Characterizing the completeness of the samples in the footprint}
\label{sec:weights}

Due to the faint, low-$S/N$ nature of the galaxies in the three tomographic bins defined in Section \ref{sec:method}, 
it is expected that their selection function will fluctuate across the survey footprint because of varying observing conditions (such as exposure time, seeing, airmass) and also due to astrophysical fluctuations (such as stellar density or extinction). These variations in the selection function will induce correlations between galaxy density and survey properties for the different tomographic bins.  Any such fluctuations will induce spurious signal in the measurement of galaxy clustering, exacerbated by patterns in e.g.~survey observing strategies or Galactic structure.
We must correct the \highz density maps for the survey selection function if we want to recover accurate measures of the \highz intrinsic galaxy clustering.

These kind of corrections due to varying observing properties have been studied extensively in DES and elsewhere \citep{Leistedt2015, BOSS_systematics, wthetapaper, Noah_weights, y3-galaxyclustering}. In many of these cases, the relationship between survey properties and galaxy selection rates was close to linear, and therefore the correction methodologies assumed a linear relationship. The samples in this work, however, present significant non-linearities in that relationship. We therefore introduce a non-linear, neural-network-based approach for characterizing the completeness of the sample with respect to the different survey properties (see \citealt{2020MNRAS.495.1613R} for a similar approach applied to the DECaLS DR7 data sample). 

In this section we describe the different survey properties we consider, the methodology used to correct for their correlations with galaxy density for the different tomographic bins, and the validation of the results. The outcome of this procedure will be a derived correction \textit{weight} for each galaxy in the different tomographic bins, inverse to the selection rate in its vicinity. This weight will then be used throughout the analysis, for the characterization of redshift distributions and uncertainties in Section~\ref{sec:redshift_uncertainty}, for the estimation of lens magnification in Section~\ref{sec:magnification}, and for the calculation of correlation functions in Section~\ref{sec:clustering}. 

\subsection{Maps of survey properties (SP)}
\label{sec:sps}

The DES collaboration develops spatial templates for different observing conditions and potential contaminants in the survey footprint by creating \healpix \citep{Gorski2005} sky maps (at \texttt{NSIDE} = 4096, corresponding to a pixel resolution of 0.86 arcmins; see \citealt{Leistedt2016} for details on the implementation). We will refer to these maps as survey property maps (or ``SP maps") and we will use them to characterize and remove any possible correlations with the observed density fields of each tomographic bin. In particular, in this analysis we consider maps of the following survey observing properties, each of them having a different map for each observed photometric band $griz$:

\begin{itemize}
    \item \textbf{Depth:} Mean survey depth, computed as the mean magnitude for which galaxies are detected at $S/N = 10$.
    \item \textbf{Sky variance:} Estimated sky brightness, or more precisely, the standard deviation of sky pixels due to shot noise and read noise, measured in units of electrons/second/pixel.
    \item \textbf{Exposure time:} Total exposure time at a given point in the survey footprint, measured in seconds. 
    \item \textbf{Airmass:} Mean airmass, computed as the optical path length for light from a celestial object through Earth’s atmosphere (in the secant approximation), relative to that at the zenith for the altitude of the telescope site.
    \item \textbf{Seeing:} Mean seeing, measured in arcseconds, computed as the full width at half maximum of the flux profile. 
\end{itemize}

Those make 20 SP maps of observing properties. Additionally, we consider two maps of potential contaminants:
\begin{itemize}
    \item \textbf{Galactic extinction}: We use the SFD dust extinction map from \citet{sfd98}, which measures the $E(B-V)$ reddening, in magnitudes.  
    \item \textbf{Stellar density:} We use a map of stellar density, in deg$^{-2}$, using stellar sources from Gaia EDR3 \citep{gaia_edr3}. 
\end{itemize}

This amounts to a total of 22 survey property maps that we will use in this analysis. For a technical description of these survey observing properties, please see \citet{y3-gold,y3-galaxyclustering,Leistedt2016}.  In principle, these SP's should be a complete list of all factors that could affect galaxy detectability.  The images themselves should be completely specified by the passband (which is constant, with very minor airmass variation), the background noise level of the images (a.k.a. sky brightness), the PSF (primarily seeing FWHM), and the shot noise from the sources (primarily exposure time).  The Galactic dust and stellar background are the two astrophysical effects expected to alter the detectability of background galaxies. The depth map should be redundant but we include it to perhaps ease the task of training the neural network.

\begin{figure*}
 \centering
 \includegraphics[width=0.98\textwidth]{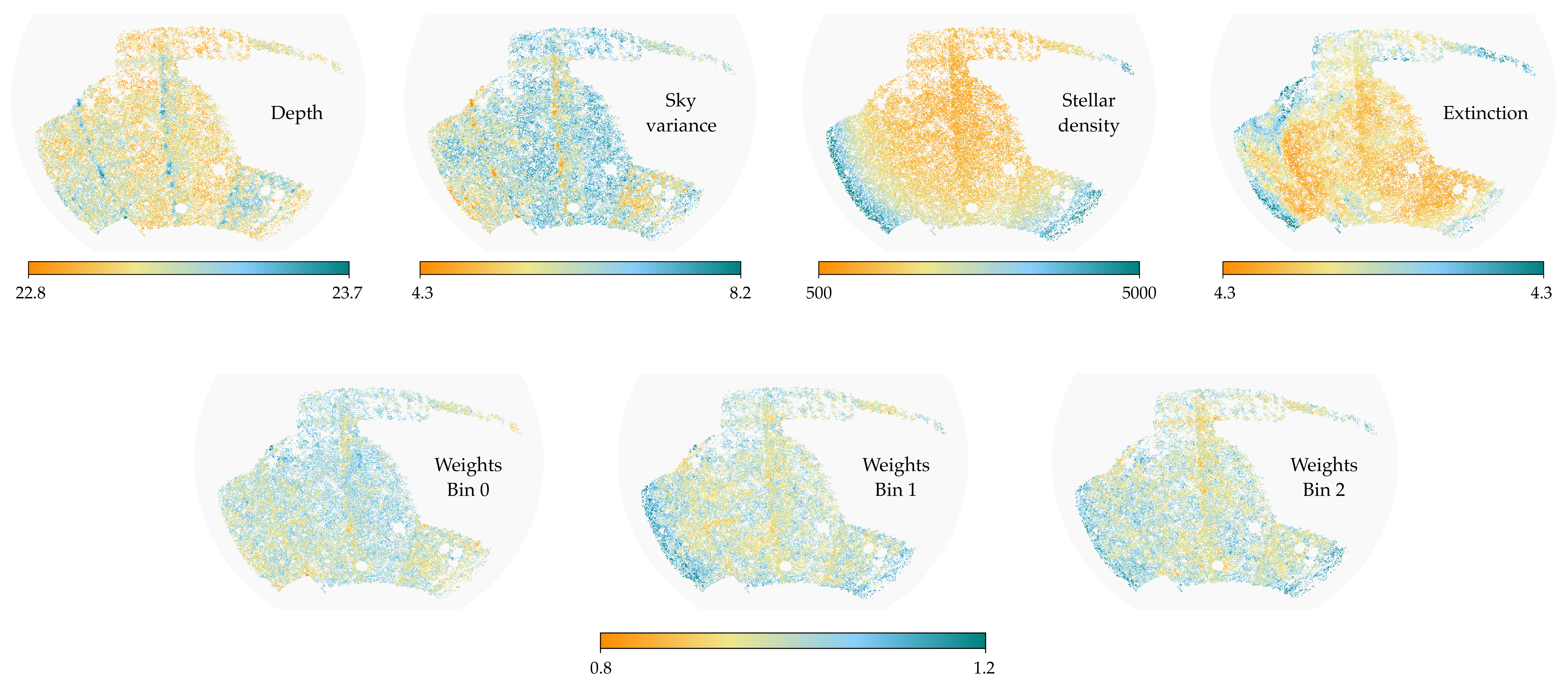}
 \caption{\textit{Upper row:} Examples of four of the survey property (SP) maps described in \S\ref{sec:sps}. In particular, we show the depth and sky variance maps in the $i$-band, and the maps of stellar density and dust extinction. \textit{Lower row:} Maps of the derived weight maps using the neural network approach described in \S\ref{sec:nn}, for the three tomographic bins in this work. }
 \label{fig:weights_maps}
\end{figure*}

\subsection{Correction method}
\label{sec:nn}

We aim to model the relationship between the survey property maps defined above and the observed galaxy count maps for each of the tomographic bins defined in \S\ref{sec:method}. For this, we will use a neural network (NN), with the 22 SP maps being the \textit{features} and the observed galaxy count maps being the \emph{label}. Naturally, the network will be able to model a nonlinear relationship between the SP maps and the raw galaxy counts. It is important to note, however, that we do not include any spatial information in the process, since we do not want the network to learn about the clustering of galaxies.

The neural network is asked to predict whether or not a particular Healpixel (at the same \texttt{NSIDE} = 4096 resolution) contains \emph{any} galaxies [that is, $p(n\ge1)$] based on the SP values for that pixel.  Note that this ignores any distinctions between Healpixels with $n=1$ vs $n=2$ or more galaxies. This helps prevent the network from learning any intrinsic galaxy clustering, since Healpixels with large number of galaxies are likely to be due to intrinsic density peaks rather than survey observing properties. At the resolution of \texttt{NSIDE} = 4096, most pixels contain either zero or one galaxies (the average number of galaxies per pixel for bins 0, 1 and 2 is 0.307, 0.200 and 0.187, respectively). The loss function for the network is the binary cross-entropy between the predicted pixel occupancy and the occupancy of the training set.

The architecture of the network is based on our guess that the selection function scales primarily as some power law combination of the SPs.  To this end, the input SP values are all logarithmically scaled (except those, such as depth, which are already logarithmic quantities), and the output of the network is exponentiated to form the selection probability.  The network output is a sum of two branches: the first branch is a simple linear combination of the 22 scaled SP's, since we expect this to capture most of the functional variation.  The second branch is intended to capture departures from a simple power law: it takes the input layer of 22 dimensions through 3 hidden layers of 64, 32 and 4 fully connected neurons, respectively, and a single neuron on the output layer, each with \texttt{relu} activation. The output of the network, for each tomographic bin, consists of a single value for each Healpixel within our mask, which will be used to weight the galaxies accordingly. Figure \ref{fig:weights_maps} shows the resulting weight maps for each tomographic bin, as well as four examples of survey property maps. 

To prevent the network from overfitting, it is constructed with $k$-fold cross-validation, which works in the following way: The \texttt{NSIDE} = 4096 maps are re-binned into a coarser grid of \texttt{NSIDE\_split} = 16 (with a resolution of about 4 degrees). We then randomly divide these cells into $k$ equal-area groups. To derive the weights for a given fold \texttt{k}, we train the NN on the other folds, using fold \texttt{k} as a validation sample (the training halts when the training metric no longer improves on the validation set). This cross-validation scheme will only work to prevent overfitting on scales below the resolution defined by \texttt{NSIDE\_split}, in this case around 4 degrees. A test using the corrected and uncorrected galaxy clustering of log-normal mock catalogs demonstrated no overfitting from the method at scales below 1 degree, and an impact of around 5\% overfitting at scales of 2 degrees. Being conservative, we keep the galaxy clustering analysis in this work to angular scales below one degree. 

\subsection{Validation of the derived correction weights}

\begin{figure*}
 \centering
 \includegraphics[width=0.98\textwidth]{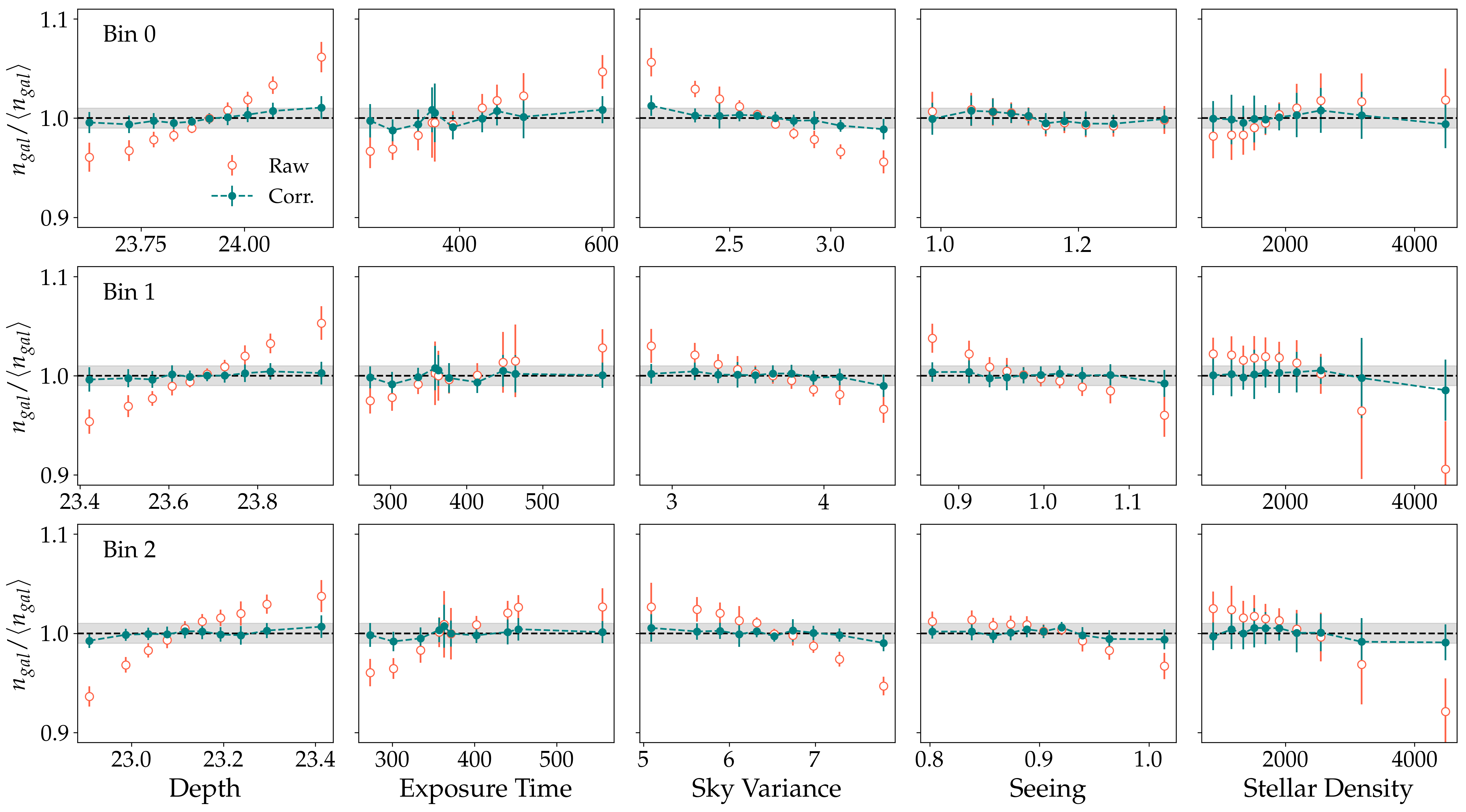}
 \caption{Visualization of the correlation between survey properties (SP) and the observed galaxy density (relative to the mean galaxy density over the full footprint), before (red) and after (blue) the correction using the galaxy weights described in \S\ref{sec:nn}. We show this relationship for depth, exposure time (in seconds), sky variance (in electrons/s/pixel) and seeing (in arcseconds), all estimated in the $i$-band, and also with stellar density (in stars/deg$^2$), in 10 bins of equal area. The uncertainties come from jackknife resampling, and the gray shaded region in the plot corresponds to a 1\% deviation. The distribution of the null $\chi^2$ values for these relationships, including all the 22 SP maps and for each of the tomographic bins, can be found in Fig.~\ref{fig:weight_chis}. }
 \label{fig:weights_correction}
\end{figure*}

\begin{figure}
 \centering
 \includegraphics[width=0.4\textwidth]{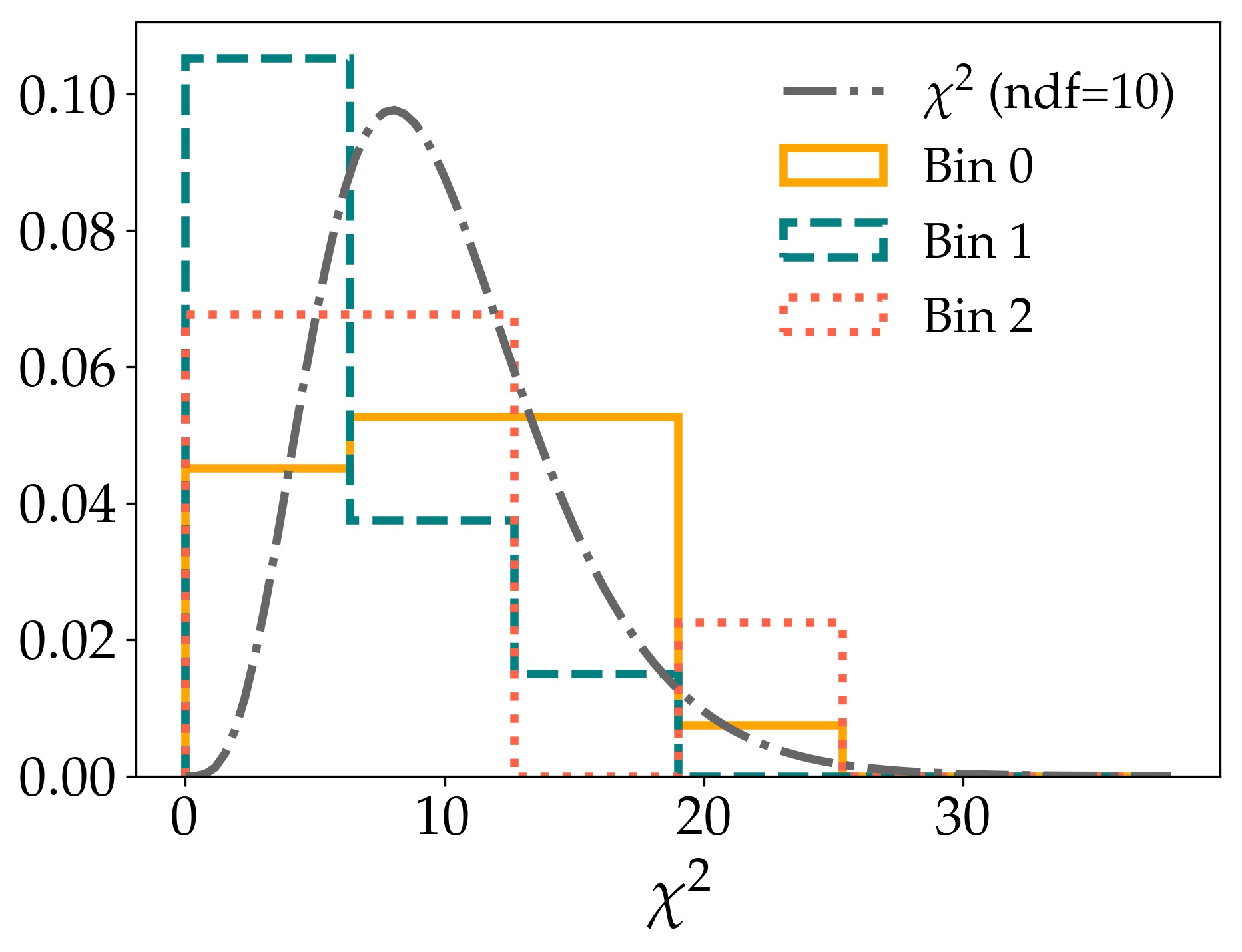}
 \caption{Distribution of the null hypothesis $\chi^2$ values for the relationship between survey property maps and the corrected (weighted) galaxy density, including all the 22 SP maps and for each of the tomographic bins. The median null $\chi^2$ values in the three tomographic bins are 11.6, 3.4 and 7.5 for 10 degrees of freedom. For the raw, uncorrected case the median null $\chi^2$ values for the three bins are 92.1, 35.0 and 51.6 for 10 degrees of freedom. }
 \label{fig:weight_chis}
\end{figure}

\begin{figure}
 \centering
 \includegraphics[width=0.48\textwidth]{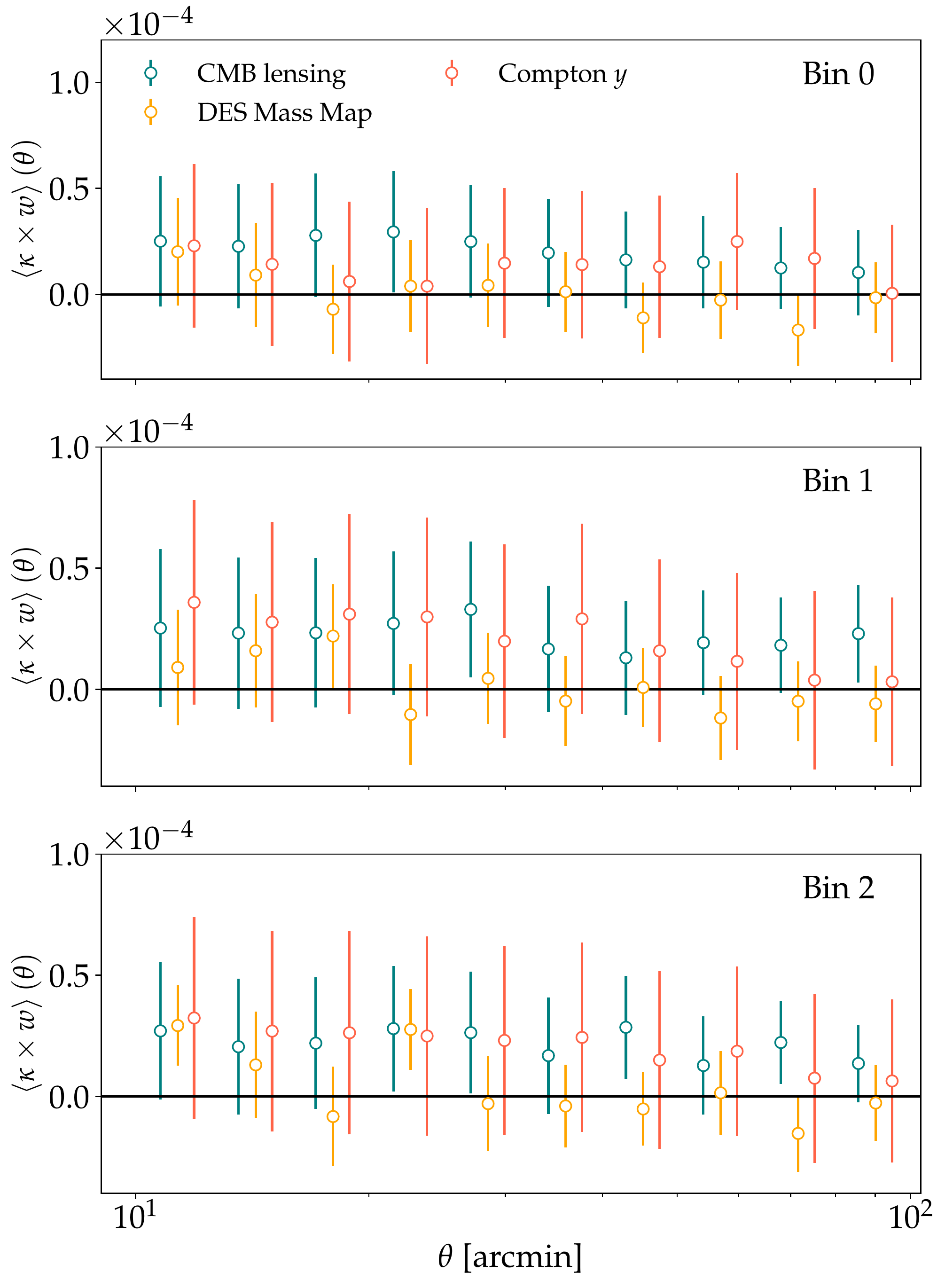}
 \caption{Cross-correlation of weight maps of the three tomographic bins and convergence field from three tracers of the large-scale structure: CMB lensing (from the Planck satellite), DES Y3 mass maps and Planck Compton $y$. Uncertainties come form jackknife resampling. The null hypothesis $\chi^2$ values can be found in Table \ref{tab:weights_chi}, all consistent with no correlation. }
 \label{fig:weights_lss}
\end{figure}

Different survey property maps show significant correlations with the raw galaxy density in each of the tomographic bins. Using the neural network implementation described above, Figure \ref{fig:weights_correction} shows these correlations, and how the derived set of weights is able to correct for any correlations between SP maps and galaxy density. Figure \ref{fig:weights_correction} shows only a limited number of examples of these correlations, for easier visualization, but we also compute the $\chi^2$ for the null hypothesis for all correlations between the 22 SP maps and the corrected galaxy density, using a jackknife approach to estimate the corresponding uncertainties. The distribution of these null $\chi^2$ values, for each of the tomographic bins, can be found in Fig.~\ref{fig:weight_chis}, and we do not find evidence of significant correlations between the SP maps and the corrected (weighted) galaxy density. The median null $\chi^2$ values for the corrected case in the three tomographic bins are 11.6, 3.4 and 7.5 for 10 degrees of freedom. On the other hand, for the raw, uncorrected case the median null $\chi^2$ values for the three bins are 92.1, 35.0 and 51.6 for 10 degrees of freedom, clearly inconsistent with the null hypothesis. 

Beyond being successful at correcting for all the correlations between galaxy density and survey property maps, we need to ensure the derived neural network weights did not learn any physical galaxy clustering at the training phase. For that purpose, we compute the cross-correlation between the weight maps as shown in Fig.~\ref{fig:weights_maps} and several tracers of the large-scale structure of the Universe. In particular, in this work we perform the correlation of the three weight maps with the convergence field estimated from CMB lensing (using both Planck, \citealt{2020A&A...641A...8P}, and SPT \citealt{2022arXiv220312439O}), the \highz mass map from the DES Year 3 analysis \citep{y3-massmapping} and the Planck Compton $y$ map \citep{2016A&A...594A..22P}. These are all tracers of the physical large-scale structure (LSS) and hence they should not present correlations with SP maps or the derived weight maps. A significant correlation would mean there has been some undesired leakage of LSS into our weights. Figure \ref{fig:weights_lss} shows these correlations between weight maps and tracers of the LSS, and Table \ref{tab:weights_chi} shows the $\chi^2$ values for the null hypothesis, demonstrating no significant correlations between weight maps and LSS tracers. 

At this point we have now tested for the correlation of the weighted galaxy density with SP maps and the correlation of weight maps with known tracers of structure, and found a null signal in both cases. These tests are necessary, but not sufficient, to show that our corrections are not imposing a significant bias on the clustering measurements, as it is still possible that the residuals in the estimation of the weight maps could affect the clustering measurements. To account for this potential effect in the clustering analysis, we will marginalize over an additive constant in the correlation function, as done in \emph{e.g.} \citet{Kwan2016} (see also \citealt{Ross2011}). This procedure, which will be described in Section \ref{sec:analysis}, will account for a potential spurious systematic effect in the clustering at first order, and it is a conservative way to marginalize over this uncertainty in the analysis. In that section we will also explore the impact of the choice of maximum angular scale in the galaxy clustering measurements. 

\begin{table}
\begin{center}
\begin{tabular}{cccc}

           & Bin0  & Bin1  & Bin2               \\ \hline
Planck CMB lensing & 9.6/9 & 6.4/10 & 6.8/10 \\
DES Mass Map & 8.2/9 & 9.9/10 & 16.1/10 \\
Planck Compton $y$ & 7.3/9 & 5.9/10 & 2.7/10 \\
\end{tabular}
\caption{Values of $\chi^2_{\mathrm{null}}$/dof for different correlations between galaxy weights and tracers of the large-scale structure of the Universe, for the three \highz bins defined in this work. We find no significant correlations between weight maps and LSS tracers. }
\label{tab:weights_chi}
\end{center}
\end{table}


\section{Characterizing redshift uncertainties}
\label{sec:redshift_uncertainty}



In this section we will describe the various sources of uncertainty in the distributions of redshift $N(z)$ within each of the three bins defined in \S\ref{sec:tomo_selection}, and how we will propagate them into cosmological analyses. We will follow a similar procedure to that in \citet{y3-sompz}, and propagate uncertainty arising from: (i) sample variance (SV) and shot noise (SN) from the finite area covered by the deep fields; (ii) biases in the individual redshift estimates of deep-field galaxies having multi-band photometry (\cosmos and \paucosmos) but no spectroscopic redshift (PZ); (iii) uncertainty in the photometric calibration (zero-point) of deep-field galaxies (ZP); and (iv) uncertainties from the ``bin conditionalization" approximation in Eq.~(\ref{eq:wide_nz_approx1}) (BCE).

To model SV and SN,  we use the approximate \sdir model (a product of three Dirichlet distributions), first presented in \citet{Sanchez2020} and then further developed in \citet{y3-sompz}. Mathematically the model describes $p(\{f_{zc}\}|\{N_{zc}\})\approx \sdir$, where $N_{zc}$  are the number counts of galaxies that have been observed to be at redshift bin $z$ and colour phenotype $c$, and with $\{f_{zc}\}$ a finite set of coefficients indicating the probability in the redshift bin $z$ and color phenotype $c$, where $\sum_{zc}f_{zc}=1$ and $0\leq f_{zc} \leq 1$. For extensive details of the model we refer the interested reader to Appendices D and E in \citet{y3-sompz}. The \sdir method yields realizations of the $f_{zc},$ which then can be summed into Eq.(\ref{eq:wide_nz_approx1}) to yield $N(z)$ estimates. The mean of these realizations is the fiducial $N(z)$.

We smooth the fiducial $N(z)$ distribution with a Savitzky–Golay filter: sample variance and shot noise from the small area of the calibration deep fields manifests in the $N(z)$ as rapid fluctuations in redshift and enter squared in the galaxy clustering signal, while the true redshift distribution over a larger area is smoother as these variations average out. We try different smoothing lengths and find compatible constraints on the main parameters of interest (see Appendix~\ref{sec:smoothing}).

Deviations from the nominal $N_i(z)$ will be modeled with three parameters: a shift $\Delz{i}$, a stretch parameter $\Sigz{i}$, and an adjustment $\Alowz{i}$ of the low-redshift tail of $N_i(z)$.  The main peak of the distribution is altered according to
\begin{equation}
N(z) \rightarrow N(\Sigz{i} (z - \Delz{i} - \bar{z}) + \bar{z})
\end{equation}
and the fraction of galaxies at low redshift ($z<0.5$) is altered as
\begin{equation}
n(z) \rightarrow
\begin{cases} 
n(z) \, \Alowz{i}   & z \leq 0.5 \\
n(z) \, (1-\Alowz{i}) & z>0.5
\end{cases}
\end{equation}
Details of this transformation are in Appendix~\ref{sec:appendix_pz_param}.  Figure~\ref{fig:nz_para} illustrates the effects of each of these parameters.

Priors on the $N(z)$ alteration parameters $\theta_i = \{\Delz{i}, \Sigz{i}, \Alowz{i}\}$ are chosen to represent the potential effects of the systematic errors by:
\begin{itemize}
    \item Quantifying the possible effects of the PZ, BCE and ZP systematic errors on the input catalogs to the redshift calibration process, as detailed in Appendices~\ref{sec:appendix_pz_biases}, \ref{sec:appendix_pz_bce}, and \ref{sec:appendix_pz_zpe}, respectively.
    \item Creating realizations of the input catalogs drawing from these systematic errors and realizing the SV and SN variations with the \sdir process.
    \item Measuring the mean, width, and low-$z$ fractions of each realized $N_i(z)$.
    \item Creating a prior for the $\theta_i$ based on the distribution of these properties of the realizations.
\end{itemize}
Figure~\ref{fig:pzpriors} shows the resultant distributions of the $N_i(z)$ recalibration parameters when various sources of systematic errors are included, and values of their means and standard deviations are listed in Table~\ref{tab:redshift}.  Sample variance/shot noise, redshift biases and zero point uncertainty all contribute significantly to the uncertainty in the mean redshift. On the other hand, the stretch uncertainty is dominated by sample variance at low redshift (Bin 0), with the zero point uncertainty significantly increasing its importance in in the highest redshift bin. Finally, the low redshift probability uncertainty is primarily dominated by sample variance and shot noise.  Similar results for redshift uncertainties are found from the North and South subsets of the data.


\begin{figure}
 \centering
 \includegraphics[width=0.45\textwidth]{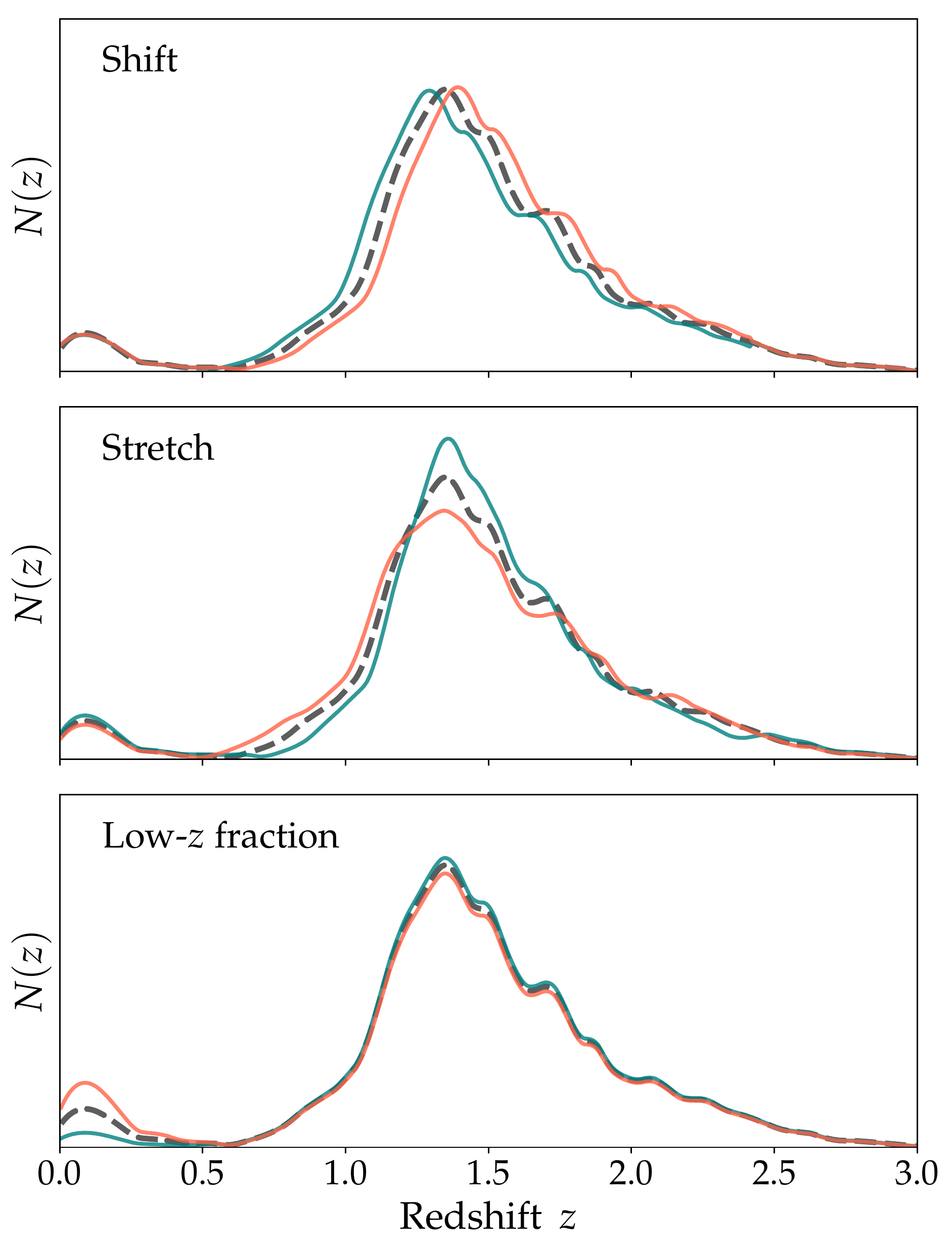}
 \caption{Visualization of the parametrization of redshift uncertainties, using the third tomographic bin as an example. The different rows show examples of how we account for shifts, stretches and variations in the low redshift fraction of the redshift distributions. }
 \label{fig:nz_para}
\end{figure}

\begin{table}
\centering
\caption{Estimates of the parameters describing our uncertainties on the redshift distributions, as described in \S\ref{sec:redshift_uncertainty}, for the three tomographic bins defined in this work. The parametrization is described visually in Fig.~\ref{fig:nz_para}. We also show the estimates for the entire footprint we use, and for the independent splits of North and South regions, which will be used in \S\ref{sec:clustering} for consistency tests.  }
\label{tab:redshift}
\vspace{1mm}
Entire footprint (All) \\
\vspace{0.5mm}
\renewcommand{\arraystretch}{1.02}
\begin{tabular}{cccc}
\hline
\hline          
\text{$z$-bin} &   \text{$\Delz{}$} &  \text{$\Sigz{}$} & \text{$\Alowz{}$} \\
\hline  
0 & \, $0.0\pm0.0051$  \, & \, $0.997\pm0.068$   & \, $0.0044\pm0.0013$ \\
1 & \, $0.0\pm0.0075$ \, & \, $0.999\pm0.041$ & \, $0.0091\pm0.0023$\\
2 & \, $0.0\pm0.0208$ \, & \, $0.998\pm0.044$ & \, $0.0383\pm0.0.0059$\\
\hline  
\end{tabular}
\vspace{1mm}
\\ North region (\emph{Planck}) \\
\vspace{0.5mm}
\renewcommand{\arraystretch}{1.02}
\begin{tabular}{cccc}
\hline
\hline          
\text{$z$-bin} &   \text{$\Delz{}$} &  \text{$\Sigz{}$} & \text{$\Alowz{}$} \\
\hline  
0 & \, $0.0\pm0.0054$  \, & \, $0.995\pm0.068$   & \, $0.0043\pm0.0015$ \\
1 & \, $0.0\pm0.0078$ \, & \, $0.999\pm0.041$ & \, $0.008\pm0.0023$\\
2 & \, $0.0\pm0.0223$ \, & \, $0.998\pm0.044$ & \, $0.038\pm0.0.0065$\\
\hline  
\end{tabular}
\vspace{1mm}
\\ South region (SPT) \\
\vspace{0.5mm}
\renewcommand{\arraystretch}{1.02}
\begin{tabular}{cccc}
\hline
\hline          
\text{$z$-bin} &   \text{$\Delz{}$} &  \text{$\Sigz{}$} & \text{$\Alowz{}$} \\
\hline  
0 & \, $0.0\pm0.0052$  \, & \, $0.998\pm0.051$   & \, $0.0041\pm0.0015$ \\
1 & \, $0.0\pm0.0114$ \, & \, $0.996\pm0.081$ & \, $0.009\pm0.0027$\\
2 & \, $0.0\pm0.0224$ \, & \, $0.998\pm0.048$ & \, $0.0337\pm0.0.0065$\\
\hline  
\end{tabular}
\end{table}

\begin{figure*}
 \centering
 \includegraphics[width=0.9\textwidth]{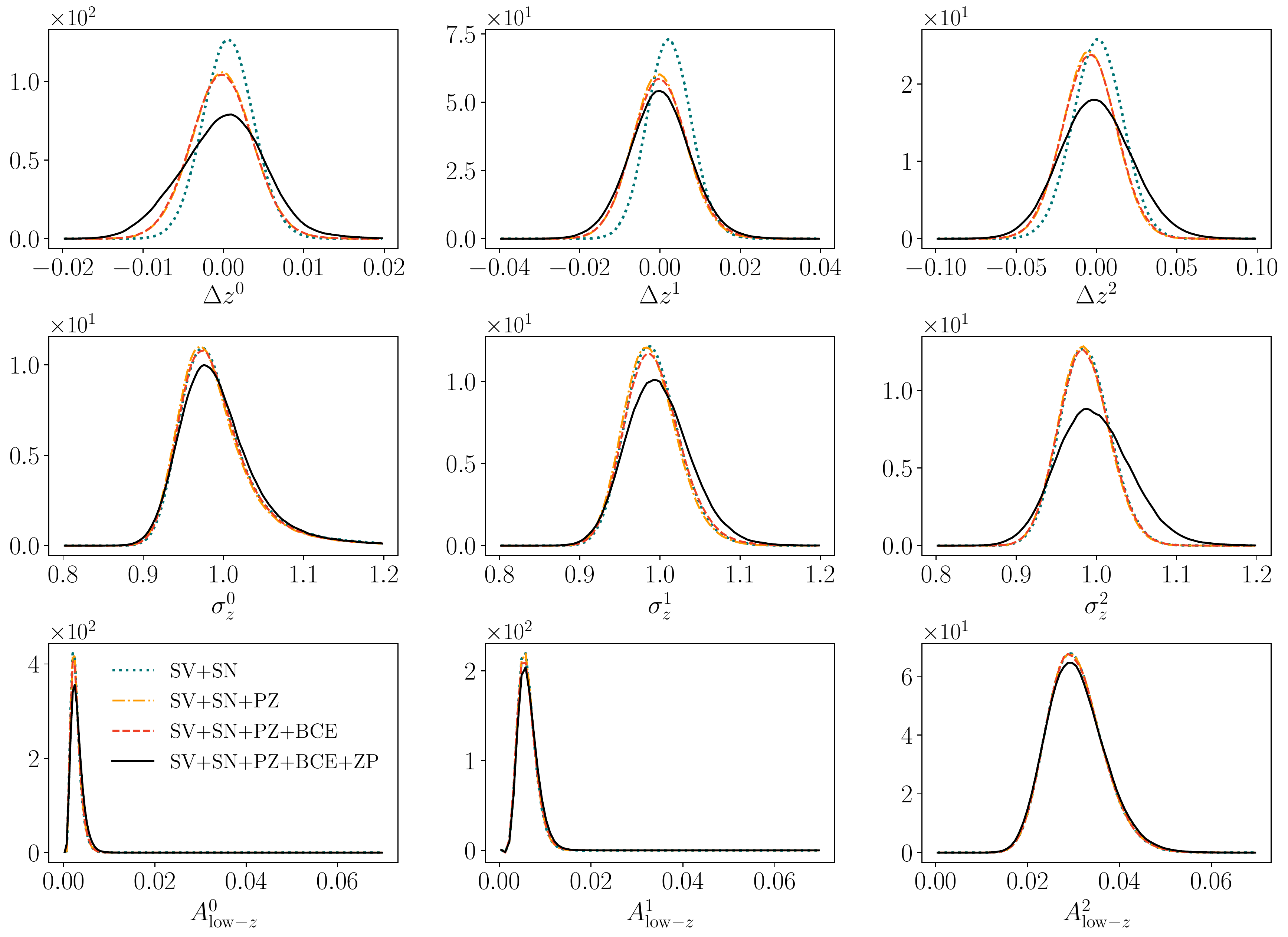}
 \caption{Prior distributions for each redshift uncertainty parameter. Each column shows the parameters for each tomographic bin (left: Bin 0; middle: Bin 1, right: Bin 2). Each row shows a different parameter  (top: $\Delz{i}$; center: $\Sigz{i}$, bottom: $\Alowz{i}$). The different lines show the cumulative uncertainty on each parameter from considering different effects. The dotted line shows the uncertainty from Sample Variance and Shot Noise in the calibration fields (SV+SN). The dot-dashed line adds the uncertainty from redshift biases in the redshift calibration samples (PZ). The dashed line adds uncertainty from redshift selection effects (BCE). The solid lines adds the zero-point photometric uncertainty in the deep field photometry (ZP). The distributions are measured from individual N(z) samples generated to include these uncertainties. For $p(\Delz{i})$ we measure the mean redshift of individual samples and subtract the mean redshift of the fiducial N(z). For $p(\Sigz{i})$ we measure the N(z) width of individual samples and divide by the width of the fiducial N(z). For $p(\Alowz{i})$ we measure the integral of each individual sample at $z<0.5$. See Section~\ref{sec:redshift_uncertainty} and Appendix~\ref{sec:appendix_pz} for details.
}
 \label{fig:pzpriors}
\end{figure*}

\section{Characterizing weak lensing magnification}
\label{sec:magnification}

In this section we study the impact of lensing magnification on the observed angular correlations of our \highz galaxy samples. 
On top of distorting the image shapes, gravitational lensing from the foreground large scale structure of the Universe also magnifies the images without changing the surface brightness, creating two effects: (i) a dilution of the source density due to the locally stretched image; and (ii) an increased flux of individual galaxies making them more likely to be detected \citep{BartelmannSchneider01,Menard2003magnification, Hildebrandt2009_magnification,Garcia_Fernandez_2018,Gaztanaga2021,kidsmagnification2021, Euclidmagnification2021}. This effect creates an additional clustering signal of the background sample which contaminates estimates of its intrinsic density fluctuations. Following the approach used in the fiducial DES Year 3 analysis \citep{y3-3x2ptkp}, we model the observed projected (lens) galaxy density contrast of tomographic bin $i$, $\delta^i_{\mathrm{obs}}$, as a combination of the projected galaxy density contrast $\delta^i_{\mathrm{g}}$ and the modulation by lens magnification $\delta^i_{\mu}$ and redshift-space distorsions (see \S\ref{sec:model} for more details): 

The change in density contrast due to magnification can be shown to be proportional to the convergence experienced by the lens galaxies $\kappa^i_l$ \citep{y3-2x2ptmagnification}:
\begin{equation}
    \delta^i_{\mu} (\theta) = C^i \kappa^i_l (\theta)
\end{equation}
The constant of proportionality $C^i$ is given by the response of the number of selected galaxies per unlensed area, and it can be split in two terms, one fixed term corresponding to the change of area and another term corresponding to changes in the light flux distribution of galaxies, which will affect their selection in different samples:
\begin{equation}
    C^i = C_{\mathrm{area}} + C^i_{\mathrm{sample}},
\end{equation}
where $C_{\mathrm{area}} = -2$ regardless of the sample selection. In this way, the characterization of lens magnification amounts to estimating the $C^i_{\mathrm{sample}}$ term for each tomographic bin. This term can be estimated empirically by artificially magnifying a galaxy sample and measuring the change in number density with respect to the applied magnification. In particular, if we apply some extra convergence $\delta\kappa$ to the images, the proportionality constant can be written as:
\begin{equation}
\label{eq:c_estimate}
    C_{\mathrm{sample}} = \frac{\delta n}{n \: \delta\kappa}, 
\end{equation}
where $\delta n / n$ corresponds to the fractional change in number density of a given sample meeting selection criteria due to the applied magnification. In this work we will follow the approach of \citet{y3-2x2ptmagnification} and estimate $C_{\mathrm{sample}}$ in two different ways, using the \textsc{Balrog} sample and directly perturbing the measured fluxes in the data. 

\subsection{Estimate from artificial galaxy injections}
\label{sec:balrog_mag}

A number of \balrog catalogs were produced for the DES Year 3 analysis \citep{y3-balrog}. In this analysis we have already used \textsc{Balrog} to estimate the transfer function between the deep and wide photometric spaces (parametrized with SOMs), as described in \S\ref{sec:method}. In this part we use an additional \textsc{Balrog} run, in which the exact same deep field objects are injected at the same coordinates as in the fiducial run, but now with a 2\% magnification applied to each galaxy image, $\mu_0=1.02$ ($\kappa_{0}\sim0.01$). For all cases, we account for the galaxy correction weights defined in \S\ref{sec:weights} and shown in Fig.~\ref{fig:weights_maps}.    


We apply the tomographic bin selections described in \S\ref{sec:tomo_selection} on both the fiducial $\kappa=0$ \balrog run (label $i$, for \emph{intrinsic}) and the $\kappa=\kappa_{0}$ run (label $o$, for \emph{observed}). In order to estimate $C_{\mathrm{sample}}$, we need, for each tomographic bin selection: 
\begin{enumerate}
    \item $N_{i}$: Selected number of galaxies in the \balrog $\kappa=0$ run. Accounting for galaxy weights $w^j_i$, it becomes $N_i = \sum_j w^j_i$, where $j$ runs over all selected galaxies. 
    \item $N_{o}$: Selected number of galaxies in the magnified \balrog run, which applies a constant magnification to the galaxy images. Accounting for galaxy weights $w^j_o$, it becomes $N_o = \sum_j w^j_o$. 
\end{enumerate}
At this point, the estimate is simply the fractional difference between the two:
\begin{equation}
    C_{\mathrm{sample}} = \frac{N_o - N_i}{\kappa_0 N_i}.
\end{equation}
This estimate should capture the impact of magnification on the specific color selection of the \highz bins defined in \S\ref{sec:tomo_selection}, and also include possible contributions due to size selections such as the star - galaxy separation cuts. We compute the uncertainties on these estimates by following a jackknife approach, splitting the footprint over 150 regions. 

\begin{table}
\centering
\caption{Estimates of the lens magnification coefficients $C_{\mathrm{sample}}$ using the \balrog and data-based methods described in \S\ref{sec:magnification}, for the three tomographic bins defined in this work. The last column shows the final estimates of the coefficients from the combination of the two different methods. We also show the estimates for the entire footprint we use, and for the independent splits of North and South regions, which will be used in \S\ref{sec:clustering} for consistency tests.  }
\label{tab:magnification}
\vspace{1mm}
Entire footprint (All) \\
\vspace{0.5mm}
\renewcommand{\arraystretch}{1.02}
\begin{tabular}{cccc}
\hline
\hline          
\text{$z$-bin} &   \text{$C^{\mathrm{Data}}_{\mathrm{sample}}$} &  \text{$C^{\mathrm{Balrog}}_{\mathrm{sample}}$} & \text{$C^{\mathrm{Final}}_{\mathrm{sample}}$} \\
\hline  
0 & \, $-0.21\pm0.03$  \, & \, $0.32\pm0.40$   & \, $0.05\pm0.48$ \\
1 & \, $2.20\pm0.04$ \, & \, $3.02\pm0.63$ & \, $2.61\pm0.75$\\
2 & \, $3.88\pm0.04$ \, & \, $4.70\pm0.59$ & \, $4.29\pm0.72$\\
\hline  
\end{tabular}
\vspace{1mm}
\\ North region (\emph{Planck}) \\
\vspace{0.5mm}
\renewcommand{\arraystretch}{1.02}
\begin{tabular}{cccc}
\hline
\hline          
\text{$z$-bin} &   \text{$C^{\mathrm{Data}}_{\mathrm{sample}}$} &  \text{$C^{\mathrm{Balrog}}_{\mathrm{sample}}$} & \text{$C^{\mathrm{Final}}_{\mathrm{sample}}$} \\
\hline  
0 & \, $-0.19\pm0.03$  \, & \, $0.29\pm0.46$  & \,  $0.05\pm0.52$ \\
1 & \, $2.15\pm0.04$ \, & \, $2.67\pm0.66$ & \, $2.41\pm0.71$\\
2 & \, $3.79\pm0.05$ \, & \, $4.85\pm0.65$ & \, $4.32\pm0.83$\\
\hline  
\end{tabular}
\vspace{1mm}
\\ South region (SPT) \\
\vspace{0.5mm}
\renewcommand{\arraystretch}{1.02}
\begin{tabular}{cccc}
\hline
\hline          
\text{$z$-bin} &   \text{$C^{\mathrm{Data}}_{\mathrm{sample}}$} &  \text{$C^{\mathrm{Balrog}}_{\mathrm{sample}}$} & \text{$C^{\mathrm{Final}}_{\mathrm{sample}}$} \\
\hline  
0 & \, $-0.23\pm0.04$  \, & \, $0.34\pm0.43$   & \, $0.05\pm0.52$ \\
1 & \, $2.23\pm0.06$ \, & \, $3.33\pm0.97$ & \, $2.78\pm1.12$\\
2 & \, $3.95\pm0.04$ \, & \, $4.54\pm0.69$ & \, $4.25\pm0.75$\\
\hline  
\end{tabular}
\end{table}

\subsection{ Estimate from perturbing  measured fluxes }
\label{sec:data_method}

The second method we consider uses the data itself to estimate the flux gradient of the samples. In this case, we add a constant offset $\Delta m$ to all photometric magnitudes in our sample:
\begin{equation}
    \label{eq:delta_magnitude}
    \Delta m = -2.5 \log_{10}( 1 + 2 \Delta \kappa ),
\end{equation}
where $\Delta \kappa=0.01$ is the constant magnification difference we are applying to each galaxy.

  
Using this new \emph{magnified} data sample, we repeat the assignment of the detected galaxies to the three \highz bins, and estimate $C_{\mathrm{sample}}$ from the differential in the resultant counts in each bin, directly from Eq.~(\ref{eq:c_estimate}), again accounting for individual galaxy weights from \S\ref{sec:weights}. This method provides an additional estimate of the magnification coefficients using only the magnification effect on the fluxes, hence ignoring other possible contributions from size selection or observational systematics. 
\subsection{Results}
Table \ref{tab:magnification} shows the estimates of $C_{\mathrm{sample}}$ using the \balrog and data-based methods described above, for the three tomographic bins and the North and South regions defined in this work. Since we have two independent methods to estimate these values, we use the average of the two methods as our final estimates $C^{\mathrm{Final}}_{\mathrm{sample}}$. For the associated uncertainties, we follow a conservative approach and add the uncertainties of the methods in quadrature, in addition to the standard deviation between the methods: 
\begin{equation}
    \sigma^{\mathrm{Final}}_C = \sqrt{\left(\sigma^{\mathrm{Balrog}}_C\right)^2 + \left(\sigma^{\mathrm{Data}}_C\right)^2 + (C^{\mathrm{Balrog}}_{\mathrm{sample}} - C^{\mathrm{Data}}_{\mathrm{sample}})^2/4}
\end{equation}
The derived magnification coefficients and their associated uncertainties will be used as Gaussian priors in the galaxy clustering analysis presented in the next section.


\section{Galaxy clustering and constraints on cosmology and galaxy bias}
\label{sec:clustering}

In this section we present the analysis of galaxy clustering in the tomographic bins defined in this work. We describe the model we use, the choice of scales, the measurements and covariance, and finally the constraints we obtain on the cosmological model and the galaxy bias of each tomographic bin, and their robustness under different analysis choices. 

\subsection{Model}
\label{sec:model}

Following the galaxy clustering analysis of the DES Year 3 fiducial sample \citep{y3-galaxyclustering}, we model the observed projected galaxy density contrast $\delta_{\mathrm{obs}}^{i}(\hat{\mathbf n})$ of galaxies in tomography bin $i$ at position $\hat{\mathbf n}$ as
\begin{equation}
\label{eq:deltas}
\delta_{g,\mathrm{obs}}^{i}(\hat{\mathbf n}) = \delta_{g,\mathrm{D}}^{i}(\hat{\mathbf n}) + \delta_{g,\mathrm{RSD}}^{i}(\hat{\mathbf n}) + \delta_{g,\mu}^{i}(\hat{\mathbf n})\,.
\end{equation}
The first term is the line-of-sight projection of the three-dimensional galaxy density contrast, $\delta_{g}^{(\rm 3D)}$; the other terms correspond the contributions from linear redshift-space distortions (RSD) and magnification ($\mu$), which are described in detail in \citet{y3-generalmethods}. We relate the galaxy density to the matter density assuming a local, linear galaxy bias model \citep{Fry1993}, $\delta_g({\bf{x}})=b\delta_m(\bf{x})$, with $\delta_Y \equiv (Y({\bf{x}})-{\bar{Y}})/{\bar{Y}}$. We assume the galaxy bias to be constant across each tomographic bin $b^i$, and we discuss more about this assumption later in this section. 

Given the three terms in Eq.~\ref{eq:deltas}, the angular power spectrum $C^{ii}_{\delta_{g,\mathrm{obs}}\delta_{g,\mathrm{obs}}}(\ell)$ has six different components, corresponding to the auto- and cross-power spectra of galaxy density, RSD, and magnification. For the accuracy of the DES Year 3 analysis, it was shown by \citet{y3-generalmethods} that the commonly-used Limber approximation is insufficient to estimate these terms, and therefore we use the non-Limber algorithm of \citet{Fang:2020}\footnote{\url{https://github.com/xfangcosmo/FFTLog-and-beyond}}. Using the full expressions for the angular power spectrum, including RSD and magnification, from \citet{Fang:2020}, the angular correlation function is given by:
\begin{align}
\label{eq:wtheta_model}
    w^i(\theta) =& \sum_\ell \frac{2\ell+1}{4\pi}P_\ell(\cos\theta) C^{ii}_{\delta_{g,\mathrm{obs}}\delta_{g,\mathrm{obs}}}(\ell)~,
\end{align}
where $P_\ell$ are the Legendre polynomials. For the implementation of these calculations, we use the \textsc{CosmoSIS} framework\footnote{\url{https://bitbucket.org/joezuntz/cosmosis}} \citep{Zuntz2015}, which in turn uses \textsc{CAMB} \citep{LewisBridle:2002} to obtain the evolution of linear density fluctuations and \textsc{Halofit} \citep{Takahashi2012} to convert to a non-linear matter power spectrum. The modeling of redshift uncertainties has been described in detail in \S\ref{sec:redshift_uncertainty}, and that parametrization has been implemented in \textsc{CosmoSIS} for this analysis. 

In addition, as explained in \S\ref{sec:weights}, we marginalize over an additive constant parameter, parametrized by $R^i$, in the galaxy angular correlation function: 
\begin{equation}
w^i(\theta) \rightarrow w^i(\theta) + 10^{R^i}.     
\label{eq:Ri}
\end{equation}
This parametrization accounts for potential residuals in the calculation of galaxy weights affecting the galaxy clustering measurements \citep{Kwan2016}. Later in \S\ref{sec:analysis} we will explore the impact of the choice of maximum angular scale in the galaxy clustering measurements.


\subsubsection{Choice of scales}
\label{sec:scale_cuts}
Given the fact that we assume a linear galaxy bias model for this analysis, we are required to remove small-scale information that can potentially be affected by non-linearities. We follow the approach of the DES Year 3 fiducial analysis \citep{y3-3x2ptkp} and we remove all galaxy clustering information below 8$h^{-1}$Mpc \citep{y3-generalmethods} (corresponding to a minumum angular scale of  12.9, 10.5 and 9.0 arcmins for the three tomographic bins in this work, respectively). We also test for the robustness of the results to a minimum scale of 12$h^{-1}$Mpc. The maximum angular scale we use is set to 60 arcmins for all measurements. This choice is driven by the correction method of obtaining galaxy weights, described in \S\ref{sec:weights}, in particular by the cross-validation scheme to avoid overfitting, which shows no signs of overfitting at angular scales below 1 degree.


\subsection{Measurements and covariance}
\label{sec:measurements}

\begin{figure*}
 \centering
 \includegraphics[width=0.98\textwidth]{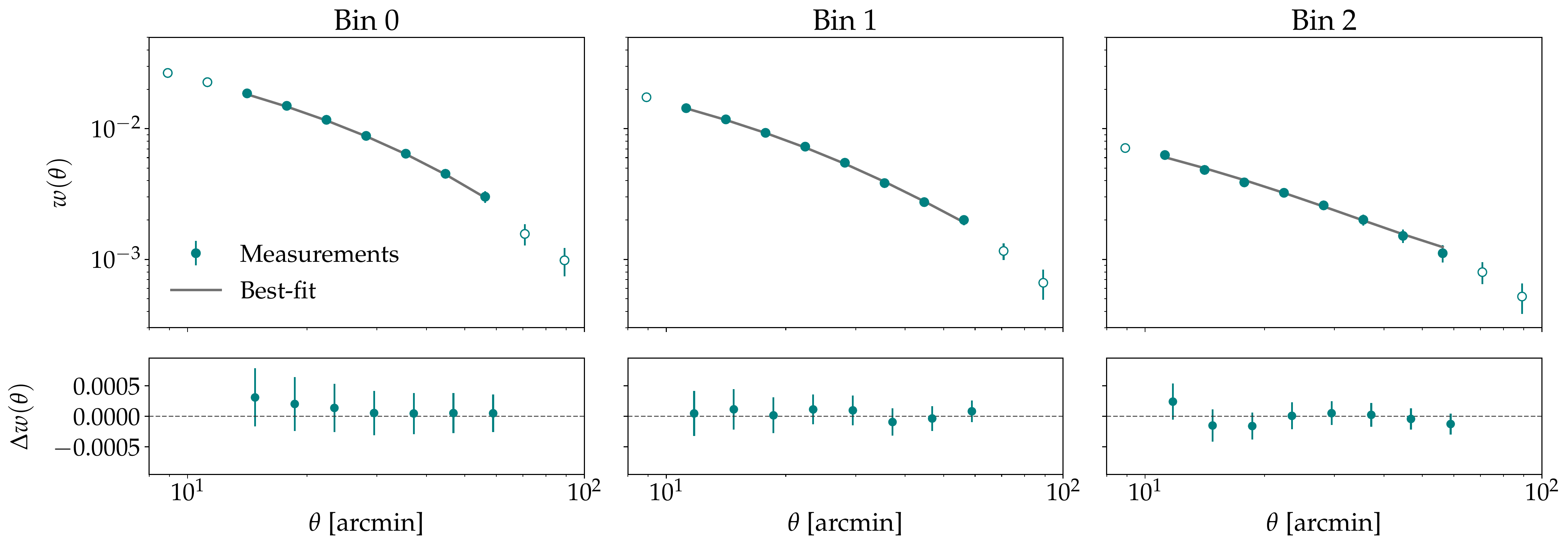}
 \caption{\emph{(Upper panels:)} Measurements of the auto-correlations of angular galaxy clustering for the three redshift bins ($0,1,2$) defined in this work, for the entire DES Y3 footprint we use. Filled colored points correspond to the measurements passing the scales cuts defined in \S\ref{sec:scale_cuts}. The methodology for the measurements and covariance, and the calculation of the corresponding signal-to-noise, can be found in \S\ref{sec:measurements}. The solid lines show the best-fit theory for the fiducial analysis choices, as described in \S\ref{sec:analysis}. The goodness of fit in that case corresponds to $\chi^2$/dof = 8.3/8.6. Error bars are smaller than the symbols, if not indicated. \emph{(Lower panels:)} Residuals of the measurements given the best-fit theory model shown in the upper panels. } 
 \label{fig:measurements}
\end{figure*}

Equation (\ref{eq:wtheta_model}) shows the modeling of the galaxy angular 2-point correlation function, $w(\theta)$. For the measurement of this galaxy clustering observable, we use \textsc{Healpix} maps (\texttt{nside} = 4096) of the corrected galaxy density contrast for each tomographic bin, including the correction weights described in \S\ref{sec:weights}, and then use a pixel-based version of the Landy-Szalay estimator \citep{LS}, following the notation of \citet{2016MNRAS.455.4301C}:
\begin{equation}
\hat{w}(\theta) = 
\sum_{i=1}^{N_{pix}}\sum_{j=1}^{N_{pix}} \frac{(N_i-\Bar{N})\cdot(N_j-\Bar{N})}{\Bar{N}^2} \,\omega_i\,\omega_j\, \Theta_{i,\, j}(\theta) \, ,
\end{equation} 
where $N_i$ is the galaxy number density in pixel $i$, and $\omega_i$ is the weight of each pixel $i$ (see \S\ref{sec:weights}). $\Bar{N}$ is the corrected mean galaxy number density over all pixels within the footprint and $\Theta_{i, \, j}$ is a top-hat function which is equal to $1$ when pixels $i$ and $j$ are separated by an angle $\theta$ within the bin size $\Delta \theta$. In practice, these correlation functions are computed using \textsc{TreeCorr}\footnote{\url{https://rmjarvis.github.io/TreeCorr}} \citep{Jarvis2004}. Figure \ref{fig:measurements} shows the $w(\theta)$ measurements for the galaxy auto-correlations of the three redshift bins considered in this work. 


We estimate the covariance matrices using two complementary methods: using Gaussian simulations, and using Jackknife. The Gaussian simulations are generated following the procedure described in~\citet{Giannantonio2008} (see Appendix~\ref{sec:appendix_cov} for details). We generate 100 realizations of a set of four correlated maps via \textsc{HEALPix} \textsc{anafast} routine. These maps, three for galaxy overdensity and one for CMB $\kappa$, are generated using the non-linear (\textsc{Halofit}) power spectrum with our fiducial cosmology. Each map includes its respective (uncorrelated) noise contribution. The advantage of this simulation-based approach is that it allows us to have an accurate estimation of the effects of the mask, and angular binning. The main downside is that this approach does not account for the non-Gaussian terms of the covariance. In order to cross-check the validity of this approach, we also estimate the covariance using the Jackknife technique, defining 150 subsamples for the measurements in \textsc{TreeCorr}. We find that both approaches are in good agreement within the range of scales used for this work, pointing to a negligible contribution of the non-Gaussian terms for this particular study. A detailed comparison can be found in Appendix~\ref{sec:appendix_cov}.  

Defining these data measurements as $\hat{\mathbf D} \equiv\{\hat{w}^{ij}(\theta)\}$ and the covariance $\mathbf{C}$, we use the following expression to compute the signal-to-noise of the measurements:
\begin{equation}
    S/N = \sqrt{\hat{\mathbf D} \: \mathbf{C}^{-1} \: \hat{\mathbf D}^T - \mathrm{ndf}},
\end{equation}
where ndf is the number of degrees of freedom, which equals the number of data points passing the scale cuts defined in \S\ref{sec:scale_cuts}. For reference, the fiducial DES Year 3 analysis had a galaxy clustering $S/N = 63$ \citep{y3-galaxyclustering}. For the sample in this work, the total $S/N$, including the 3 auto-correlations after applying scale cuts, is $S/N = 70$. Breaking this into the individual measurements, the auto-correlations for bins 0, 1 and 2 get $S/N = 43$, $49$, and $37$, respectively. 

\subsection{Analysis and results}
\label{sec:analysis}

\begin{table}
\caption{The model parameters and their priors used in the fiducial flat $\Lambda$CDM analysis, using the entire DES Y3 footprint. The parameters are defined in Sec.~\ref{sec:analysis}.}
\begin{center}
\renewcommand{\arraystretch}{1.05}
\begin{tabular*}{\columnwidth}{ l  @{\extracolsep{\fill}} c  c}
\hline
\hline
Parameter & \multicolumn{2}{c}{Prior}  \\  
\hline 
\multicolumn{2}{l}{{\bf Cosmology}} \\
$\Omega_{\mathrm{m}}$  &  Flat  & (0.1, 0.9)  \\ 
$10^{9}A_{\mathrm{s}}$ &  Flat  & ($0.5,5.0$)  \\ 
$n_{\mathrm{s}}$ &  Flat  & (0.87, 1.07)  \\
$\Omega_{\mathrm{b}}$ &  Flat  & (0.03, 0.07)  \\
$h$  &  Flat  & (0.55, 0.91)   \\
$10^{3}\Omega_\nu h^2$  & Flat  & ($0.60$, $6.44$) \\
\hline
\multicolumn{2}{l}{{\bf Galaxy Bias} } \\
$b^{i} (i\in[0,2])$   & Flat  & (0.8, 3.0) \\
\hline
\multicolumn{2}{l}{{\bf Weight residuals} } \\
$R^0 $ & Flat &  ($-8, -2$) \\
$R^1 $ & Flat &  ($-8, -2$) \\
$R^2 $ & Flat &  ($-8, -2$) \\
\hline
\multicolumn{2}{l}{{\bf Lens magnification} } \\
$C^0 $ & Gaussian &  ($0.0275, 0.24$) \\
$C^1 $ & Gaussian &  ($1.305, 0.375$) \\
$C^2 $ & Gaussian &  ($2.145, 0.36$) \\
\hline
\multicolumn{2}{l}{{\bf Redshifts } } \\
$\Delz{0}$   & Gaussian  & ($0.0, 0.0051$) \\
$\Delz{1}$  & Gaussian  & ($0.0, 0.0075$) \\
$\Delz{2}$  & Gaussian  & ($0.0, 0.0208$) \\
$\Sigz{0}$  & Gaussian  & ($0.997, 0.068$) \\
$\Sigz{1}$  & Gaussian  & ($0.999, 0.041$) \\
$\Sigz{2}$  & Gaussian  & ($0.998, 0.044$) \\
$\Alowz{0}$  & Gaussian  & ($0.0044, 0.0013$) \\
$\Alowz{1}$  & Gaussian  & ($0.0091, 0.0023$) \\
$\Alowz{2}$  & Gaussian  & ($0.0383, 0.0059$) \\
\hline
\hline
\end{tabular*}
\end{center}
\label{tab:params}
\end{table}

\subsubsection{Parameter inference}

In this part we are interested in placing model constraints given the measured two-point functions of galaxy clustering shown in Fig.~\ref{fig:measurements}. In general, given our model $M$, we want to infer parameters $\mathbf p$ from the set of measured two-point correlation functions in our data, $\hat{\mathbf D}$.
The theoretical model prediction for the two-point correlation functions, computed using the parameters $\mathbf p$ of the model ${M}$, is $\mathbf T_M(\mathbf{p}) \equiv \{w^{ij}(\theta,\mathbf p)\}$. We compare the measurements and model predictions using a Gaussian likelihood, using the data covariance, $\mathbf{C}$, defined above:
\begin{equation}
\mathcal{L}(\hat{\mathbf D}|\mathbf p, M) \propto e^{-\frac{1}{2}\left[\left(\hat{\mathbf D}-\mathbf T_M(\mathbf p)\right)^{\mathrm{T}} \mathbf{C}^{-1}\left(\hat{\mathbf D}-\mathbf T_M(\mathbf p)\right)\right]}.
\end{equation}
In this way, the posterior probability distribution for the parameters $\mathbf p$  of the model $M$ given the data $\hat{\mathbf D}$ is given by
\begin{equation}
P(\mathbf p |\hat{\mathbf D}, M) \propto \mathcal{L}(\hat{\mathbf D}|\mathbf{p}, {M})P(\mathbf{p}|{M}),
\end{equation}
where $P(\mathbf{p}|{M})$ is the prior probability distribution on the parameters.


We sample the posterior of the galaxy clustering measurements in the flat $\Lambda$CDM model, using the same parameter space as the DES Year 3 fiducial analysis \citep{y3-3x2ptkp}. The six cosmological parameters we vary are listed in Table \ref{tab:params}, together with their respective uniform priors. These prior ranges are chosen to encompass at least five times the 68\% C.L. from relevant external constraints. Also, even though we sample the amplitude of primordial scalar density perturbations $A_{\mathrm{s}}$, sometimes we will refer to the amplitude of density perturbations at $z=0$ in terms of the RMS amplitude of mass on scales of $8h^{-1}$ Mpc in linear theory, $\sigma_8$. In addition to these cosmological parameters, our fiducial analysis includes 18 nuisance parameters to describe: galaxy bias (see \S\ref{sec:model}),  potential residuals in the galaxy weight calculation (see \S\ref{sec:weights}), lens magnification (see \S\ref{sec:magnification}) and uncertainties in the redshift distribution of our three redshift bins (see \S\ref{sec:redshift_uncertainty}), all of them described in Table \ref{tab:params}.

\subsubsection{DES Y3 High-$z$ results and robustness tests}
\label{sec:robustness}

Next we analyze the model constraints from the measurements of galaxy clustering. In this case, there exists a strong degeneracy between galaxy bias and the amplitude of matter fluctuations, $\sigma_8$, and therefore the analysis presented here is not sensitive to $\sigma_8$. The combination of clustering and weak gravitational lensing can be used to break these degeneracies, and that will be presented in a companion paper (\emph{in preparation}), using CMB lensing from the South Pole Telescope (SPT) and \emph{Planck}. However, for the clustering-only case analyzed here, the shape of the galaxy clustering measurements is sensitive to the scale of matter-radiation equality in the matter power spectrum, which in turn depends on a combination of the matter density $\Omega_m$ and the Hubble constant $h$, close to the direction $\Omega_m h$ (see e.g.~\citealt{2021PhRvD.103b3538P}). 

\begin{figure*}
 \centering
 \includegraphics[width=0.98\textwidth]{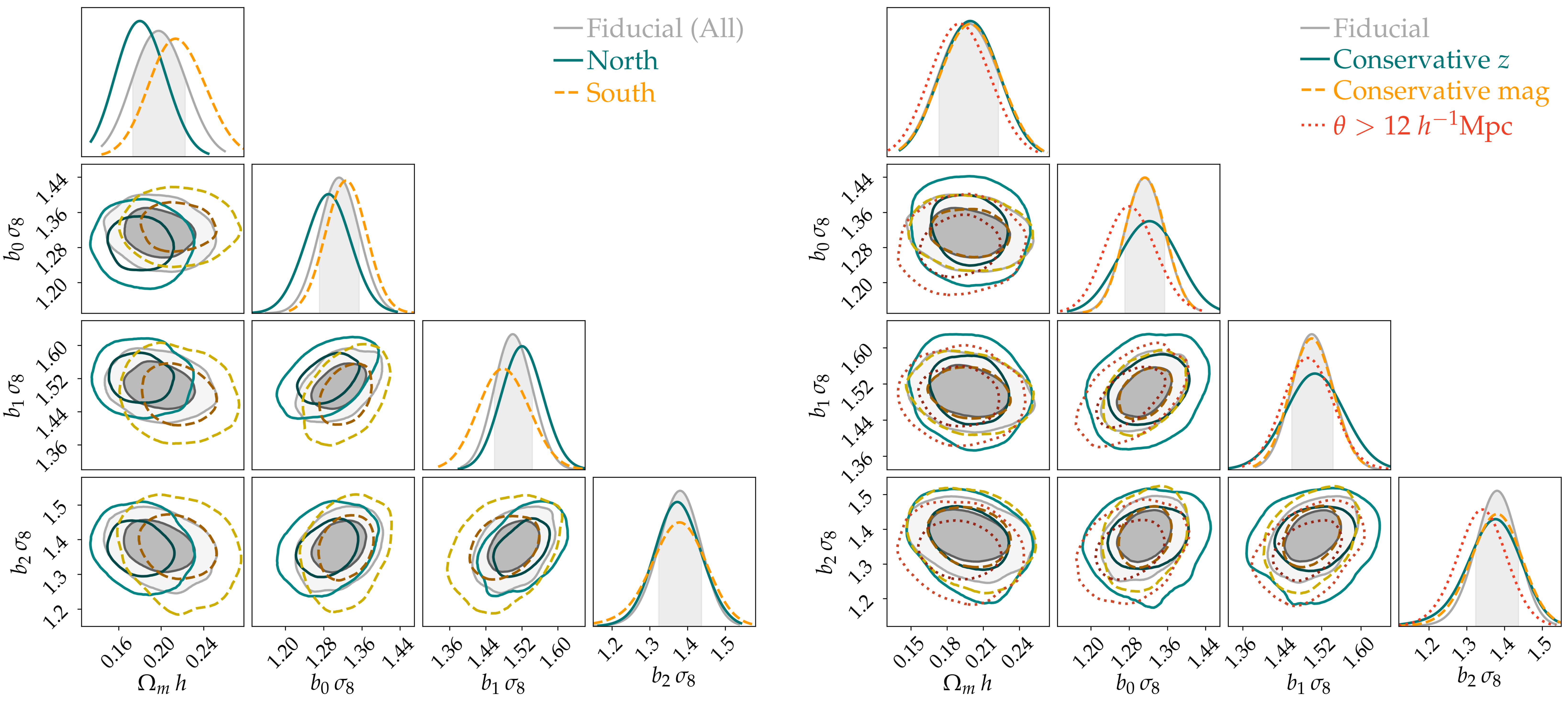}
 \caption{Constraints on the combination of cosmological parameters and galaxy bias derived from out measurements of galaxy clustering, for various analysis configurations. The left panel shows the fiducial constraints using the entire footprint (All), compared to the constraints using the independent splits in North and South regions. The right panel shows the comparison between the fiducial constraints and three analysis variations, one with conservative redshift priors ($\times 2$ width in all redshift parameter priors), one with conservative magnification priors ($\times 2$ width in all magnification parameter priors), and larger minimum angular scales. }
 \label{fig:chains}
\end{figure*}

\begin{figure}
 \centering
\includegraphics[width=0.48\textwidth]{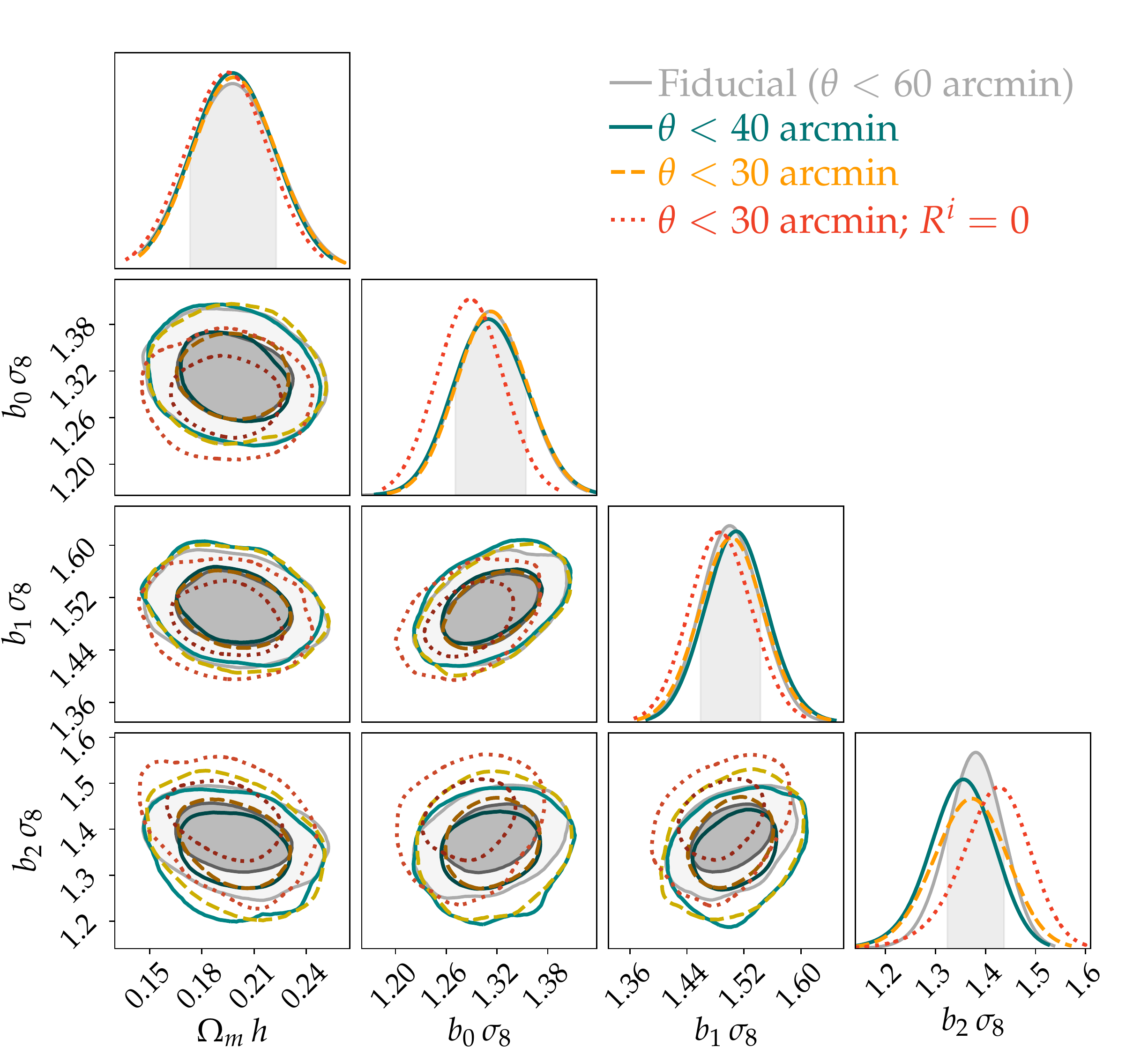}
 \caption{Comparison of the parameter constraints from galaxy clustering using different choices for the maximum angular scale, as well as not marginalizing over an additive constant in the galaxy clustering measurements.  }
 \label{fig:theta_max}
\end{figure}

Figure \ref{fig:chains} shows the constraints we obtain for the parameters we are sensitive to, namely $\Omega_m h$ and the product of $\sigma_8 b^i$ for the three redshift bins we use. The fiducial constraints use the entire survey footprint, the auto-correlations shown in Fig.~\ref{fig:measurements}, the scale cuts described in \S\ref{sec:scale_cuts} and the priors shown in Table \ref{tab:params}, and they result in constraints on a combination of the fraction of matter in the Universe $\Omega_m$ and the Hubble parameter $h$, $\Omega_m h = 0.195^{+0.023}_{-0.018}$, and 2-3\% measurements of the amplitude of the galaxy clustering signals for the three redshift bins, probing galaxy bias and the amplitude of matter fluctuations, $b \sigma_8$. The best-fit theory model for this fiducial case is shown together with the measurements in Fig.~\ref{fig:measurements}, and the corresponding $\chi^2$/ndf is 8.3/8.1, where ndf is the estimated effective number of degrees of freedom. Using the UDM (Update Difference in Means) tension metric from \citet{y3-tensions}, we find the posterior constraints to be compatible with the redshift prior, with a tension of 0.34$\sigma$, and also compatible with the magnification prior, with a tension at 0.03$\sigma$.

In addition, to assess the robustness of the results, in Fig.~\ref{fig:chains} we show constraints for various alternative cases. First, we analyze the constraints we obtain from the independent North and South regions, where we split the data into two independent patches: "North" (DEC $> -39^{\rm o}$) and "South" (DEC $< -40^{\rm o}$). This is motivated by the fact that we will combine the clustering measurements shown here with CMB lensing measurements from \emph{Planck} and SPT in a companion paper \emph{(in prep.)}. Since SPT only covers the South region in this split, we do this test to check for the consistency of the clustering measurements. In this test, the  redshift and magnification priors are computed specifically for each region, although they are largely consistent (see Tables \ref{tab:redshift} and \ref{tab:magnification}), and the galaxy clustering measurements are also performed separately for the two regions. The analysis of the North and South regions yields best-fit theory models with $\chi^2$/ndf is 10.1/8.6 and $\chi^2$/ndf is 15.7/8.6, respectively. When using the entire parameter space, the constraints from the two independent regions are in agreement, with an estimated tension of 0.65$\sigma$, using the non-Gaussian parameter difference tension metric from \citet{2021PhRvD.104d3504R,y3-tensions}. When restricting the set of parameters to $\Omega_m$, $\Omega_m \: h$, $b^0 \: \sigma_8$, $b^1 \: \sigma_8$, $b^2 \: \sigma_8$, the constraints from the independent North and South regions are also in agreement, with an estimated tension of 0.41$\sigma$.

Figure \ref{fig:chains} also shows the galaxy clustering constraints under some different analysis choices. In particular, we study the impact of redshift and magnification priors, both described in Table \ref{tab:params}, by studying the conservative case of doubling the width these priors. When broadening the width of redshift priors by a factor of 2, the constraints on $b^0 \sigma_8$, $b^1 \sigma_8$ and $b^2 \sigma_8$ widen by a factor of 1.47, 1.41 and 1.27, respectively. When broadening the width of magnification priors by a factor of 2, the constraints on $b^2 \sigma_8$ broaden by a factor of 1.20. Therefore, redshift priors are relevant for all bins, especially for bins 0 and 1, while lens magnification is only relevant in bin 2, at higher redshift. None of these changes has an important effect on $\Omega_m h$, which shows very robust constraints under all different analysis choices. Using larger minimum angular scales, corresponding to $12h^{-1}$ Mpc, as opposed to the fiducial $8h^{-1}$ Mpc, broadens the constraints on $b^i \sigma_8$ by a factor of 1.28, 1.21 and 1.17 for bins $i=$ 0, 1 and 2, while having no significant effect on $\Omega_m h$. 

We also explore the impact of the choice of maximum angular scale on the clustering analysis. The fiducial value for the maximum angular scale is 60 arcmins, driven by the method used to correct for correlations between galaxy density and survey properties. In order to account for any residuals coming from that method, we also marginalize over an additive constant parameter for each tomographic bin $R^i$ (see Eq.~\ref{eq:Ri}). Figure \ref{fig:theta_max} shows the galaxy clustering constraints when limiting the maximum angular scale to 40 and 30 arcmins, and also, for the latter case, when not marginalizing over additive constants (setting $R^i = 0$). The figure shows how the galaxy clustering constraints are robust to these choices. The constraints on $\Omega_m h$ are not sensitive to the variations, and the main impact of limiting the maximum angular scale is a $\sim$20\% decrease in constraining power for $b^2 \sigma_8$. Regarding the posterior values of $R^i$, we find $R^0 = -5.13 ^{+0.84}_{-1.93}$, $R^1 = -3.42 ^{+0.31}_{-0.65}$, $R^2 = -3.21 ^{+0.06}_{-0.09}$. We can see how this parameter is constrained to be very small for the first bin, and its importance grows with redshift (and $i$-band magnutude) of the tomographic bin.

\subsubsection{Blinding procedure}

In order to minimize a potential impact of experimenter bias, we have adopted a blinding procedure throughout this work. In that way, we have kept the results on the main parameters constrained in this analysis (those depicted in Figs.~\ref{fig:chains} and \ref{fig:theta_max}) blinded to the analysis until the robustness tests performed in \S\ref{sec:robustness} satisfied the tension metrics reported there. An internal review committee set up by the DES collaboration was in charge of over-viewing this procedure and allowing for the unblinding of the constraints. 

\subsubsection{Comparison with other DES Y3 clustering analyses}

\begin{figure}
 \raggedleft
\includegraphics[width=0.48\textwidth]{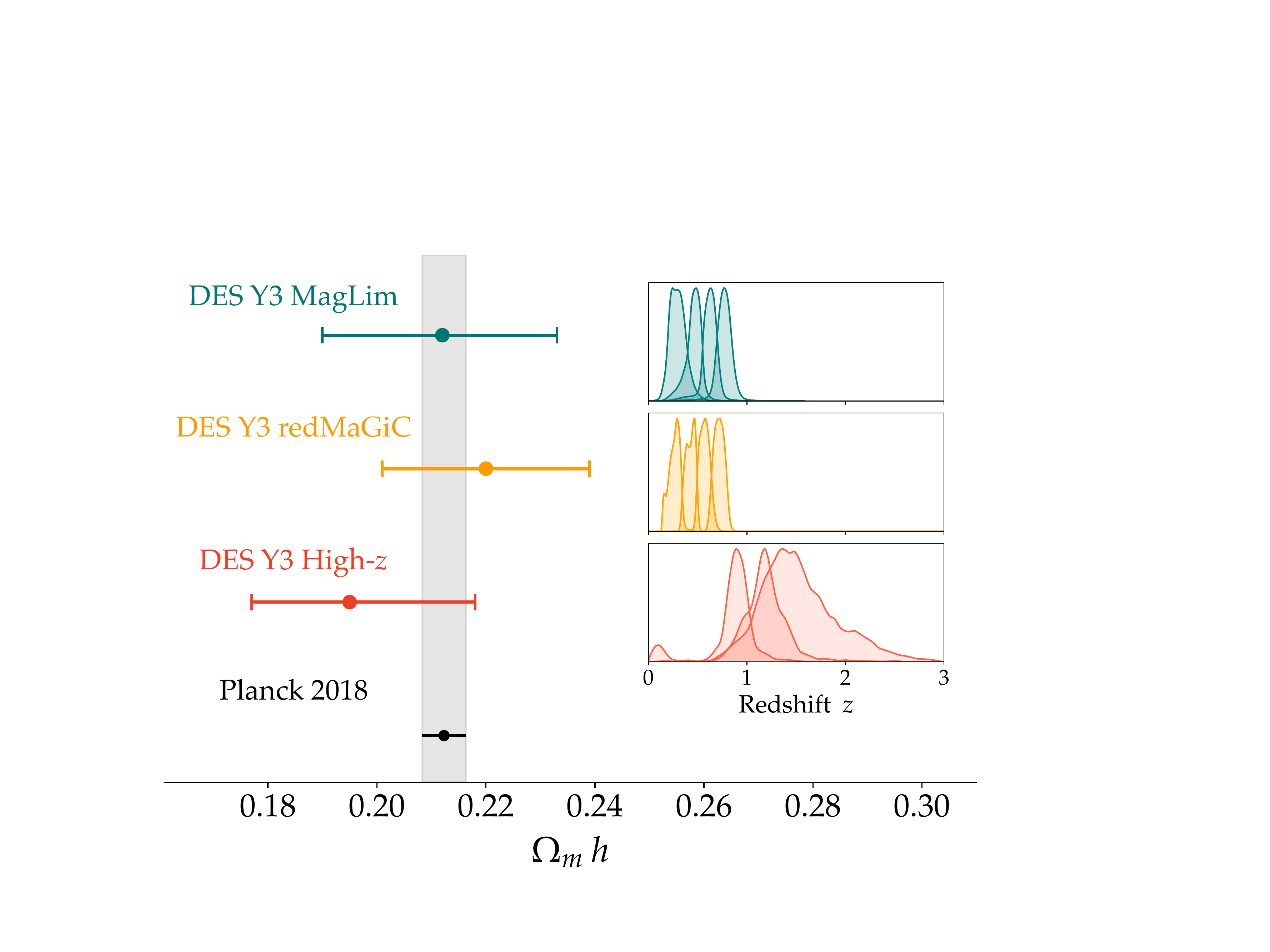}
 \caption{Comparison of the constraints on the parameter combination $\Omega_m h$ from galaxy clustering analyses using three different lens samples in DES Y3. The constraints from the DES Y3 Fiducial sample, also called MagLim sample \citep{y3-2x2ptaltlensresults}, are shown in blue; the constraints from the redMaGiC sample are shown in yellow and the constraints from the High-$z$ sample described in this work are shown in red. The Planck 2018 constraint is shown in black. The inset panel on the right of the plot depicts the different redshift range probed by the three DES Y3 lens samples. }
 \label{fig:cosmo_comparison}
\end{figure}

Given the parameter constraints obtained in the analysis of galaxy clustering with the DES Y3 High-$z$ sample presented in this work, we can now compare how these constraints compare with the corresponding clustering analyses of the other DES Y3 lens samples already defined and used in other works. The fiducial DES Y3 lens sample is the so-called \textsc{MagLim} sample \citep{y3-2x2maglimforecast}, while the alternative lens sample is \textsc{redMaGiC} \citep{y3-2x2ptbiasmodelling} (see Table \ref{tab:samples} for a comparison of the number densities of the three samples). Figure \ref{fig:cosmo_comparison} shows the constraints on the cosmological parameter combination of $\Omega_m h$ provided by each of the three DES Y3 lens samples, together with the Planck 2018 constraint. The figure shows the DES Y3 constraints to be in agreement between the three samples, and with the Planck result, and also having similar constraining power. However, while the constraints from \textsc{MagLim} and \textsc{redMaGiC} probe similar redshift ranges, the High-$z$ constraints come from significantly higher redshifts, extending the redshift range probed by the DES Y3 data. This results demonstrate the robustness of the clustering measurements in this work and our ability to produce a well-characterized high-redshift sample, which is complimentary to the DES fiducial analysis in terms of the redshift range it probes.  Note that the upcoming analyses combining the High-$z$ galaxy clustering presented in this work with cross-correlation with weak gravitational lensing will be able to break the degeneracy between galaxy bias and the amplitude of matter fluctuations, $\sigma_8$, allowing us to place constraints on the latter at higher redshifts than probed in the fiducial DES analysis.

\section{Summary and outlook}
\label{sec:summary}

The cosmological analysis of imaging galaxy surveys
provides powerful measurements of the amplitude of matter fluctuations in the late time Universe. In recent years, the analyses of different surveys like DES, KiDS and HSC, probing the regime at $z<1$, have reported persistent tensions with the predicted value from the CMB, a problem known as the $S_8$ tension. Measurements at a higher redshift regime ($1 < z < 3$) would be crucial for understanding the origin of this tension. In addition, such measurements would probe the matter-dominated epoch and would shed light on dynamical dark energy models that can mimic a cosmological constant at late times but differ substantially during the matter-dominated era.

In this work we describe the selection and characterization of three galaxy samples covering the approximate redshift range $0.8 < z < 2.5$ (see Figure \ref{fig:nz_comparison}) using data from the third year of the Dark Energy Survey (DES Y3). To enable the selection and characterization of these high-$z$ samples, which push the limits of DES Y3 data, we introduce several changes with respect to the fiducial DES Y3 lens galaxy sample: 

\begin{enumerate}
    \item We start from a fainter galaxy selection, excluding all lens galaxies used in the DES Y3 fiducial analysis. The average $i$-band magnitude of the three High-$z$ redshift bins is 22.6, 22.9 and 23.1, respectively, while all four redshift bins used in the fiducial analysis had average $i$-band magnitudes brighter than $i=22$. 
    \item Both the selection and redshift characterization of the samples are based on a principled, Bayesian scheme using a novel \emph{Self-Organizing Map} (SOM) algorithm better suited for the characterization of lower S/N galaxies \citep{Sanchez2020}. 
    \item We use a redshift marginalization scheme that explicitly accounts for uncertainties in the tails of redshift distributions. 
    \item We use a non-linear, machine-learning-based approach to correct for correlations between galaxy number density and survey observing properties like depth, stellar density and sky noise.
\end{enumerate}

Out of this list of changes with respect to the fiducial analysis, steps (i) and (ii) are responsible for the selection of high redshift galaxies, and steps (iii) and (iv) are required due to the faint, high redshift selection. The procedure results in the definition of three redshift bins with mean redshifts around $z = 0.9, 1.2$ and $1.5$, which
significantly extend the redshift coverage of the fiducial DES Year 3 analysis. In addition, these samples contain a total of about 9 million galaxies, resulting in a galaxy density that is more than 2 times higher than those in the DES Year 3 fiducial case \citep{y3-2x2ptaltlensresults}. 

After the selection and characterization of the High-$z$ galaxy samples, we perform an analysis of their galaxy clustering auto-correlation measurements. The analysis provides robust constraints on the product of the fraction of matter in the Universe $\Omega_m$ and the Hubble parameter $h$, $\Omega_m h = 0.195^{+0.023}_{-0.018}$, and 2-3\% measurements of the amplitude of the galaxy clustering measurements for the three redshift bins, probing galaxy bias times the amplitude of matter fluctuations, $b \sigma_8$. The constraints on $\Omega_m h$ are compatible and show comparable uncertainties to the clustering analyses on the fiducial and alternative lens galaxy samples using DES Y3 data \citep{y3-2x2ptaltlensresults,y3-2x2ptbiasmodelling}, but probing a complementary, much higher redshift range. This part also showcases the robustness of the galaxy clustering analysis, which is highly non-trivial when using galaxy samples going as faint as $i\sim23$ in DES Y3 data. 

The definition and characterization of high redshift galaxy samples in this work represents the first step to analyze the $0.8 < z < 2.5$ redshift range made by DES and other Stage III surveys. It therefore develops the tools that will enable similar analyses with other data sets, including Rubin LSST and Euclid, and it  opens the door to a range of scientific analyses exploiting the unique nature of the selections. In subsequent publications, we will explore this set of applications using the samples defined in this work. We will present the cross-correlation of High-$z$ galaxies with CMB lensing maps from SPT and \emph{Planck}, providing crucial constraints on $S_8$ at high redshift \citep{2020A&A...641A...8P,2022arXiv220312439O}. We will also study their cross-correlations with galaxy lensing, probing $S_8$, lensing magnification and intrinsic alignments at high redshifts, and the clustering cross-correlations with lower-redshift galaxies, probing lensing magnification and the redshift evolution of galaxy bias. The redshift regime of these samples is also well suited to study the star formation history using cross-correlations with the Cosmic Infrared Background (CIB) \citep{Jego22clustering,Jego22shear}. The outcome of these analyses will provide important information about this particularly unexplored period in the Universe, and will set the tools and expectations for future analyses with more powerful data sets. 

\section*{Acknowledgements}

CS is supported by a Junior Leader fellowship from the ”la Caixa” Foundation (ID 100010434), with code LCF/BQ/PI22/11910018. CS and GMB are supported by grants AST-2009210 from the U.S. National Science Foundation, and DE-SC0007901 from the U.S. Department of Energy (DOE). Argonne National Laboratory's work was supported by the U.S. Department of Energy, Office of High Energy Physics. Argonne, a U.S. Department of Energy Office of Science Laboratory, is operated by UChicago Argonne LLC under contract no. DE-AC02-06CH11357.

Funding for the DES Projects has been provided by the U.S. Department of Energy, the U.S. National Science Foundation, the Ministry of Science and Education of Spain, 
the Science and Technology Facilities Council of the United Kingdom, the Higher Education Funding Council for England, the National Center for Supercomputing 
Applications at the University of Illinois at Urbana-Champaign, the Kavli Institute of Cosmological Physics at the University of Chicago, 
the Center for Cosmology and Astro-Particle Physics at the Ohio State University,
the Mitchell Institute for Fundamental Physics and Astronomy at Texas A\&M University, Financiadora de Estudos e Projetos, 
Funda{\c c}{\~a}o Carlos Chagas Filho de Amparo {\`a} Pesquisa do Estado do Rio de Janeiro, Conselho Nacional de Desenvolvimento Cient{\'i}fico e Tecnol{\'o}gico and 
the Minist{\'e}rio da Ci{\^e}ncia, Tecnologia e Inova{\c c}{\~a}o, the Deutsche Forschungsgemeinschaft and the Collaborating Institutions in the Dark Energy Survey. 

The Collaborating Institutions are Argonne National Laboratory, the University of California at Santa Cruz, the University of Cambridge, Centro de Investigaciones Energ{\'e}ticas, 
Medioambientales y Tecnol{\'o}gicas-Madrid, the University of Chicago, University College London, the DES-Brazil Consortium, the University of Edinburgh, 
the Eidgen{\"o}ssische Technische Hochschule (ETH) Z{\"u}rich, 
Fermi National Accelerator Laboratory, the University of Illinois at Urbana-Champaign, the Institut de Ci{\`e}ncies de l'Espai (IEEC/CSIC), 
the Institut de F{\'i}sica d'Altes Energies, Lawrence Berkeley National Laboratory, the Ludwig-Maximilians Universit{\"a}t M{\"u}nchen and the associated Excellence Cluster Universe, 
the University of Michigan, NSF's NOIRLab, the University of Nottingham, The Ohio State University, the University of Pennsylvania, the University of Portsmouth, 
SLAC National Accelerator Laboratory, Stanford University, the University of Sussex, Texas A\&M University, and the OzDES Membership Consortium.

Based in part on observations at Cerro Tololo Inter-American Observatory at NSF's NOIRLab (NOIRLab Prop. ID 2012B-0001; PI: J. Frieman), which is managed by the Association of Universities for Research in Astronomy (AURA) under a cooperative agreement with the National Science Foundation.

The DES data management system is supported by the National Science Foundation under Grant Numbers AST-1138766 and AST-1536171.
The DES participants from Spanish institutions are partially supported by MICINN under grants ESP2017-89838, PGC2018-094773, PGC2018-102021, SEV-2016-0588, SEV-2016-0597, and MDM-2015-0509, some of which include ERDF funds from the European Union. IFAE is partially funded by the CERCA program of the Generalitat de Catalunya.
Research leading to these results has received funding from the European Research
Council under the European Union's Seventh Framework Program (FP7/2007-2013) including ERC grant agreements 240672, 291329, and 306478.
We  acknowledge support from the Brazilian Instituto Nacional de Ci\^encia
e Tecnologia (INCT) do e-Universo (CNPq grant 465376/2014-2).

This manuscript has been authored by Fermi Research Alliance, LLC under Contract No. DE-AC02-07CH11359 with the U.S. Department of Energy, Office of Science, Office of High Energy Physics.

This research used resources of the National Energy Research Scientific Computing Center (NERSC), a U.S. Department of Energy Office of Science User Facility located at Lawrence Berkeley National Laboratory, operated under Contract No. DE-AC02-05CH11231.

We thank the developers of \textsc{NumPy} \citep{numpy}, {\sc SciPy} \citep{scipy}, \textsc{Jupyter} \citep{jupyter}, \textsc{IPython} \citep{ipython}, \textsc{conda-forge} \citep{conda_forge_community_2015_4774216}, \textsc{Matplotlib} \citep{matplotlib} and \textsc{Keras} \citep{chollet2015keras} for their extremely useful free software.

\section*{Data Availability}

A general description of DES data releases is available on
the survey website at \href{https://www.darkenergysurvey.org/
the-des-project/data-access/}{https://www.darkenergysurvey.org/
the-des-project/data-access/}. DES Y3 cosmological data
has been partially released on the DES Data Management website
hosted by the National Center for Supercomputing Applications at
\href{https://des.ncsa.illinois.edu/releases/y3a2}{https://des.ncsa.illinois.edu/releases/y3a2}.


\bibliography{library}
\bibliographystyle{mnras_2author}



\appendix

\section{Removing stars from the Deep sample}
\label{sec:appendix_stars}
\begin{figure}
 \centering
\includegraphics[width=0.49\textwidth]{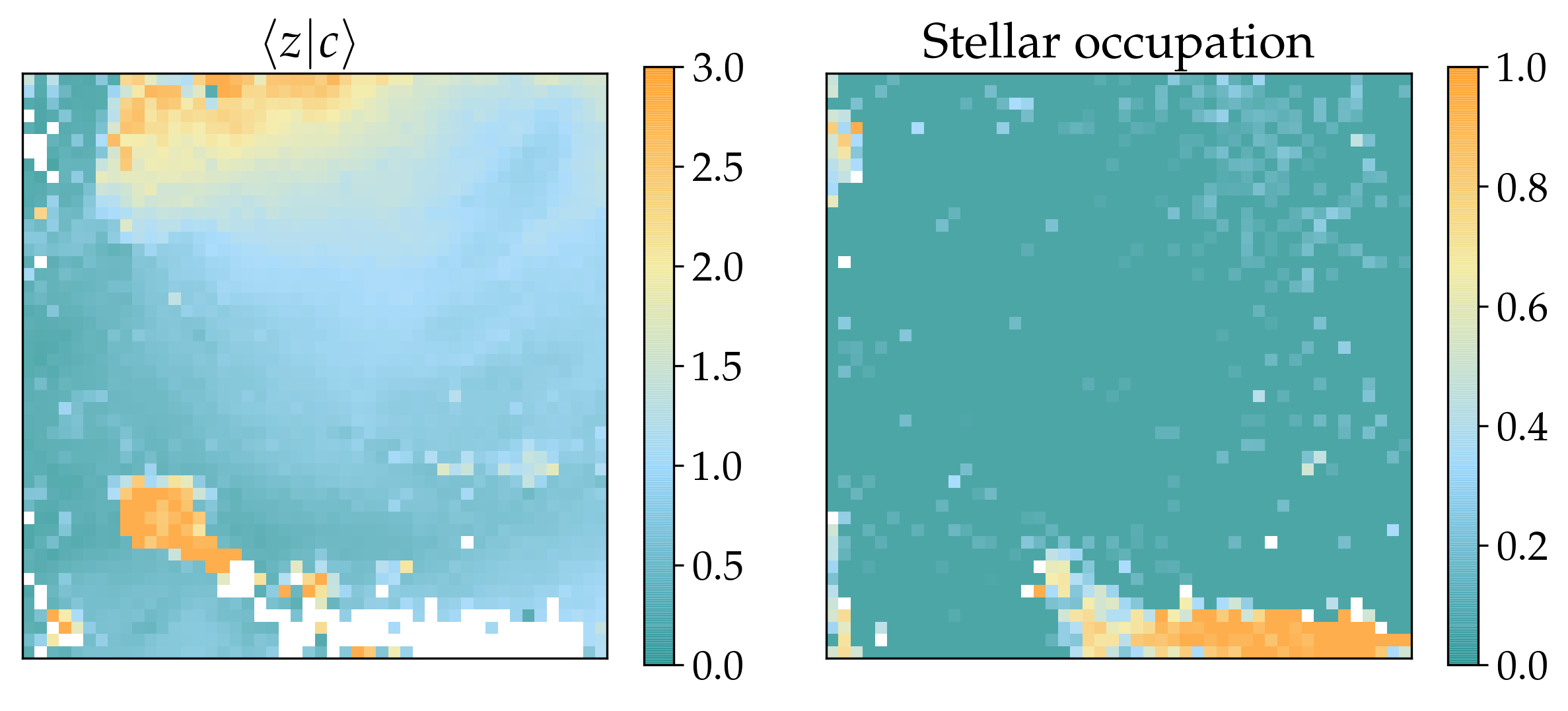}
 \caption{\emph{(left panel):} Redshift mapping of a previous version of the Deep SOM, showing significant areas with no redshift information (white cells). \emph{(right panel):} Stellar occupation of the same Deep SOM, using star-galaxy classification from \citet{Laigle2016}. One can see how the regions without redshift information correspond to the regions of high stellar occupation, which we proceed to remove from the sample to then re-train the SOM on the cleaned sample. }
 \label{fig:stars}
\end{figure}

In a previous version of the Deep SOM, we found that a significant fraction of Deep SOM cells did not have redshift information, i.e.~no deep galaxies with spectroscopic or high-quality redshift information were matched to any of those SOM cells. These regions with no redshift information were also clustered together and placed at the edges of the Deep SOM (see left panel in Fig.\ref{fig:stars}). When investigating the source of this issue, we found that these regions were mainly populated by stars in the \citet{Laigle2016} catalog (see right panel in Fig.\ref{fig:stars}). To correct for the contamination of stars into our Deep sample, we remove all the deep objects falling into Deep SOM cells with majority of stellar occupation. Since those regions are very well clustered, this only removes 0.3\% of the galaxies in the sample, according to the classification from \citet{Laigle2016}. After this selection to remove stars, we re-train the Deep SOM and find an excellent coverage of the entire SOM with redshift information (see \S\ref{sec:deep_som} and Fig.~\ref{fig:deep_som}), and we use that SOM as the fiducial for this analysis. 

\section{Redshift distribution uncertainties}
\label{sec:appendix_pz}

In this section we go over the redshift calibration presented in Section~\ref{sec:redshift_uncertainty} in detail.

\subsection{Redshift uncertainty parametrization}
\label{sec:appendix_pz_param}

We can express the parametric N(z) error model as:
\begin{eqnarray}\label{eq:nzerrormodel}
N_i(z,\theta^i, \Alowz{i}) &=& C_{N_i}\times
\begin{cases} 
G_i \, \Alowz{i}   & z \leq 0.5 \\
G_i \, (1-\Alowz{i}) & z>0.5
\end{cases}\nonumber\\
G_i(z, \theta^i)&=& C_{G_i}\times
\begin{cases} 
F_i(y) & |z-\bar{z}_i| \leq 2\Sigma_{z_i} \\
F_i(z) & |z-\bar{z}_i| > 2\Sigma_{z_i}
\end{cases}\nonumber\\
y &=& \Sigz{i} (z - \Delz{i} - \bar{z}) + \bar{z}\nonumber \\
\bar{z}_i &=& \int z\,  F_i(z) \, dz\nonumber \\
\Sigma_{z_i} &=& \sqrt{\int (z-\bar{z}_i)^2\,  F_i(z)\, dz} \nonumber\\
F_i(z) &=& i\mathrm{-th\, Fiducial\, redshift\, distribution}\nonumber\\
\theta^i &=& \{ \Delz{i},\, \Sigz{i} \}\nonumber\\
(C_{N_i})^{-1} &=&  \int N_i(z,\theta^i, \Alowz{i})\, dz  \nonumber\\
(C_{G_i})^{-1} &=&  \int G_i(z, \theta^i)\, dz  \nonumber\\
\end{eqnarray}
with $\Delz{i}$ the shift, $\Sigz{i}$ the stretch and $\Alowz{i}$ the low redshift fraction free parameters of the model. 

A visualization of the shift, stretch and $\mathrm{low-}z$ fraction parameters can be seen in Figure~\ref{fig:nz_para}. On the one hand, the galaxy clustering signal cares both about the mean redshift of the distribution but also of its spread in redshift, as the more spread out galaxies are the less physically correlated they become, reducing the clustering signal. On the other hand, the majority of the selected galaxies live primarily at high redshift, but with $griz$ colors a population of $\mathrm{low-}z$ galaxies leaks into the selection, especially in our highest redshift bin, producing a distinct clustering signal than that of the high redshift galaxies.  Furthermore, we smooth the fiducial redshift distribution with a Savitzky–Golay filter: sample variance and shot noise from the small area of the calibration deep fields manifests in the $N(z)$ as rapid fluctuations in redshift and enter squared in the galaxy clustering signal, while the true redshift distribution over a larger area is way more smooth as these variations average out. We try different smoothing lengths and find compatible constraints on the main parameters of interest (see Appendix~\ref{sec:smoothing}).

\subsection{Redshift biases}
\label{sec:appendix_pz_biases}

To measure the color-redshift relation in the deep fields we build our redshift sample from a combination of the redshift information that we have available from spectroscopic and multi-band photometric redshifts, SPC (see Section~\ref{sec:redshiftdata}). Whenever a galaxy has spectroscopic measurements, we use them. Alternatively, we use photometric redshifts from the \paucosmos, and when that is not available we use redshifts from \cosmos. After removing color regions with significant stellar contamination and retraining the deep SOM (see Section~\ref{sec:method}), we find that only 9 out of 2304 cells ($0.4\%$) do not have any overlapping redshifts, but relative to the probability of finding galaxies in these cells $p(c)$, they amount to only $0.1\%$ of the probability. Each tomographic bin relates with different probability to each deep cell, and when we take that into account the relative probability without redshift information in each tomographic bin is $0.1\%$, $0\%$ and $0\%$.

We only use  high-quality spectroscopic redshifts, therefore we assume the spectroscopic redshifts are accurate and precise. However, the photo-z from \cosmos and \paucosmos are estimated from multi-band photometric band data, with band filters spanning a wide range in wavelength and with multiple intermediate and narrow bands. The individual $p(z)$ from these catalogs are broader, but their width is still negligible compared to the redshift resolution from noisier wide field observations with $griz$ broad bands, and so we simply stack the individual $p(z)$. Stacking the $p(z)$ is statistically incorrect, and for galaxies where the $p(z)$ is degenerate between two different redshift values, or if the the $p(z)$ were wider, then a more correct technique should be used \citep[e.g.][]{Leistedt2016, Sanchez2018, HBM_clustering, Malz2022, Rau2022}. We defer the application of such techniques for future work.

\begin{figure}
 \centering
\includegraphics[width=0.49\textwidth]{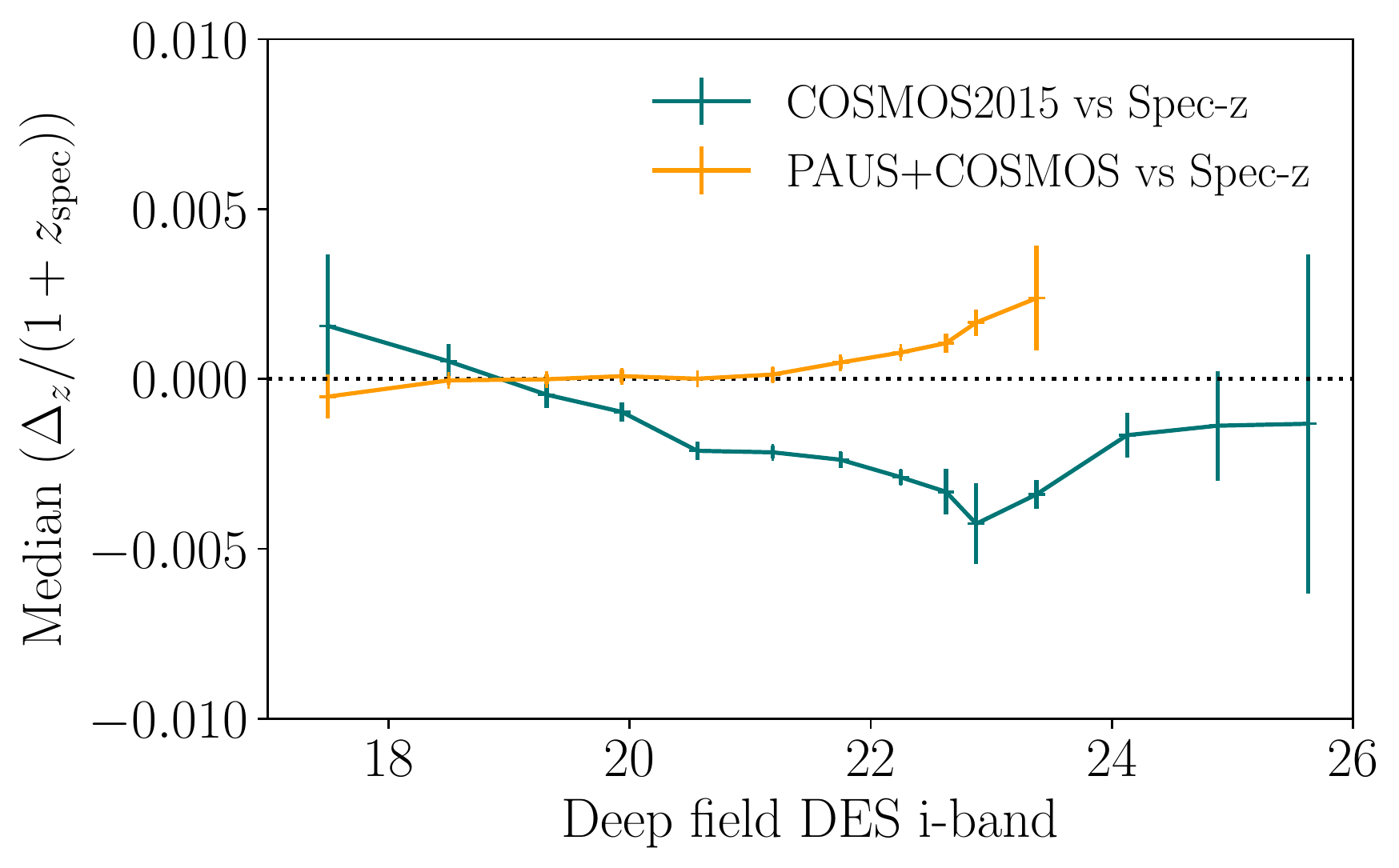}
 \caption{Median photo-z bias as a function of the deep field DES i-band. The bias is calculated for \cosmos and \paucosmos for galaxies where a spectroscopic measurement also exists, with $\Delta z = z_{\rm phot} - z_{\rm spec}$. This measured bias is used to estimate the redshift bias of this catalogs and is marginalized over in our analysis. See \S\ref{sec:appendix_pz_biases} for more details. }
 \label{fig:pzbias}
\end{figure}

An additional concern is whether the photo-z estimates from these catalogs are systematically biased from an incorrect modeling of the galaxy SEDs \citep[e.g.][]{KidsDEScomb, y3-sompz, vandenBusch2022}. Here we measure the bias by comparing the photo-z estimates of individual objects in both catalogs to overlapping spectroscopic measurements (described in Section~\ref{sec:redshiftdata}). For each of these objects, we calculate $(z_{\rm phot} - z_{\rm spec})/(1+z_{\rm spec})$, with $z_{\rm phot}$ the mode of the $p(z)$, and we plot the distributions. By visual inspection we find that the distributions of \cosmos and \paucosmos are generally unimodal, but sometimes slightly biased. We define the median bias as a function of the DES deep field $i{\rm - band}$ magnitude as
\begin{equation}
  b(i) = {\rm Median}\left(\frac{z_{\rm phot} - z_{\rm spec}}{1+z_{\rm spec}} \mid i\right)  
\end{equation}
Figure~\ref{fig:pzbias} shows $b(i)$ from both catalogs: we find a slight positive bias $b(i)\sim 0.002$ at faint magnitudes in the \paucosmos catalog, while the \cosmos catalog presents a negative bias reaching a minimum value of $b(i=22.5)\sim -0.005$. We model the redshift bias uncertainty in these samples with a parameter $\alpha$ that shifts the individual $p(z)$ of \cosmos or \paucosmos galaxies (one $\alpha$ parameter for each catalog). This $\alpha$ parameter shifts $p(z) \rightarrow p(z-\delta(\alpha, i)\cdot(1+z)$) by an amount $\delta$ that is proportional to the median bias of a galaxy of magnitude $i$: 
\begin{equation}
    \delta(\alpha, i) = \alpha\, b(i)
\end{equation}
We place a Gaussian prior on this parameter and marginalize over it, $p(\alpha)=\mathcal{N}(\mu=1, \sigma=1)$. Therefore, our most likely guess for the systematic bias is centered at the measured median bias $b(i)$, but we assign an uncertainty equal to the magnitude of $b(i)$. Note that the value $\alpha$ is the same for all galaxies in the same catalog, but the magnitude of the shift to the $p(z)$ ultimately depends on both the redshift and magnitude of each galaxy: $\delta(\alpha,i)\cdot(1+z)$.

\subsection{Selection biases}
\label{sec:appendix_pz_bce}

We empirically measure the prior on the color-redshift relation from the galaxies in the deep field that have overlapping redshifts. Since we do not parametrize this prior and let the parameters update hierarchically with wide field galaxies, it is crucial to include all selection effects for the final estimate to be unbiased. Balrog injects versions of these galaxies into the wide field and allows us to measure the probability they will be selected into each of our tomographic bins, and therefore to correct for these selection effects. However, due to the limited number of Balrog injections, we cannot always measure these effects accurately, leading to several approximations to the SOMPZ methodology described in section~\ref{sec:method}. In this section we explain these approximations and their validity, and provide a way to marginalize over the potential systematic biases that they might introduce.

\begin{figure*}
 \centering
\includegraphics[width=0.99\textwidth]{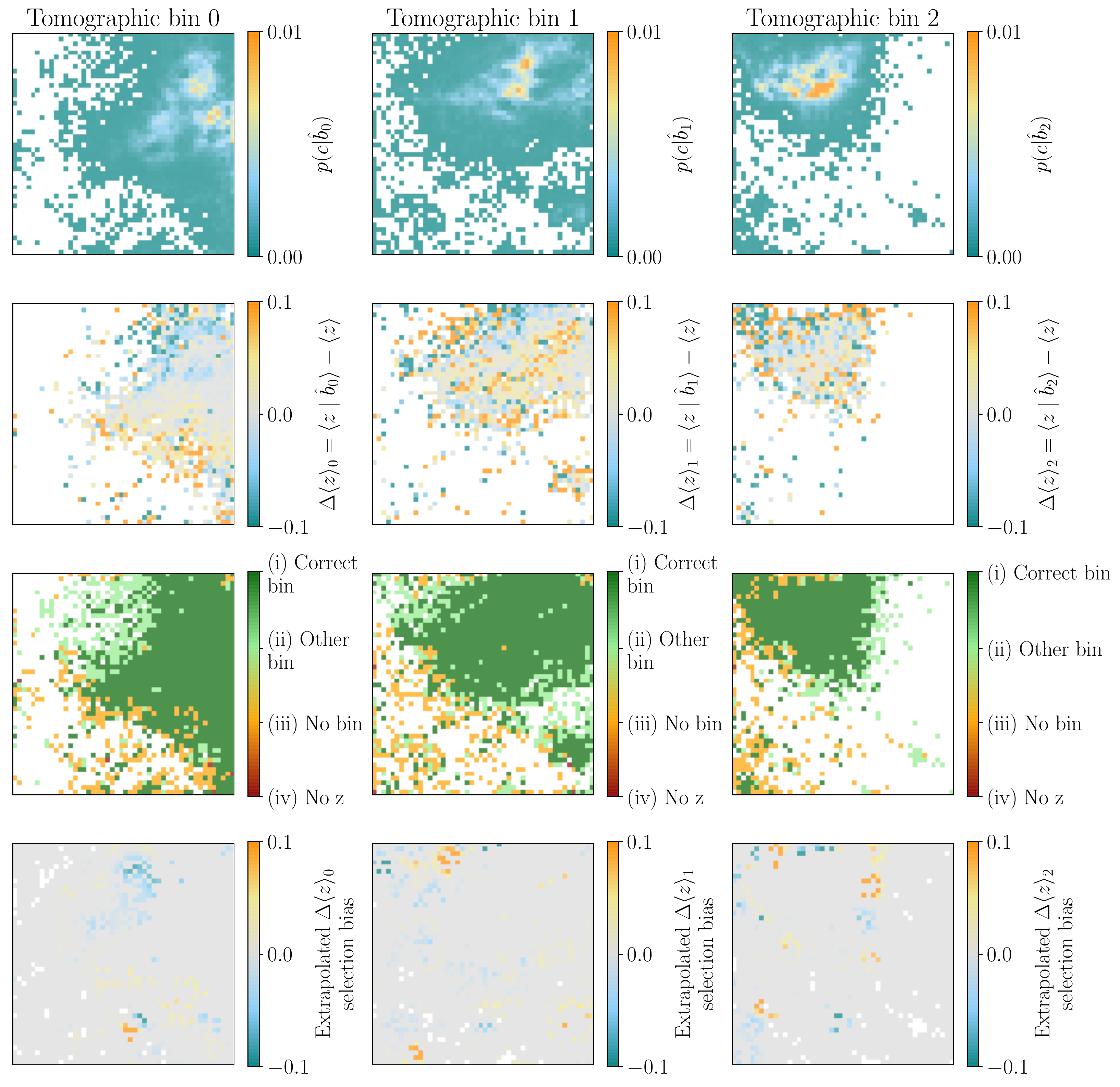}
 \caption{The redshift selection effects and the extrapolated $\Delz{}$ selection effect bias. Each column shows a different tomographic bin. The first row of panels shows the pdf of deep field cells conditioned on each tomographic bin, $p(c|\hat{b}_i)$. The second row of panels shows the mean redshift difference of deep field cells when galaxies are additionally conditioned to be observed by Balrog into our each tomographic bin. The third row of panels shows which cells have some galaxy with redshift information selected into the bin by Balrog (i),
 which do not (ii)-(iii), and also which do not have any z information (iv) (only five cells for bin 0, four for bin 1 and one for bin 2).
 The fourth row of panels show an extrapolated redshift bias. The redshift bias due to the additional selection of galaxies into the bin is extrapolated from (i) cells that have galaxies selected into the bin to cells (ii)-(iii)-(iv) that do not. See \S\ref{sec:appendix_pz_bce} for more details. 
 }
 \label{fig:bce}
\end{figure*}

\begin{figure}
 \centering
\includegraphics[width=0.49\textwidth]{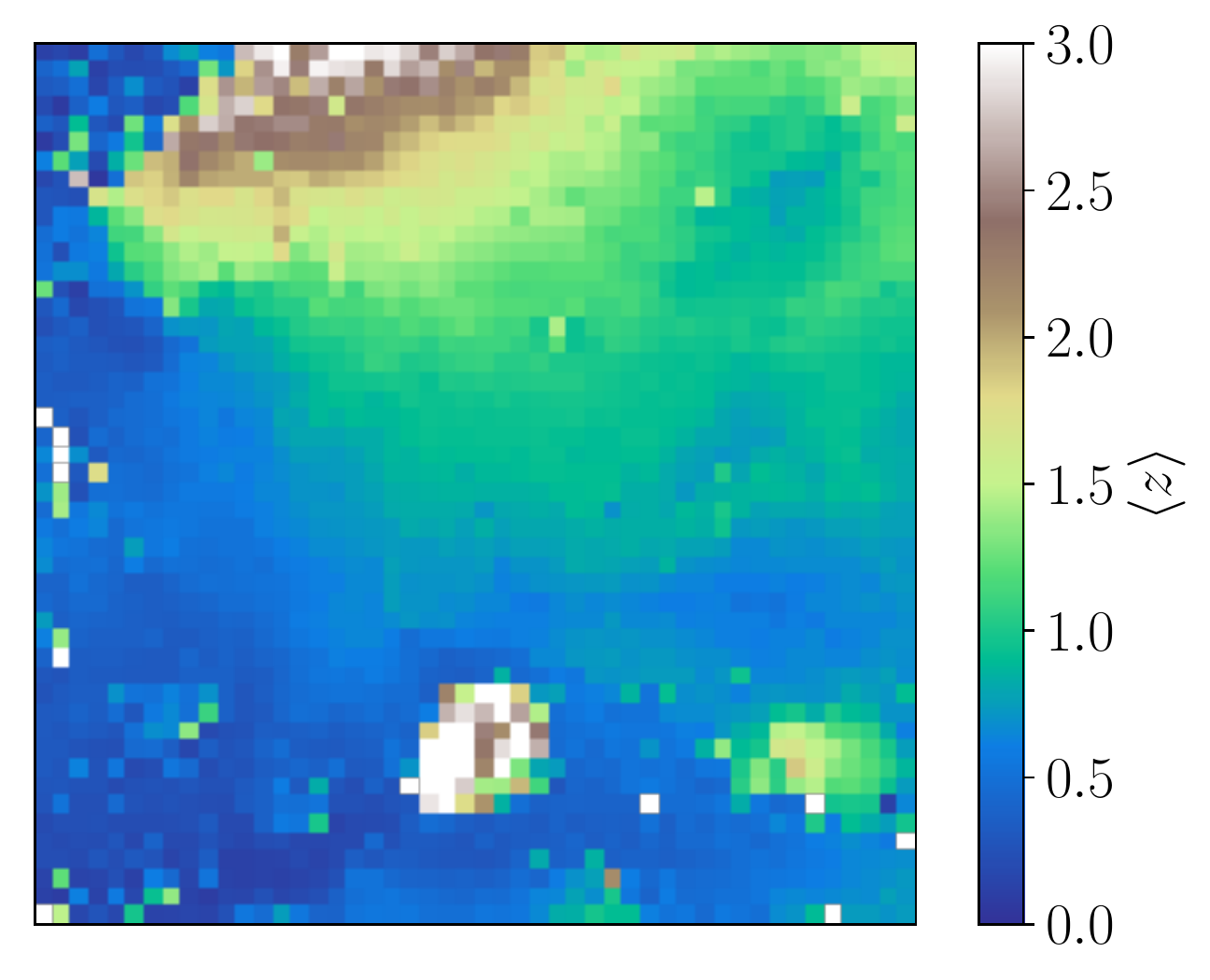}
 \caption{Deep SOM mean redshift. The modified terrain colormap highlights the different mean redshift levels, with the flooded area roughly showing redshifts below our samples and our high-z galaxies lifting out of the oceans of low-z galaxies. The grassy area roughly shows the redshifts of our first two tomographic bins, while the north-northwest hill shows the area of our highest redshift bin. Going south we find the snowed peaky island showing the area of very high redshift Lyman-break galaxies, with very low redshift Balmer-break galaxies lurking below the icy glaciers of Lyman-break galaxies.}
 \label{fig:meanc_terrain}
\end{figure}

The first row of panels (from the top) of Figure~\ref{fig:bce} show the distribution of deep field galaxies in the deep SOM weighted by their probability of being selected in each tomographic bin as measured by Balrog. This distribution is different than the one presented in Figure~\ref{fig:deep_som}, where we show the distribution of deep field galaxies weighted by their probability of being selected at $22\leq i \leq 23.5$ according to Balrog. Note how in each panel the distribution peaks around deep SOM cells with high redshift and has little to no overlap with cells at lower redshift, as expected (compare to Figure~\ref{fig:meanc_terrain} for the distribution of mean redshift in the deep SOM).

The redshift distribution of each deep SOM cell formally depends on the pre-selections $\hat{s}$ and on the wide SOM cell where galaxies are selected, $p(z|c, \hat{c}, \hat{s})$, see Equation~\ref{eq:wide_nz_exact}:
\begin{align}
    p(z|\hat{b}, \hat{s}) &= \sum_{\hat{c} \in \hat{b}}\, p(z|\hat{c}, \hat{s}, \,\hat{b})\, p(\hat{c}|\hat{s}, \hat{b}) \\
    &= \sum_{\hat{c} \in \hat{b}} \sum_{c} p(z|c, \hat{c}, \hat{s}) \,p(c|\hat{c},\hat{s})\,p(\hat{c}|\hat{s}, \,\hat{b}) \label{eq:wide_nz_exact} \\
    &\approx \sum_{\hat{c} \in \hat{b}} \sum_{c} p(z|c, \hat{b}, \hat{s}) \,p(c|\hat{c},\hat{s})\,p(\hat{c}|\hat{s}, \,\hat{b})\label{eq:wide_nz_approx1}\\
    &\approx \sum_{\hat{c} \in \hat{b}} \sum_{c} p(z|c, \hat{B}, \hat{s}) \,p(c|\hat{c},\hat{s})\,p(\hat{c}|\hat{s}, \,\hat{b})\label{eq:wide_nz_approx2}\\    
    &\approx \sum_{\hat{c} \in \hat{b}} \sum_{c} p(z|c, \hat{s})\, p(c|\hat{c},\hat{s})\,p(\hat{c}|\hat{s}, \,\hat{b}).\label{eq:wide_nz_approx3}    
\end{align}

%
Using Balrog we can empirically measure how often deep field galaxies $c$ will get through our pre-selections $\hat{s}$ and also how often they get selected in the different wide field cells $\hat{c}$. However, due to the limited number of Balrog injections it is not possible to accurately measure the relation between all $(z,\, c,\, \hat{c})$. Following \citet{y3-sompz}, we use the approximation shown in Eq.~\ref{eq:wide_nz_approx1} for our fiducial estimation of the redshift distribution of deep cells using $p(z|c,\hat{c})\approx p(z|c,\hat{b})$, with $\hat{b}$ representing the set of $\hat{c}$ of a tomographic bin. When no redshift galaxy satisfies both $c$ and $\hat{b}$ then we use $p(z|c, \hat{B}, \hat{s})$ (Eq.~\ref{eq:wide_nz_approx2}) using redshift information from galaxies that are selected into any of the tomographic bins $\hat{B}\equiv\{\hat{b}_0,\,\hat{b}_1,\,\hat{b}_2\}$, or else $p(z|c, \hat{s})$ (Eq.~\ref{eq:wide_nz_approx3}), using redshift information from any galaxies satisfying our pre-selection $\hat{s}$. 

The second row of panels of Figure~\ref{fig:bce} shows the difference in the mean redshift of each cell from including the tomographic bin selection, showing:
\begin{equation}
\begin{split}
    \Delta\langle z \rangle_i &\equiv \int z\, p(z|c,\hat{b}_i,\hat{s})\, dz  - \int z\, p(z|c,\hat{s})\, dz \\
    &\equiv \langle z \mid \hat{b}_i \rangle - \langle z \rangle
\end{split}
\end{equation}
Note how the $\Delta\langle z \rangle_i$ values tend to be close to 0 where the distribution of $p(c\mid b_i)$ peaks (top panels), as most galaxies from these cells get selected very often into that tomographic bin. However, note that $\Delta\langle z \rangle_i$ shows larger differences at the tails of the $p(c\mid b_i)$ distribution. In such cells, generally speaking, galaxies with a redshift that is closer to the average redshift of the tomographic bin get preferentially selected, and consequently cells with a $\meanz$ smaller than the average redshift of the bin tend to have a positive $\Delta\langle z \rangle_i$, and vice-versa. This effect is very clear in bin 0, where cells at the lower part of the SOM have a $\meanz$ that is smaller than the typical redshift of galaxies in bin 0, and they show a positive $\Delta\langle z \rangle_0$, implying that additionally conditioning on the tomographic bin tends to increase the mean redshift of these cells. We find the contrary for cells at the top of the SOM, which have a $\meanz$ that is larger than the typical redshift of galaxies in this bin and they present a negative $\Delta\langle z \rangle_0$ that lowers the average redshift of the cell when we condition their selection to the bin.

This highlights how important it is to at least include the so-called \textit{bin conditionalization}\footnote{We follow the notation introduced in \citet{y3-sompz}.}, i.e. using $p(z|c,\hat{b},\hat{s})$ instead of just $p(z|c,\hat{s})$. Otherwise one will introduce important selection effect biases, as those found by \citet{y3-sompzbuzzard}, where they found a positive bias for low redshift bins relative to the average redshift and a negative bias for high redshift bins, as a result of just using $p(z|c,\hat{s})$. 

The third row of panels in Figure~\ref{fig:bce} shows with a color code which cells have redshift estimates that include accurate tomographic bin selection effects. The color code goes as:
\begin{enumerate}
    \item \textit{Dark green}: cells that have at least one redshift galaxy that has been selected by Balrog into the corresponding tomographic bin, we use Eq.~\ref{eq:wide_nz_approx1}, $p(z|c, \hat{b}, \hat{s})$.
    \item \textit{Light green}: cells that have do not have any galaxy selected into the corresponding tomographic but at least one redshift galaxy that has been selected by Balrog into one of the other two tomographic bins, we use Eq.~\ref{eq:wide_nz_approx2}, $p(z|c, \hat{B}, \hat{s})$.
    \item \textit{Light red}: cells that have do not have any galaxy selected into any tomographic bin, but at least some galaxy satisfying our pre-selection $\hat{s}$. We use Eq.~\ref{eq:wide_nz_approx3}, $p(z|c, \hat{s})$.
    \item \textit{Dark red}: cells that have do not have any redshift galaxy satisfying our pre-selection $\hat{s}$. We do not have direct redshift information for these cells.
\end{enumerate}

Note how the $\Delta\langle z \rangle_i$ from the second row of panels can only be calculated for (i)/Dark Green cells in the third row of panels. The remaining cells do not have any galaxy selected by Balrog into the corresponding tomographic bin, and bin conditionalization cannot be estimated directly, which is a source of potential systematic uncertainty. We test this effect by calculating the mean redshift bias in Dark Green cells from neglecting the bin conditionalization, and extrapolating it to other nearby cells using a Gaussian smoothing. The last row of panels in Figure~\ref{fig:bce} shows the bias values from extrapolation for every deep cell, showing that certain groups of cells have under-/over-estimated mean redshifts. We parametrize this possible systematic bias with the same parameter $\epsilon$ that shifts the $p(z|c)\rightarrow p(z-\epsilon(\beta, c)|c)$ of each deep cell; with $\epsilon(\beta, c) = \beta \, b(c)$; and $b(c)$ the estimated systematic bias from the last row of panels in Figure~\ref{fig:bce}. We place a Gaussian prior on this parameter and marginalize over it, $p(\beta)=\mathcal{N}(\mu=1, \sigma=1)$. Figure~\ref{fig:pzpriors} shows that this missing selection effect (labelled as BCE in the figure) has a very negligible effect to all the $N(z)$ parameters relative to the other sources of uncertainty.

\subsubsection{Cell conditionalization}
\label{sec:appendix_cellcond}

\begin{figure}
 \centering
\includegraphics[width=0.49\textwidth]{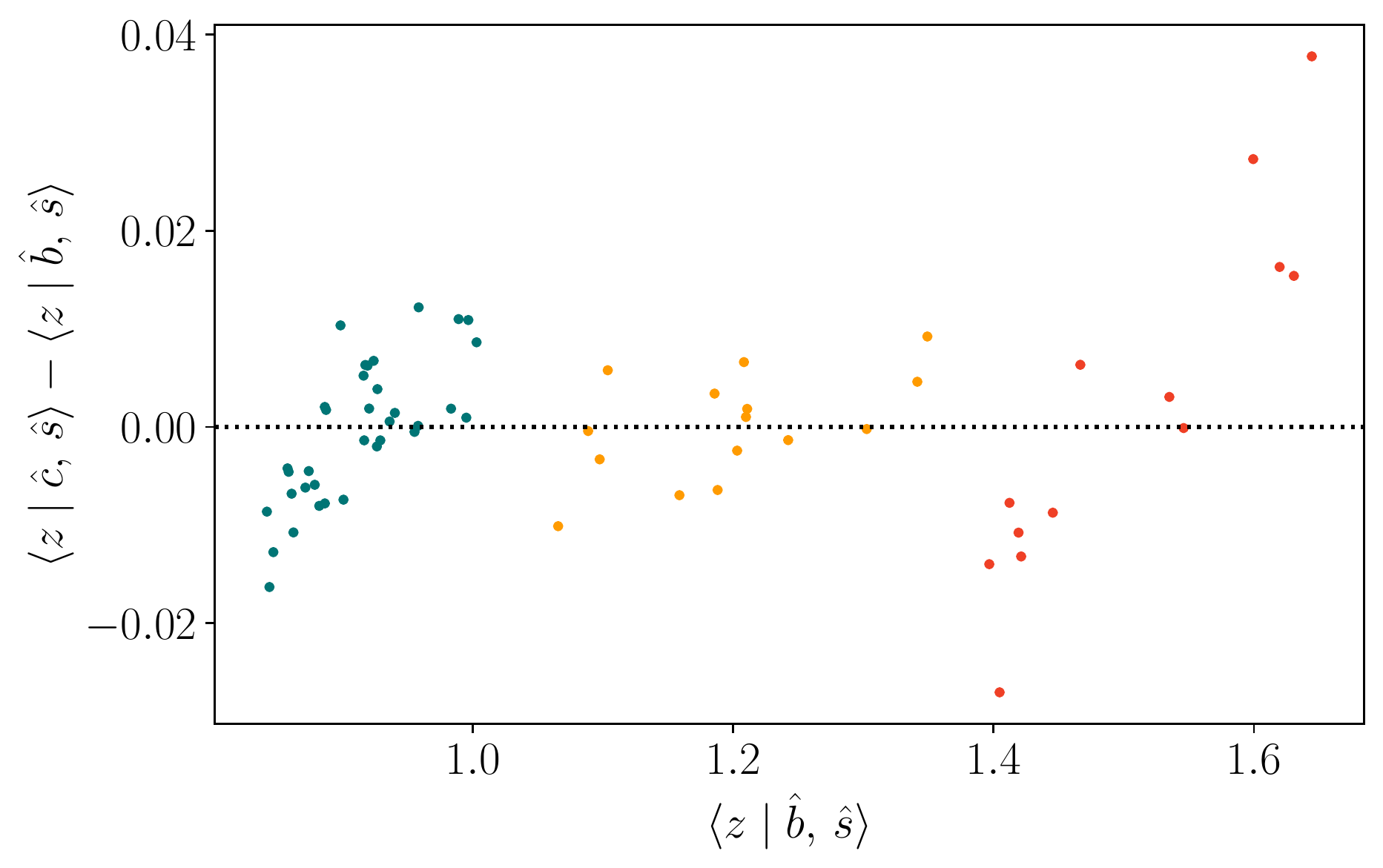}
 \caption{Mean redshift difference for each wide field cell between using cell conditionalization (cellcond) or bin conditionalization (bincond). Cellcond refers to conditioning the redshift distribution of deep SOM cells to galaxies that are selected into each wide SOM cell (i.e. using Equation~\ref{eq:wide_nz_exact}). In contrast, Bincond only requires galaxies to have been selected into any wide SOM cell belonging to the tomographic bin (i.e. using Equation~\ref{eq:wide_nz_approx1}). As expected, wide SOM cells with lower redshift within the bin have a lower estimated mean redshift when we additionally require deep field galaxies to be selected into that particular wide SOM cell. See \S\ref{sec:appendix_cellcond} for more details.}
 \label{fig:cellcond}
\end{figure}

An additional source of systematic error comes from the approximation of using bin conditionalization (or bincond, Equation~\ref{eq:wide_nz_approx1}) instead of the exact \textit{cell conditionalization} (or cellcond, Equation~\ref{eq:wide_nz_exact}). Figure~\ref{fig:cellcond} explores the difference in mean redshift for (i)/Dark Green cells  between using cellcond and bincond. We find a clear (but somewhat noisy) trend, where cells within a tomographic bin with a lower than average mean redshift have an over-estimated mean redshift, and viceversa, as expected. The overall trend within the same tomographic bin is centered around 0, as bincond already corrects for most of the overall redshift selection effect bias. 

We have calculated the $N(z)$ using cellcond, and despite the large biased trend seen in Figure~\ref{fig:cellcond}, we have found the resulting $n(z)$ from using cellcond or bincond to have very negligible differences to its mean redshift, width and low redshift fraction. Upon closer inspection, the $p(z | c,\hat{c})$ and $p(z | c,\hat{b})$ distributions differ at their tails, which produces significant changes to their mean redshifts $\langle z | c,\hat{c}\rangle$ and $\langle z | c,\hat{b}\rangle$, but this effect ends up cancelling out after adding up the contributions from each deep field cell to calculate the final $N(z)$ for each bin.

\subsection{Zero-point uncertainty}
\label{sec:appendix_pz_zpe}

As measured in \citet{y3-deepfields}, the deep field photometry has some residual photometric zero point error. This error is largest in the $u$-band (0.055), and much smaller in the other bands: 0.005 in $griz$ and 0.008 in $JHK$ (Table 5 in \citet{y3-deepfields}). This in principle impacts our analysis in two ways. First, most of the redshift information is in the COSMOS fields, while X3, C3, E2 have little or no redshift information. Therefore, we are extrapolating the redshift information measured in one field to the colors of all fields, and measuring the color abundance from all fields. The zero-point uncertainty affects the accuracy of this extrapolation, as well as the measured deep color abundance. On the other hand, a zero-point error on the deep field fluxes introduces an error in the input injected model fluxes used by Balrog, which in turn will induce a slight error on the distribution of recovered wide field Balrog fluxes. Since the error in the u-band is the largest, there is no u-band in the wide field, and the zero-point errors in $griz$ are small, we assume the former is the only form of zero-point error we need to worry about. 

Since the zero-point photometric uncertainty is mainly measured from the variance of the stellar and red galaxy loci between each band and field \citep[for full details see][]{y3-deepfields}, we perturb the zero-point magnitude of each deep field (X3, C3, E2, COSMOS) and band by an amount drawn from a Gaussian distribution with zero mean and variance equal to the measured variance from \citet{y3-deepfields}. Since only the relative zero-point matters, we fix the zero-point of one of the fields (COSMOS) and perturb the zero-point of the remaining fields (X3, C3, E2). We marginalize over this uncertainty by (i) drawing 3 zero-point shifts for each X3, C3, E2 field, (ii) we modify the fluxes and flux errors by the corresponding amount, (iii) we reassign each galaxy to the deep SOM based on the perturbed fluxes, and (iv) we re-calculate the n(z) based on this new assignment.

\begin{figure}
 \centering
\includegraphics[width=0.49\textwidth]{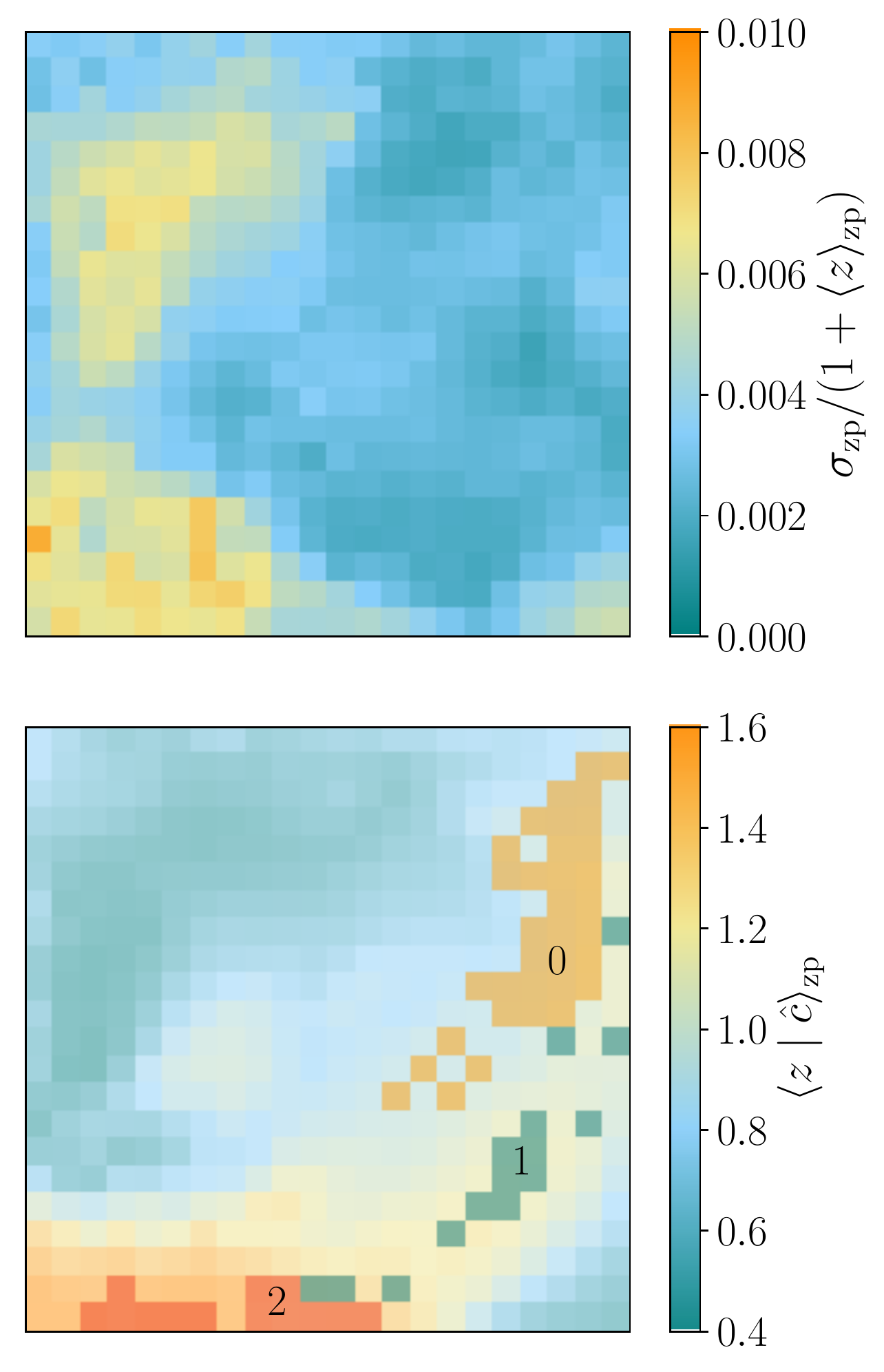}
 \caption{Wide SOM mean redshift variance from zero-point photometric uncertainty. The top panel shows the scatter in the mean redshift of each wide SOM cell $\sigma_{\rm zp}$ from perturbing the fluxes of deep SOM galaxies with the zero-point photometry uncertainty, weighted by $1+\langle z\rangle_{\rm zp}$, with the latter being the average mean redshift from the same variations. The bottom panel shows $\langle z\rangle_{\rm zp}$ for each wide SOM cell for reference. The wide SOM cells pertaining to each tomographic bin are highlighted in the bottom panel with the same color. See \S\ref{sec:appendix_pz_zpe} for more details.}
 \label{fig:zpu_meanzchat}
\end{figure}

Figure~\ref{fig:zpu_meanzchat} shows the resulting variance in mean redshift for each wide SOM cell in the top panel, as a result of perturbing the fluxes of the deep field galaxies. The average mean redshift shown in the bottom panel for reference, with the cells pertaining to each tomographic bin indicated with different colors. As expected, we find a large effect in cells with a low redshift, as the u-band uncertainty is the largest, which affects the classification of low redshift galaxies. We also find a large effect in some of the wide cells that have a high mean redshift but that are next to wide cells with low redshift, i.e. cells that are near color-redshift degeneracies.

\subsection{Redshift uncertainty parameter priors}
\label{sec:appendix_pz_priors}

To estimate priors $p(\Delz{i})$, $p(\Sigz{i})$ and $p(\Alowz{i})$ on these parameters we draw $N(z)$ samples from the sources of uncertainty described in this appendix and calculate the spread in the mean redshift, $N(z)$ width and $\mathrm{low-}z$ fraction ($z<0.5$) values of the individual distribution samples. Summarizing, we have 24 zero point systematic shifts (8 bands and 3 out of 4 fields), two redshift systematic shifts (one for \cosmos and one for \paucosmos) and one selection effect bias parameter. We draw 100 samples in quantile space using Latin hypercube sampling, a stratified random sampling technique for generating near-random samples of parameter values that is more efficient than a pure random sampling. For each of these 100 samples we shift the $p(z)$ of individual galaxies, we shift the deep fluxes of galaxies and reassign them to deep cells. Then for each of these 100 samples we generate 5,000 N(z) samples using \sdir. We properly weight deep field galaxies injected by Balrog by the clustering weight (Section~\ref{sec:weights}) of the spot where they were injected. We produce samples for all the area, and the North (\emph{Planck}) and South (SPT) regions.

The fiducial redshift distribution $F(z)$ of Equation~\ref{eq:nzerrormodel} is the average $N(z)$ of the distribution samples with an additional smoothing. We apply a Savitzky-Golay filter on the average $N(z)$, using a 0.21 smoothing length in redshift for Bins 0 and 1, while for Bin 2 we use a combination of two smoothing lengths: we use a length of 0.21 at $z<0.5$ and a length of 0.45 for $z>0.5$.

\subsection{Smoothing of the redshift distributions}
\label{sec:smoothing}

\begin{figure}
 \centering
\includegraphics[width=0.45\textwidth]{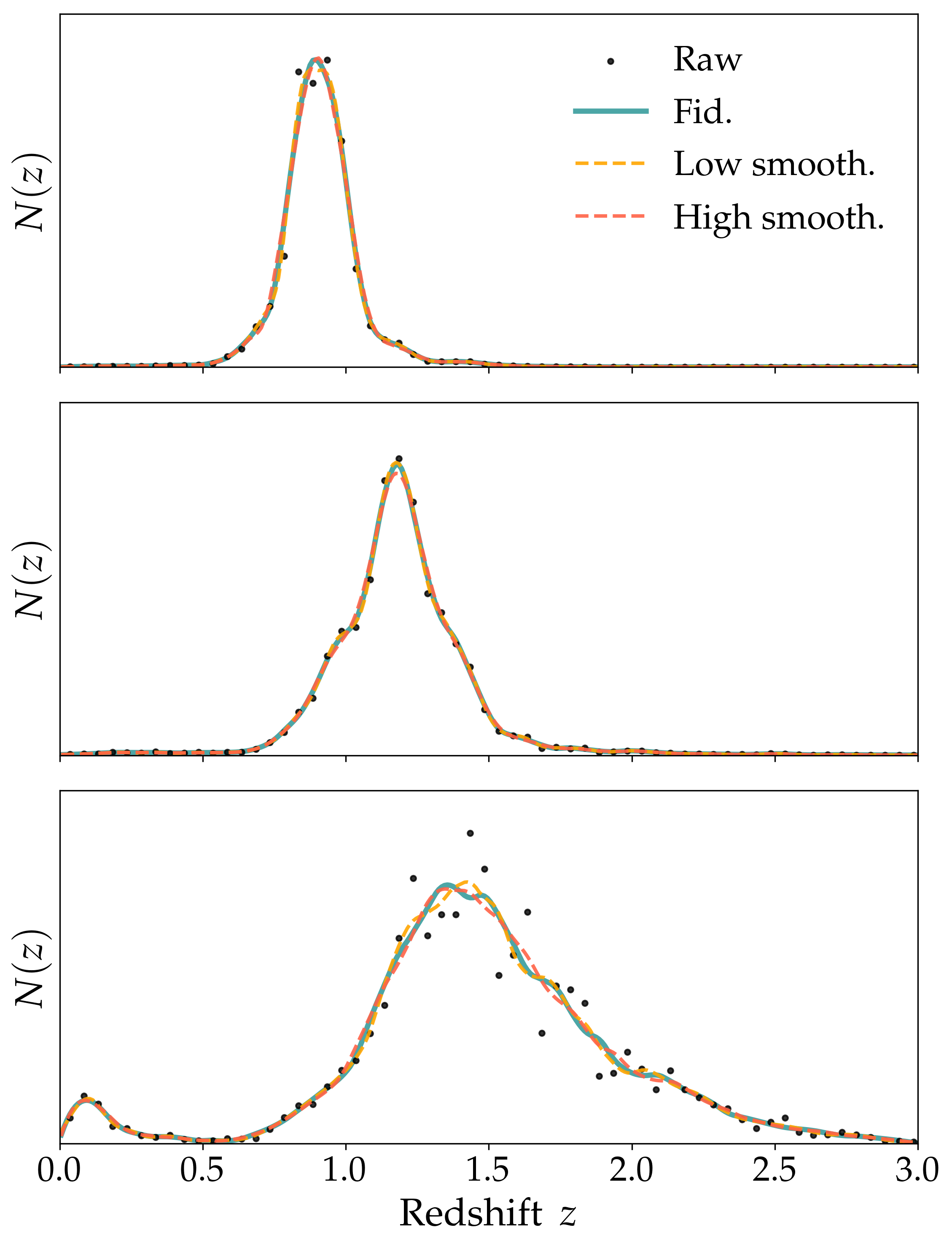}
 \caption{Visualization of the smoothing procedure applied to the raw redshift distributions using a Savitzsky Golay (SG) filter. A fiducial set of distributions is presented, along with two alternative sets using lower (higher) amounts of smoothing, as described in Appendix \ref{sec:smoothing}. }
 \label{fig:smoothing}
\end{figure}

\begin{figure}
 \centering
\includegraphics[width=0.48\textwidth]{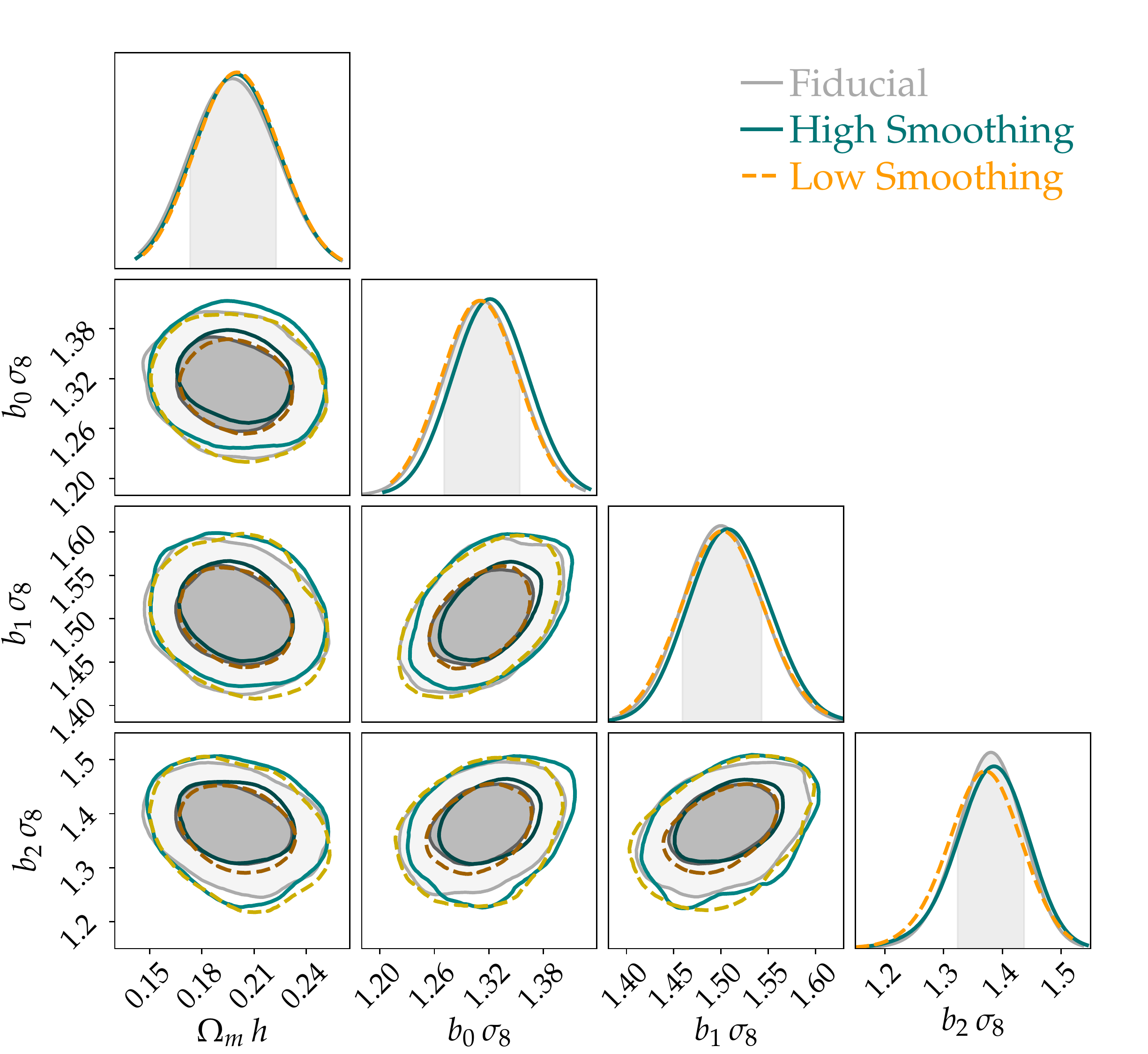}
 \caption{Comparison of the parameter constraints from galaxy clustering using higher and lower amounts of $N(z)$ smoothing, as described in Appendix \ref{sec:smoothing}, demonstrating the small impact of the choice of smoothing scale in the analysis.  }
 \label{fig:smoothing_chains}
\end{figure}

The redshift inference methodology described in \S\ref{sec:method} is subject to effects of shot noise and especially sample variance in the redshift samples \citep{Sanchez2020}, which result in noisy estimates of the redshift distributions of our tomographic bins. The uncertainties coming from these effects are properly taken into account in \S\ref{sec:redshift_uncertainty}. In addition, we also apply a smoothing procedure to the redshift distributions used in this work, since noise in the redshift distributions can cause instabilities in the analysis of galaxy clustering. For that purpose, we apply a Savitzsky Golay (SG) filter with a third-order polynomial to the raw redshifts distributions, as depicted in Fig.~\ref{fig:smoothing}. In our fiducial case, the length of the filter window is set to 0.21 in redshift for the low redshift part of the distributions ($z<0.5$), and 0.45 in redshift for the higher redshift part of the distributions ($z>0.5$). In order to test the stability of our results to the particular smoothing filter choices, we define two alternative sets of smoothed redshift distributions, corresponding to lower (higher) smoothings, using SG filters with window lengths of 0.15 (0.27) in redshift for the low redshift part of the distributions ($z<0.5$), and 0.27 (0.55) in redshift for the higher redshift part of the distributions ($z>0.5$). The comparison between the raw estimates and the smoothed versions of the redshift distributions for the three tomographic bins is shown in Fig.~\ref{fig:smoothing}, and the negligible impact on parameter constraints from galaxy clustering is shown in Fig.~\ref{fig:smoothing_chains}. 

\section{Comparison between JK and theory covariance}
\label{sec:appendix_cov}
\begin{figure}
 \centering
 \includegraphics[width=0.5\textwidth]{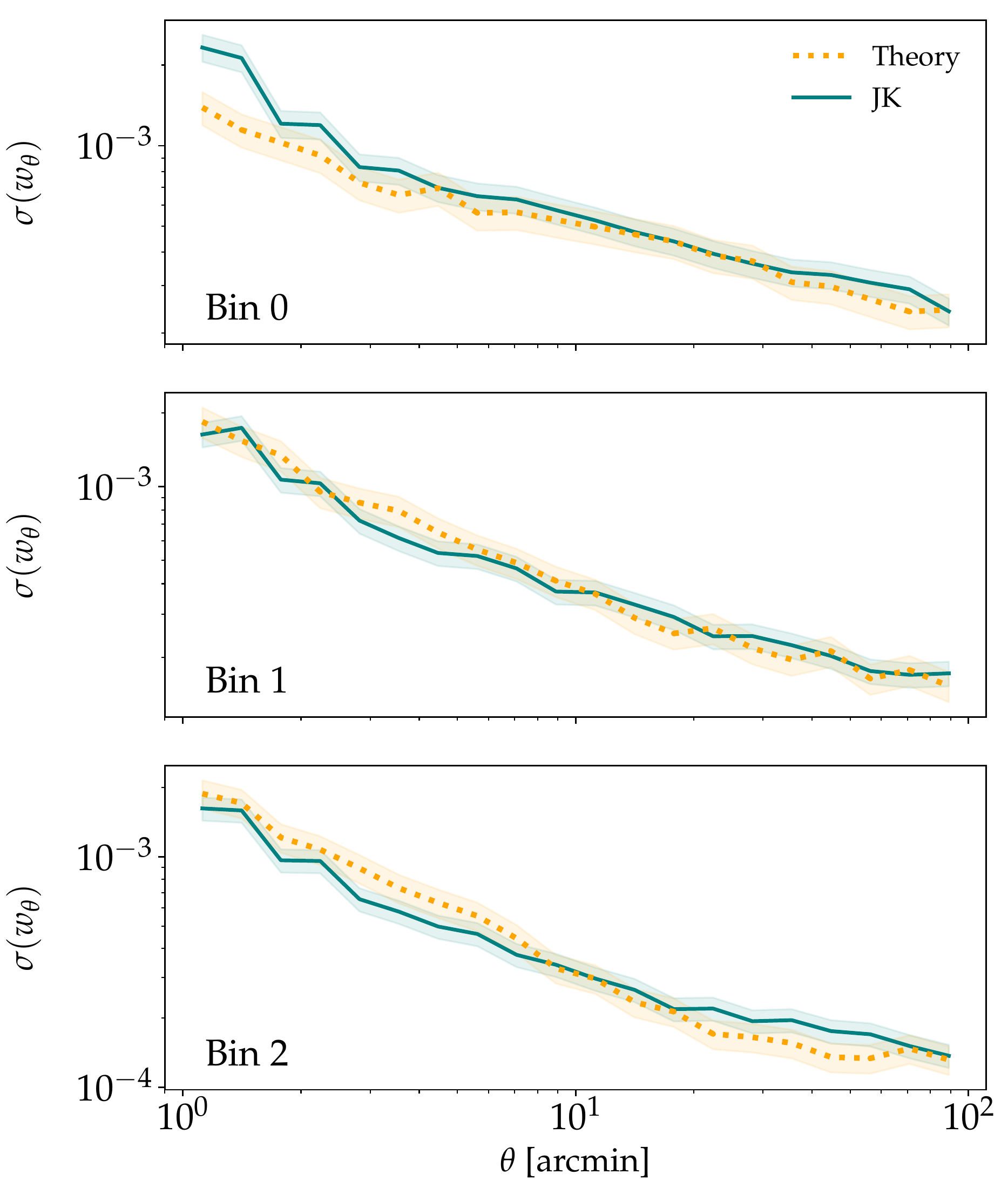}
 \caption{Comparison of the diagonal elements of the theory and jackknife covariance matrices for the auto-correlations of angular galaxy clustering for the three redshift bins ($0,1,2$) defined in this work. The methodology for the measurements and covariance can be found in \S\ref{sec:measurements} and Appendix \ref{sec:appendix_cov}. }
 \label{fig:cov_comparison}
\end{figure}

In this section we compare the two covariance estimates (based on Gaussian simulations, and based on Jackknife estimates) presented in section~\ref{sec:clustering}. In order to generate each realization of the Gaussian simulations, we generate a set of four maps following the procedure detailed in~\citet{Giannantonio2008}. In order to obtain correlated maps with the correct power spectrum, we have to generate a set of correlated (in-phase) screens with an amplitude $T_{i,k}$, where the subindex $i$ refers to the final map, and $k$ to the phase. So we add all contributions with the same index $i$ to get the $i$-th map, and all screens that have the same index $k$ are generated using the same random seed (are in-phase). Each screen is generated using \texttt{hp.anafast(T**2\_{ij}, nside)}. The amplitudes $T_{ik}$ are calculated as follows: 
\begin{eqnarray}
T_{1a} = \sqrt{C^{00}_{\ell}} \\
T_{2a} = \frac{C_{\ell}^{01}}{T_{1a}} \\
T_{2b} = \sqrt{C_{\ell}^{11}-T^{2}_{2a}} \\
T_{3a} = \frac{C_{\ell}^{02}}{T_{1a}} \\
T_{3b} = \frac{C_{\ell}^{12}-T_{2a}}{T_{3a}} \\
T_{3c} = \sqrt{C_{\ell}^{22}-T_{3a}^{2}-T_{3b}^{2}} \\
T_{4a} = \frac{C_{\ell}^{0\kappa}}{T_{1a}} \\
T_{4b} = \frac{C_{\ell}^{1\kappa}-T_{2a}T_{4a}}{T_{2b}} \\
T_{4c} = \frac{C_{\ell}^{2\kappa}-T_{3a}T_{4a}-T_{3b}T_{4b}}{T_{3c}} \\
T_{4d} = \sqrt{C_{\ell}^{\kappa\kappa}-T^{2}_{4a}-T^{2}_{4b}-T^{2}_{4c}}
\end{eqnarray}
We generate 100 realizations of these maps, and get their covariance. We compare the resulting covariance with the Jackknife estimate in Figure~\ref{fig:cov_comparison}. In this Figure we can see that the diagonal terms from both covariance estimates are in excellent agreement in the range of scales that we are considering.

\section*{Affiliations}

$^{1}$ Institute of Space Sciences (ICE, CSIC),  Campus UAB, Carrer de Can Magrans, s/n,  08193 Barcelona, Spain\\
$^{2}$ Department of Physics and Astronomy, University of Pennsylvania, Philadelphia, PA 19104, USA\\
$^{3}$ Argonne National Laboratory, 9700 South Cass Avenue, Lemont, IL 60439, USA\\
$^{4}$ Space Telescope Science Institute, 3700 San Martin Drive, Baltimore, MD  21218, USA\\
$^{5}$ Department of Physics, University or Genova and INFN, Genova division, Via Dodecaneso 33, 16146, Genova, Italy\\
$^{6}$ Department of Astronomy and Astrophysics, University of Chicago, Chicago, IL 60637, USA\\
$^{7}$ Kavli Institute for Cosmological Physics, University of Chicago, Chicago, IL 60637, USA\\
$^{8}$ Department of Physics, University of Michigan, Ann Arbor, MI 48109, USA\\
$^{9}$ Lawrence Berkeley National Laboratory, 1 Cyclotron Road, Berkeley, CA 94720, USA\\
$^{10}$ Centro de Investigaciones Energ\'eticas, Medioambientales y Tecnol\'ogicas (CIEMAT), Madrid, Spain\\
$^{11}$ Institute for Astronomy, University of Hawai'i, 2680 Woodlawn Drive, Honolulu, HI 96822, USA\\
$^{12}$ Department of Physics, Stanford University, 382 Via Pueblo Mall, Stanford, CA 94305, USA\\
$^{13}$ Kavli Institute for Particle Astrophysics \& Cosmology, P. O. Box 2450, Stanford University, Stanford, CA 94305, USA\\
$^{14}$ Institute of Astronomy, University of Cambridge, Madingley Road, Cambridge CB3 0HA, UK\\
$^{15}$ Kavli Institute for Cosmology, University of Cambridge, Madingley Road, Cambridge CB3 0HA, UK\\
$^{16}$ Physics Department, 2320 Chamberlin Hall, University of Wisconsin-Madison, 1150 University Avenue Madison, WI  53706-1390\\
$^{17}$ Department of Physics, Northeastern University, Boston, MA 02115, USA\\
$^{18}$ NASA Goddard Space Flight Center, 8800 Greenbelt Rd, Greenbelt, MD 20771, USA\\
$^{19}$ Department of Physics, Carnegie Mellon University, Pittsburgh, Pennsylvania 15312, USA\\
$^{20}$ Instituto de Astrofisica de Canarias, E-38205 La Laguna, Tenerife, Spain\\
$^{21}$ Laborat\'orio Interinstitucional de e-Astronomia - LIneA, Rua Gal. Jos\'e Cristino 77, Rio de Janeiro, RJ - 20921-400, Brazil\\
$^{22}$ Universidad de La Laguna, Dpto. Astrofísica, E-38206 La Laguna, Tenerife, Spain\\
$^{23}$ Center for Astrophysical Surveys, National Center for Supercomputing Applications, 1205 West Clark St., Urbana, IL 61801, USA\\
$^{24}$ Department of Astronomy, University of Illinois at Urbana-Champaign, 1002 W. Green Street, Urbana, IL 61801, USA\\
$^{25}$ Institut d'Estudis Espacials de Catalunya (IEEC), 08034 Barcelona, Spain\\
$^{26}$ Fermi National Accelerator Laboratory, P. O. Box 500, Batavia, IL 60510, USA\\
$^{27}$ NSF AI Planning Institute for Physics of the Future, Carnegie Mellon University, Pittsburgh, PA 15213, USA\\
$^{28}$ Department of Astronomy/Steward Observatory, University of Arizona, 933 North Cherry Avenue, Tucson, AZ 85721-0065, USA\\
$^{29}$ Jet Propulsion Laboratory, California Institute of Technology, 4800 Oak Grove Dr., Pasadena, CA 91109, USA\\
$^{30}$ Department of Physics and Astronomy, University of Waterloo, 200 University Ave W, Waterloo, ON N2L 3G1, Canada\\
$^{31}$ Department of Astronomy, University of California, Berkeley,  501 Campbell Hall, Berkeley, CA 94720, USA\\
$^{32}$ University Observatory, Faculty of Physics, Ludwig-Maximilians-Universit\"at, Scheinerstr. 1, 81679 Munich, Germany\\
$^{33}$ School of Physics and Astronomy, Cardiff University, CF24 3AA, UK\\
$^{34}$ Department of Astronomy, University of Geneva, ch. d'\'Ecogia 16, CH-1290 Versoix, Switzerland\\
$^{35}$ Department of Physics, University of Arizona, Tucson, AZ 85721, USA\\
$^{36}$ Department of Applied Mathematics and Theoretical Physics, University of Cambridge, Cambridge CB3 0WA, UK\\
$^{37}$ SLAC National Accelerator Laboratory, Menlo Park, CA 94025, USA\\
$^{38}$ Center for Cosmology and Astro-Particle Physics, The Ohio State University, Columbus, OH 43210, USA\\
$^{39}$ Department of Physics, The Ohio State University, Columbus, OH 43210, USA\\
$^{40}$ Institute for Astronomy, University of Edinburgh, Edinburgh EH9 3HJ, UK\\
$^{41}$ \\
$^{42}$ Brookhaven National Laboratory, Bldg 510, Upton, NY 11973, USA\\
$^{43}$ Department of Physics, Duke University Durham, NC 27708, USA\\
$^{44}$ Cerro Tololo Inter-American Observatory, NSF's National Optical-Infrared Astronomy Research Laboratory, Casilla 603, La Serena, Chile\\
$^{45}$ CNRS, UMR 7095, Institut d'Astrophysique de Paris, F-75014, Paris, France\\
$^{46}$ Sorbonne Universit\'es, UPMC Univ Paris 06, UMR 7095, Institut d'Astrophysique de Paris, F-75014, Paris, France\\
$^{47}$ Department of Physics \& Astronomy, University College London, Gower Street, London, WC1E 6BT, UK\\
$^{48}$ Institut de F\'{\i}sica d'Altes Energies (IFAE), The Barcelona Institute of Science and Technology, Campus UAB, 08193 Bellaterra (Barcelona) Spain\\
$^{49}$ Physics Department, William Jewell College, Liberty, MO, 64068\\
$^{50}$ Jodrell Bank Center for Astrophysics, School of Physics and Astronomy, University of Manchester, Oxford Road, Manchester, M13 9PL, UK\\
$^{51}$ University of Nottingham, School of Physics and Astronomy, Nottingham NG7 2RD, UK\\
$^{52}$ Astronomy Unit, Department of Physics, University of Trieste, via Tiepolo 11, I-34131 Trieste, Italy\\
$^{53}$ INAF-Osservatorio Astronomico di Trieste, via G. B. Tiepolo 11, I-34143 Trieste, Italy\\
$^{54}$ Institute for Fundamental Physics of the Universe, Via Beirut 2, 34014 Trieste, Italy\\
$^{55}$ Hamburger Sternwarte, Universit\"{a}t Hamburg, Gojenbergsweg 112, 21029 Hamburg, Germany\\
$^{56}$ Department of Physics, IIT Hyderabad, Kandi, Telangana 502285, India\\
$^{57}$ Universit\'e Grenoble Alpes, CNRS, LPSC-IN2P3, 38000 Grenoble, France\\
$^{58}$ Institute of Theoretical Astrophysics, University of Oslo. P.O. Box 1029 Blindern, NO-0315 Oslo, Norway\\
$^{59}$ Instituto de Fisica Teorica UAM/CSIC, Universidad Autonoma de Madrid, 28049 Madrid, Spain\\
$^{60}$ School of Mathematics and Physics, University of Queensland,  Brisbane, QLD 4072, Australia\\
$^{61}$ Santa Cruz Institute for Particle Physics, Santa Cruz, CA 95064, USA\\
$^{62}$ Center for Astrophysics $\vert$ Harvard \& Smithsonian, 60 Garden Street, Cambridge, MA 02138, USA\\
$^{63}$ Australian Astronomical Optics, Macquarie University, North Ryde, NSW 2113, Australia\\
$^{64}$ Lowell Observatory, 1400 Mars Hill Rd, Flagstaff, AZ 86001, USA\\
$^{65}$ George P. and Cynthia Woods Mitchell Institute for Fundamental Physics and Astronomy, and Department of Physics and Astronomy, Texas A\&M University, College Station, TX 77843,  USA\\
$^{66}$ Instituci\'o Catalana de Recerca i Estudis Avan\c{c}ats, E-08010 Barcelona, Spain\\
$^{67}$ Observat\'orio Nacional, Rua Gal. Jos\'e Cristino 77, Rio de Janeiro, RJ - 20921-400, Brazil\\
$^{68}$ Department of Astrophysical Sciences, Princeton University, Peyton Hall, Princeton, NJ 08544, USA\\
$^{69}$ School of Physics and Astronomy, University of Southampton,  Southampton, SO17 1BJ, UK\\
$^{70}$ Computer Science and Mathematics Division, Oak Ridge National Laboratory, Oak Ridge, TN 37831\\
$^{71}$ Institute of Cosmology and Gravitation, University of Portsmouth, Portsmouth, PO1 3FX, UK

\bsp	
\label{lastpage}
\end{document}